\documentclass[12pt]{article} 

\usepackage{calc}
\usepackage{amsmath,amssymb,amsthm,amscd}
\numberwithin{equation}{section}
\usepackage[bf]{caption}
\usepackage{longtable}
\usepackage{array}
\usepackage{enumerate}
\usepackage{float}
\usepackage{subfig}
\usepackage{multicol} 
\usepackage{epsfig}
\usepackage{graphicx}
\usepackage{multirow}
\usepackage{color}
\usepackage{cite}
\usepackage[width=1.09\textwidth]{caption}
\usepackage{tikz}
\usepackage[hidelinks]{hyperref}%\usepackage{footmisc}
\usepackage{tikz,textcomp}
\usetikzlibrary{shapes,arrows,positioning,chains,calc}

\usepackage[utf8]{inputenc}
\usepackage[T1]{fontenc}
\usepackage[english]{babel}
\usepackage{xspace}
\usepackage{pifont}

\renewcommand{\tilde}{\widetilde}

\newcommand{\fsu}{\mathfrak{su}}
\newcommand{\fso}{\mathfrak{so}}

\newcommand{\fg}{\mathfrak{g}}
\newcommand{\ff}{\mathfrak{f}}
\newcommand{\fe}{\mathfrak{e}}

\newcommand{\beq}{\begin{equation}}
\newcommand{\eeq}{\end{equation}}
\newcommand{\bea}{\begin{eqnarray}}
\newcommand{\eea}{\end{eqnarray}}

 \usepackage[OT2,T1]{fontenc}
 \DeclareSymbolFont{cyrletters}{OT2}{wncyr}{m}{n}		%Tate Shafarevich group 1
 \DeclareMathSymbol{\Sha}{\mathalpha}{cyrletters}{"58}		%Tate Shafarevich group 2

\begin{document}

\baselineskip=15pt
\begin{titlepage}

\begin{center}
\vspace*{ 2.0cm}
{\Large {\bf Twisted Fibrations in M/F-theory}}\\[12pt]
\vspace{-0.1cm}
\bigskip
\bigskip 
{
{{Lara B.~Anderson}$^{\,\text{a }}$},  {{James~Gray}$^{\,\text{a }}$} and {{Paul-Konstantin~Oehlmann}$^{\,\text{b }}$}
\bigskip }\\[3pt]
\vspace{0.cm}
{\it 
 ${}^{\text{a}}$ Physics Department,~Robeson Hall,~Virginia Tech,~Blacksburg,~VA 24061,~USA   \\
  ${}^{\text{b}}$ Department of Physics \& Department of Mathematics, Northeastern University \\
360 Huntington Avenue, Boston, MA 02115, United States   
}
\\[2.0cm]
\end{center}

\begin{abstract}
\noindent In this work we investigate 5-dimensional theories obtained from M-theory on genus one fibered threefolds which exhibit twisted algebras in their fibers. We provide a base-independent algebraic description of the threefolds and compute light 5D BPS states charged under finite sub-algebras of the twisted algebras. We further construct the Jacobian fibrations that are associated to 6-dimensional F-theory lifts, where the twisted algebra is absent. These 6/5-dimensional theories are compared via twisted circle reductions of F-theory to M-theory. In the 5-dimensional theories we discuss several geometric transitions that connect twisted with untwisted fibrations. We present detailed discussions of $\fe_6^{(2)},\, \fso_8^{(3)}$ and $\fsu_3^{(2)}$ twisted fibers and provide several explicit example threefolds via toric constructions. Finally, limits are considered in which gravity is decoupled, including Little String Theories for which we match 2-group symmetries across twisted T-dual theories. 
\end{abstract}

\end{titlepage}
\clearpage
\setcounter{footnote}{0}
\setcounter{tocdepth}{2}
\tableofcontents
\clearpage

\section{Introduction}\label{sec:Intro}

String compactifications lead to a rich array of effective supergravity (SUGRA) and quantum (and sometimes conformal) field theories in 6- and 5- dimensions which have been the focus of much recent work \cite{Kumar:2010ru,Morrison:2012np,Heckman:2015bfa,Heckman:2018jxk,Apruzzi:2019opn,Apruzzi:2019enx,Bhardwaj:2019fzv,Bhardwaj:2018yhy,Bhardwaj:2018vuu,Bhardwaj:2019jtr,Bhardwaj:2020gyu}. The close relationship between the $5$/$6-$dimensional theories under a circle ($S^1$) reduction has yielded much information about the possible structure of both theories (see e.g. \cite{Bonetti:2011mw,Grimm:2015zea,Jefferson:2018irk}) and inspired many recent classifications. In this work, we are interested in a new example of such 6D/5D relationships as they arise in compactications of M/F-theory on elliptically and genus one fibered Calabi-Yau threefolds. In particular, we are interested in circle reductions of 6-dimensional F-theory effective theories which include flux along the circle and the possibility of \emph{twisted circle reductions}\cite{Kim:2019dqn,Bhardwaj:2019fzv,Braun:2021lzt,Kim:2021cua,Bhardwaj:2022ekc,Lee:2022uiq} (leading to different ranks in the 5-dimensional gauge theories compared to their 6-dimensional uplifts). It is important to note at this point that the phrase ``twisted reduction" has been used in multiple ways in the literature (sometimes indicating only circle reductions including flux/Wilson lines). In the context of this work we will \emph{only} use the term ``twisted reduction" to refer to reductions which change the rank of the gauge group. This will include an F-/M-theory realization of automorphisms that ``fold" affine Dynkin diagrams \cite{Baume:2017hxm,Oehlmann:2019ohh,Anderson:2019kmx} to produce twisted algebras and Calabi-Yau genus-one geometries with markedly different physics (including different rank gauge groups) to their Jacobian fibrations.

Let us begin by briefly reviewing a now well-established story \cite{Grimm:2010ez,Morrison:2012ei,Cvetic:2012xn} of F-theory circle reductions with \emph{Abelian gauge groups} and M-theory on genus one fibered Calabi-Yau manifolds. We will outline a few key ideas here and leave a more detailed review for Section \ref{sec:Circle_review}. For the purposes of this discussion it should be noted that a ``genus-one fibered" manifold is one whose generic fiber is a $T^2$ and which does not in general admit a \emph{section} to that fibration, but rather a multi-section (which intersects the generic fiber $n>1$ times). In contrast, a $T^2$-fibered manifold which does admit a rational section is referred to as an ``elliptically fibered" manifold. A 5-dimensional compactification of M-theory on a genus one fibered Calabi-Yau threefold $X$ is closely related to a 6-dimensional compactification of F-theory on a particular elliptically fibered Calabi-Yau threefold known as the Jacobian of $X$ and denoted $J(X/B_2)$. In mathematical terms, the Jacobian of a genus-one fibered CY manifold is an elliptically fibered manifold which shares the same base and discriminant locus to the fibration (and the same J-function). In general, for torus-fibered CY threefolds $J(X/B_2)$ and $X$ are topologically distinct and many different genus one fibered manifolds can share \emph{the same} Jacobian. 

The set of such geometries (together with information about the form of the birational mapping takes) form the group of CY Torsors \cite{Bhardwaj:2015oru,Hajouji:2020ddi}. In many cases this reduces to the Weil-Ch$\hat{\text{a}}$talet group $WC(X)$ \cite{dolgachev1993elliptic,Gross1997EllipticTI, Bhardwaj:2015oru, Anderson:2018heq, Anderson:2019kmx}.
The Weil-Ch$\hat{\text{a}}$talet group contains the Tate-Shafarevich (TS) group $\Sha(X)$ \cite{dolgachev1993elliptic,Gross1997EllipticTI} as a subgroup and the two coincide if the genus-one fibrations do not contain isolated multiple fibres. 
In such simple situations, the number of K\"ahler moduli among all TS elements have been observed to match \cite{Cvetic:2015moa,Mayrhofer:2014opa,Anderson:2014yva,Buchmuller:2017wpe}.
Multiple fibres over smooth points\footnote{For a discussion of multiple fibres over isolated singularities in $B_2$ see \cite{Anderson:2018heq,Anderson:2019kmx} and for smooth points see Appendix B of \cite{Oehlmann:2019ohh}.}
on the other hand appear to be closely linked to the twisted affine algebras that we construct in this work and hence in these cases the $WC(X)$ group does not simply reduce to the TS group $\Sha(X)$.

Returning to the F-/M-theory physics of these manifolds, a compactification of F-theory \emph{requires the existence of a section} in addition to a $T^2$ fiber \cite{Morrison:1996pp,Morrison:1996xf} and thus, we can consider defining F-theory on a given elliptically fibered CY threefold. However, this single compactification of F-theory can give rise to a number of distinct 5-dimensional compactifications of M-theory via circle reduction with Wilson lines (and possibly matter field vevs). As an example, if we consider a purely Abelian 6D effective theory arising from F-theory, then it is possible to choose a non-trivial $U(1)$-charged vev $\langle \phi_q \rangle\neq 0$ and discrete Wilson line (i.e. discrete circle flux) along the circle
\beq\label{flux1}
\xi \sim\int_{S^1} A ~.
\eeq 
As was demonstrated in \cite{Mayrhofer:2014opa,Cvetic:2015moa,Knapp:2021vkm}, different choices of $\xi$ lead to compactifications of M-theory on (possibly) distinct genus-one manifolds which are elements of the group of CY torsors (note that the reduction with $\xi=0$ leads to an M-theory compactification on the elliptically fibered manifold $J(X/B_2)$ itself). The set of these 5-dimensional theories are connected by geometric transitions in the underlying genus one fibered manifolds \cite{Morrison:2014era,Mayrhofer:2014opa,Cvetic:2015moa,Knapp:2021vkm}. This framework has been used to gain a deep understanding of discrete symmetries in F-/M-theory. In particular the associated Jacobian fibrations admit non-trivial torsion \cite{Mayrhofer:2014opa} and terminal singularities that do not admit a crepant resolution \cite{Arras:2016evy,Grassi:2018rva}. When compactified on one more circle to 4D, one obtains type IIA strings on the same geometry but it is possible to twist along this additional cycle, which allows to resolve the singularity via a non-commutative resolution \cite{Schimannek:2021pau,Katz:2022lyl,Katz:2023zan}.

In the present paper, we will be interested in compactifications of F-theory in 6-dimensions which include not only an Abelian factor, but also a non-Abelian gauge algebra (e.g. $\mathfrak{g} \times \mathbb{Z}_n$ for the Jacobian theory or $\mathfrak{g} \times \mathfrak{u}(1)$ if we move to an enhanced loci in moduli space to simply describe the dimensional reduction). In this case, the circle reductions can involve flux from gauge fields in the Cartan subalgebra of $\mathfrak{g}$ as well as the $U(1)$ factors. Importantly, a feature that is possible in this context is the presence of boundary conditions which can mix these different fluxes together in non-trivial ways:
\beq\label{twist_action}
A^B(y)=\sigma^B(A^C(y+2\pi))
\eeq
where $\sigma$ is associated to a discrete automorphism of the gauge group. In reductions of pure gauge theories the discrete symmetry action $\sigma$ is known to be an outer automorphism of the gauge group. As we will discuss in this work, in the context of F-theory (as a theory of supergravity) this discrete symmetry is expected to be gauged and to be locally realized in the Calabi-Yau geometry. This involves a more complex interplay between the gauge theory and local discrete automorphisms of the Calabi-Yau geometry. 

Non-trivial circle boundary conditions such as in \eqref{twist_action} (and fluxes) have been implemented in a variety of other contexts under the name of ``twisted" circle reductions (see e.g. \cite{Kim:2019dqn,Bhardwaj:2019fzv,Kim:2021cua,Duan:2021ges}) but have not been previously studied in the compact setting in F-theory. As we mention above, we will refer to reductions implementing non-trivial boundary conditions such a \eqref{twist_action} as ``twisted reductions". We find that in the presence of these boundary conditions we generate a new set of 5-dimensional vacua associated to M-theory compactifications on genus-one fibered manifolds with ``twisted" torus fibrations realizing twisted Lie algebras (e.g. $\fe_6^{(2)}$) in their fibers and non-Abelian gauge algebras of the form $\mathfrak{h}_i \times \mathfrak{u}(1)$ where $\mathfrak{h}_i \subseteq \mathfrak{g}$ is a sub-group of the 6D F-theory gauge algebra. Geometrically, these geometries are intriguing in that the genus one fibered manifold and its Jacobian can realize \emph{different} non-Abelian gauge groups, despite the fact that they share a discriminant locus. This arises through a discrete ``folding" of the affine Dynkin diagrams found in the elliptic/genus-one fibers. 

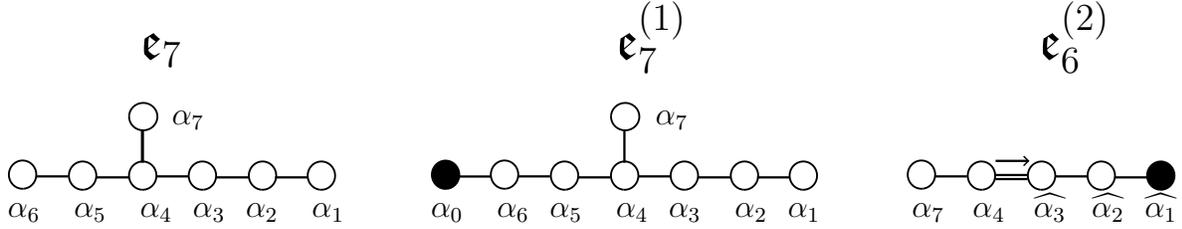
\begin{figure}[t!]
\begin{center}
\begin{picture}(-20,70)
\put(-190,60){\LARGE $\fe_7$}
\put(-10,60){\LARGE $\fe_7^{(1)}$}
\put(150,60){\LARGE $\fe_6^{(2)}$}

\put(-240,10){\includegraphics[scale=0.4]{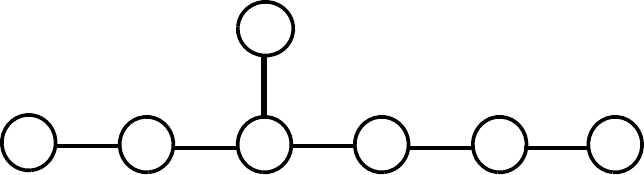}}
\put(-240,00){$\alpha_6$}
\put(-215,00){$\alpha_5$}
\put(-190,00){$\alpha_4$}
\put(-170,00){$\alpha_3$}
\put(-178,35){$\alpha_7$}
\put(-150,00){$\alpha_2$}
\put(-125,00){$\alpha_1$}
\put(-80,10){\includegraphics[scale=0.4]{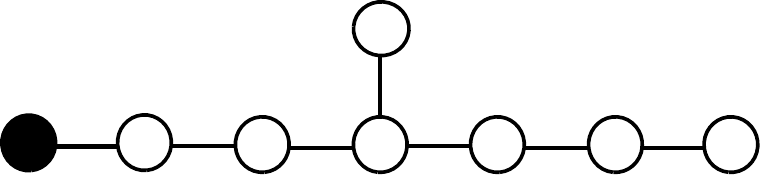}}
\put(-55,00){$\alpha_6$}
\put(-35,00){$\alpha_5$}
\put(-10,00){$\alpha_4$}
\put(10,00){$\alpha_3$}
\put(5,35){$\alpha_7$}
\put(35,00){$\alpha_2$}
\put(55,00){$\alpha_1$} 
\put(-80,00){$\alpha_0$}   
\put(100,10){\includegraphics[scale=0.4]{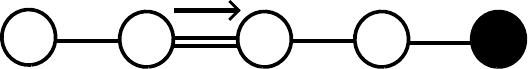}}
\put(102,00){$\alpha_7$}
\put(125,00){$\alpha_4$}
\put(148,00){$\widehat{\alpha_3}$}
\put(170,00){$\widehat{\alpha_2}$} 
\put(190,00){$\widehat{\alpha_1}$} 
\end{picture}
\caption{ \label{fig:E7E62Folding0} 
From left to right: Upon affinization, the $\fe_7$ Dynkin diagram obtains an $\mathbb{Z}_2$ outer automorphism that interchanges the affine and fundamental root $\alpha_0$ and $\alpha_1$. Quotienting by that automorphism, yields twisted affine $\fe_6^{(2)}$.
}
\end{center}
\end{figure}

In this work we will provide several examples of the geometric structure described above in which a genus-one fibered CY manifold and its Jacobian differ significantly in terms of their respective 5D/6D physics. As our central, illustrative example, we will study in detail the geometry and BPS spectra of a manifold whose fibers correspond to the twisted algebra $\fe_6^{(2)}$. The $\fe_6^{(2)}$ twisted algebra admits an $\ff_4$  finite sub-algebra, as observed from the Dynkin diagram in 
Figure~\ref{fig:E7E62Folding0}.
Upon unfolding one notices that the Dynkin diagram is obtained from an $\fe_7^{(1)}$ cover, twisted by its outer automorphism.
A key feature of this construction is that this folding symmetry only exists in the diagram upon affine extension but not the finite  $\fe_7$ gauge algebra alone. Correspondingly, if this is to be realized inside a compact CY fibration, the folding is incompatible with the existence of a single section, as that would mark the affine node $\alpha_0$  breaking the symmetry $\alpha_0 \leftrightarrow \alpha_1$. Instead, as we will review in detail below, the folded fibers appear in genus one geometries with a 2-section \cite{Bhardwaj:2019fzv}. 
As described in Section \ref{sec:main_5D_example}, such a 2-section comes with a special divisor $\overline{D}_{2}=d_8^2 + 4 d_5$
across which the two branch points are interchanged. This monodromy divisor plays a crucial role in realizing the folded fibers: 
To obtain the $\fe_6^{(2)}$ geometry, we require the $\fe_7$ divisor in the base to intersect $\overline{D}_2$ non-trivially.
As we will outline in detail in Section \ref{sec:main_5D_example} below, the singular genus-one fibration admits a quartic fiber presentation in $\mathbb{F}_2$, given as
\begin{align}\label{intro_g_one}
Z^2=&\,\, z^3  d_1   X^4 + z^3  d_2  X^3 Y + z^2  d_3  X^2 Y^2 + z d_4   X Y^3 \nonumber \\
  &\, \, \, \quad \qquad+d_5  Y^4 + d_6  z^2  X^2 Z + z d_7 X Y Z +  d_8  Y^2 Z \, ,
\end{align}
with what we call a type $III^{*}$ non-split fiber that can be resolved to  $\fe_6^{(2)}$. 

This fiber structure leads to several open questions regarding the 5D/6D effective physics which are explored in this work. The torus-fiber of the genus-one fibration indicates a 6D F-theory uplift. As the 6D F-theory geometry is simply the encoding of the monodromy data of the IIB axio-dilaton, this information can be obtained by mapping the singular genus-one model into the Jacobian fibration \cite{Klevers:2014bqa} which takes at leading order the form
\begin{align}
\begin{split}
\label{eq:discrE7U1}
f=z^3 \overline{D}_2 +\mathcal{O}( z^{4}) \, , \quad 
g=z^5 q + \mathcal{O}( z^{6}) \, , \quad 
\Delta=z^9 \overline{D}_2^3 + \mathcal{O}( z^{10}) \, .
\end{split}
\end{align}
Due to the presence of the section in the Jacobian, the Kodaira/Tate classification applies and we do not obtain an $\ff_4$ (e.g. via a $IV^{*,ns}$-singularity) but instead a type $III^*$ which yields an
  $\fe_7$ gauge algebra. Thus, we find an enhancement of the non-Abelian gauge symmetry in the 6D theory compared to that in 5D. This observation regarding the F-theory lift of such genus-one models differs from those previously studied in the literature (see e.g. \cite{Braun:2014oya}).
  In particular we find a difference of the K\"ahler moduli of the (maximally resolvable) Jacobian and the genus-one fibrations as also observed in \cite{Oehlmann:2016wsb,Baume:2017hxm,Oehlmann:2019ohh}. From the point of view of circle reductions of 6D gauge theories, this may come as a surprise, as ``twisted" dimensional reductions leading to $\fe_6^{(2)}$ fibers are realized in theories decoupled from gravity by twisted reductions of $\fe_6$ (see e.g. \cite{Bhardwaj:2019fzv,Lee:2022uiq}). In the present context, we find that the Jacobian symmetry instead reflects the geometric cover of $\fe_6^{(2)}$, that is $\fe_7$. This is further discussed in Section \ref{sec:Circle_reduc_spec}. This structure appears in each of our examples and applies also to the $\fsu_3^{(2)}$ and $\fso_8^{(3)}$ geometries that are explored in Section \ref{sec:MoreTwistedTheories}. 

  Further interesting differences between the 5/6D theories arise when considering the spectrum of light states in the two theories. Upon resolving the genus-one model, we study the geometry over an arbitrary two-fold base and compute the number of light 5D BPS states associated to the twisted fibration which are determined by the number of  $R=[z] \cdot \overline{D}_2$ ramification points  and the genus $g$ of the base curve. Computing those we find that our results are in full agreement with the multiplicities of 5D states that would arise from an $\fe_6$ twisted circle reduction (despite the fact that the apparent 6D gauge symmetry is $\fe_7$ as described above). We also discuss several conifold-type transitions of the threefold, and match it to 5D geometric transitions which change the fiber structure as $\fe_7^{(1)} \rightarrow \fe_6^{(2)}$. Field theoretically we realize this transition via motion along the Higgs/Coulomb branches of the theories. We test our constructions also in the context of little string theories, by matching higher group symmetries across T-dual pairs, as well as SCFTs.

The structure of this paper is as follows. In Section~\ref{sec:Circle_review} we give a brief introduction to circle compactifications in F-theory and M-theory on genus-one fibrations. In Section~\ref{sec:Circle_reduc_spec} we discuss twisted circle reductions of $\mathfrak{u}_1^2$ and $\mathfrak{e}_7 \times \mathfrak{u}_1 $ theories and a rich array of questions that arise in those contexts. From there we proceed to the discussion of 5D M-theory vacua and genus one fibrations with $\fe_6^{(2)}$ fibers in Section \ref{sec:main_5D_example}, where we describe the full geometry, compute BPS multiplicities and geometric transitions to untwisted theories. In Section~\ref{sec:MoreTwistedTheories} we then discuss further examples including $\fsu_3^{(2)}$ and $\fso_8^{(3)}$ twisted fibers.  In Section~\ref{sec:TwistedDuality} we employ the geometric constructions described above to discuss twisted T-dual little string theories and match their global symmetries. These considerations provide a non-trivial cross-check for our constructions. Finally, in Section~\ref{sec:TVSUT} we compare the moduli spaces of twisted and untwisted genus-one fibrations before we conclude in Section~\ref{sec:summary}. Appendix~\ref{app:Example3folds} and Appendix~\ref{app:LBData} are devoted to additional geometric data for some of the explicit examples presented in this work.  

\vspace{5pt} 
%%%%%%%%%%%%%%%%%%%%%%%%%%%%%%%%%%%%%%%%%%%%%%%%%
\section{Circle Reductions of F-theory to M-Theory}\label{sec:Circle_review}
In the following sections we provide a brief review of the relationships between 6/5 dimensional theories and the geometry of Calabi-Yau elliptic and genus-one fibrations.

\subsection{A brief review of twisted circle reductions in F-theory}
Following the analysis of \cite{Mayrhofer:2014laa,Cvetic:2015moa}, one can consider a compactification of a 6-dimensional $U(1)$ gauge theory on a circle of radius $R$. In particular, we consider the possibility of giving a field $\phi$ of charge $q$ an expectation value, possibly depending upon the circle direction, whose coordinate we denote by $y$. Denoting the five-dimensional coordinates by $x$ we can perform the usual Fourier expansion of the scalar field.
\begin{eqnarray} \label{fourier1}
\phi (x,y) = \sum_{n  \in \mathbb{Z}} \phi_n(x) e^{2 \pi i n y \tau} \, ,
\end{eqnarray}
with the inverse circle radius $\tau=\frac{1}{R}$. 
In addition to a scalar field vev, one could consider turning on a Wilson line associated to the $U(1)$ gauge field which can be denoted by $\xi = \int_{S^1} A$ as in \eqref{flux1}. In such a background, the mass term obtained for the $n$'th Kaluza-Klein (KK) mode $\phi_n$ is proportional to the following
\begin{eqnarray} \label{mass1}
m_n =|q \xi+n \tau|~~.
\end{eqnarray}
In this setting that there are several distinct possibilities for the nature of the compactified theory. The mode of $\phi$ which is given an expectation value must be massless according to (\ref{mass1}). For example, if we choose $\xi=-\tau k/q $ this is simply the mode $n=k$. One can then ask, if we give an expectation value to a KK mode $\phi_n$ then what symmetry is preserved? This is straightforward, but we write out the argument below here as we will soon need to repeat it for more complicated examples. The mode transforms under the symmetry as follows.
\begin{eqnarray}\label{transf_rule}
\phi_n \to e^{i(q \theta + n \psi)} \phi_n \, .
\end{eqnarray}
Here $\theta$ and $\psi$ are the parameters associated to $U(1)_{6D}$ and $U(1)_{KK}$ respectively. Given (\ref{transf_rule}) the expectation value is invariant if,
\begin{eqnarray} \label{solset}
q \theta + n \psi =  2 \pi \gamma\;,
\end{eqnarray}
where $\gamma \in \mathbb{Z}$. Clearly we will always have a $U(1)$ symmetry simply by taking $\gamma=0$ and $\theta= n/q \psi$. To complete the analysis we simply need to decide if any non-zero values of $\gamma$ are not identified with the $\gamma=0$ case.

The two gauge parameters are identified under
\begin{eqnarray} 
\theta \approx \theta + 2 \pi \alpha \;\; ,\;\;
\psi \approx \psi +2 \pi \beta \;.
\end{eqnarray}
Therefore,
\begin{eqnarray} \label{redef}
q \theta +n \psi \approx q\theta + n \psi + 2 \pi ( q \alpha + n \beta)\;.
\end{eqnarray}
Comparing (\ref{redef}) and (\ref{solset}), we see that values of $\gamma$ for which $\gamma =( q \alpha + n \beta)$ can be identified with zero. The question therefore becomes what is the smallest positive integer combination of $q$ and $n$. This is, of course $\textnormal{gcd}(q,n)$ and thus values of $\gamma$ can be identified under integer multiples of this greatest common divisor. Given this analysis, a dimensional reduction of this type results in a symmetry $U(1) \times \mathbb{Z}_{\textnormal{gcd}(q,n)}$.

As an example, we see from this analysis that if $q=3$ then one should obtain a $U(1)$ symmetry unless $n$ is an integer multiple of $3$ as noted in \cite{Cvetic:2015moa}. It should be noted that this result has implications of the structure of the group of Calabi-Yau Torsors under the standard F-theory framework of how such geometries are related to sets of genus one fibered geometries which share a Jacobian \cite{Braun:2014oya,Morrison:2014era,Mayrhofer:2014laa,Cvetic:2015moa,Bhardwaj:2015oru}.

\subsection{F/M-theory circle reductions and Calabi-Yau geometry}
\label{ssec:Genus1Review}
In this section we review the geometry and physics of genus-one fibrations in M-and F-theory on a compact threefold and its relation to discrete symmetries. Those properties can be deduced intrinsically from the threefolds \cite{Schimannek:2019ijf,Schimannek:2021pau,Dierigl:2022zll} however we will rather discuss them in terms of a Higgs transition following \cite{Mayrhofer:2014opa}.

In that we discuss with an elliptic threefold with a rank one MW group \cite{Grimm:2010ez,Cvetic:2012xn,Morrison:2012np} (also see the reviews \cite{Cvetic:2018bni,Weigand:2018rez} and reference therein). A well explored example is the Morrison-Park \cite{Morrison:2012np} model \cite{Morrison:2012ei} given as the elliptic curve inside of $BL_1 \mathbb{P}^2_{1,1,2}$, 
\begin{align}
\label{eq:MPmodel}
p= e_1^3 s_1 X^4 + e_1^2 s_2 X^3 Y + e_1 s_3 X^2 Y^2 + s_4 X Y^3 + 
 e_1^2 s_6 X^2 Z + e_1 s_7 X Y Z + s_8 Y^2 Z + e_1 s_9 Z^2 \, . 
\end{align} 
Here the $X,Y,Z,e_1$ can be thought of the analog of the Weierstrass coordinates, and the $s_i$ are some non-vanishing sections of the base that are specified in Appendix~\ref{app:LBData}. The fibral part of the Stanley-Reisner ideal (SRI)
is given by  
\begin{align}
\mathcal{SRI}: \{  X Z \, , e_1 Y  \} \,,
\end{align}
where both $X=0$ and $e_1=0$ are sections $S_0$ and $S_1$ respectively, that generate a MW group given by the Shioda map:
\begin{align}
\sigma(s_1) = [e_1]-[X] \, .
\end{align}
The form dual to this divisor leads to a U(1) gauge potential by an expansion of the M-theory three-form that also survives in the F-theory limit. Notably, this fibration admits two kinds of reducible curves over codimension two points in the base, where the fiber becomes of $I_2$ form. The first is over the toric locus $s_4=s_8=0$ where the fiber degenerates into $\mathbb{T}^2 \rightarrow C_1 + C_2$
and the other one over a non-toric ideal $I_{(2)}$ where the fiber becomes $\mathbb{T}^2 \rightarrow B_1 + B_2$.
The fiber structures and especially their intersections with the two set of sections are given as
 \begin{align} 
 \begin{array}{c|cc}
   & C_1 & C_2 \\ \hline
 S_0 & 1 & 0 \\
 S_1 & -1 & 2 \\ 
 \sigma(s_1)& -2 &2 \\ \hline
 vol & t_0 & 2 t_s \\ \hline
 \end{array}\, , \qquad  \begin{array}{c|cc}
   & B_1 & B_2 \\ \hline
 S_0 & 1 & 0 \\ 
 S_1 & 0 & 1 \\
 \sigma(s_1)& -1 &1 \\ \hline
 vol & t_0 + t_s & t_s \\ \hline 
 \end{array} \, .
 \end{align}
 The volumes of those effective curves were computed using the fibral part of the K\"ahler form (see \cite{Mayrhofer:2014laa}) given as
 \begin{align}
 J = t_0 S_0 + t_s (S_0 + S_1) \, .
 \end{align}
This geometry admits two conifold 
transitions that either branch towards a genus-one fibration or its Jacobian respectively. In order to do so, either of the two $\mathbb{P}^1$'s must be blown-down which yields a singularity which we can then deform. The two phases are topologically distinct geometries:
\begin{enumerate}
    \item $t_0\rightarrow 0$ yields the genus-one fibration (after deformation). Note that the residual fibers $B_{1/2}$ stay at finite volume.
    \item $t_s\rightarrow 0$ yields the Jacobian fibration (after deformation). The curve $B_2$ shrinks to zero size, which results in an $I_2$ singularity that can not be (crepantly) resolved. 
\end{enumerate}
In the genus-one fibration  we   deform   the singular curves $C_1$ by adding the terms $s_5 Y^4$ to obtain 
\begin{align}
\label{eq:GenericQuartic}
p=s_1   X^4 + s_2  X^3 Y + s_3  X^2 Y^2 + s_4   X Y^3 + 
 s_5  Y^4 + s_6   X^2 Z + s_7 X Y Z + s_8  Y^2 Z +   s_9 Z^2 \, ,
\end{align}
which yields the genus-one fibration. Shrinking $C_2$ preserves the section and yields the Jacobian fibration. From a 6D perspective, the resulting F-theory lift yields a discrete $\mathbb{Z}_2$ symmetry. This is clear once from observing that we Higgs on a non-minimal $U(1)$ charged state but also due to the fact the all elements in the TS group $\Sha(X)$ must yield the same 6D F-theory and thus a (discrete) gauge symmetry.  
 
In the absence of a section we can at most define an N-section, which refers to a divisor 
  $S_0^N$ that intersects the generic torus $N$ times
\begin{align}
    S_0^N \cdot [T^2]=N \, .
\end{align}
Those N-points make sense only as a collection, as they are permuted along monodromies in the base along the \textit{Monodromy divisor}   $\overline{D}_{N,S_0}$. This divisor can be computed from the discriminant locus of $S_0^N$. For the quartic the relevant monodromy divisor is given as
\begin{align} 
    \overline{D}_{2,X}=  s_8^2 - 4 s_9 s_5 \, .
\end{align}  
While the different elements of $\Sha(X)$ uplift to the same 6D F-theory, their 5D M-theory physics differs. As described in the previous section, since M-theory is related to F-theory by a circle compactification, the differences in 5D effective theory arise from a choice of a circle flux $\xi$ in the 6D gauge group that for generic values breaks reduces to the maximal Cartan subalgera . In the context of the geometries above, these choices can be seen by following the Higgsing in 5D and noting that M2 branes can wrap the curves $C_i$ which, on a generic point on the Coulomb branch (CB) obtain the masses 
 \begin{align}
        m_{w,n}=|w^I \xi_I + q  \xi+ n \tau    |  \, .
        \end{align}
        whereas $w^I$ denotes a weight of the 6D representation $\mathbf{R}_q$. For a singlet state under the non-Abelian gauge algebra, $\mathbf{1}_N$, the choice of circle flux $\xi=-\frac{n}{N}$ from the perspective of the 6D theory allows for massless states in the 5D theory that enable the geometric/Higgsing transitions described above. The choice of $\xi=0$ corresponds to the Jacobian geometry.
        
From the point of view of the circle reduction, the unbroken $U(1)_E$ in 5D is a linear combination of both $U(1)_{6D}$ and $U(1)_{KK}$ symmetries given as
\begin{align}
\begin{array}{|c|c|}\hline
\text{flux } & \text{5D symmetry} \\ \hline
\xi  = -\frac{n}{N}  & U(1)_{E} =  n \cdot U(1)_{6D}-N \cdot U(1)_{KK} \\ 
\xi  = 0 & \mathbb{Z}_N \times U(1)_{KK} \\ \hline
\end{array} \, 
\end{align}
(where we have taken gcd($q,n)=1$ here for simplicity). Also note that the Jacobian vacuum with $\xi =0$ preserves the discrete gauge algebra factor in 
5D\footnote{See \cite{Mayrhofer:2014opa,Schimannek:2021pau} for a discussion of the torsion cycles in the Jacobian that yield the discrete symmetry. 
}
The  choice with a discrete holonomy will be relevant in the next section in conjunction with additional non-trivial 6d gauge symmetries to lead to twisted affine type of fibers.
 
 Note that at the origin of the Coulomb branch in the Jacobian, with $\xi_{Disc}=0$ there are
KK zero-modes which yield massless 5D states. Note also that discrete charged singlets stay massless on a generic point of the Coulomb branch moduli space.
These states correspond to terminal $I_2$ fibral singularities, that do not admit any crepant resolution \cite{Arras:2016evy,Grassi:2018rva}. 
 In the genus-one fibration on the other hand every massless 6D state with representation $\mathbf{R}_q$ and $\mathbb{Z}_N$ remnant charge $q  \mod N$  admits shifted 5D mass
\begin{align}
m_{w,n}=|w^I \xi_I +   (n - q \frac{n_0}{N})\tau  |  \, .
\end{align}
The total contribution $q_E=(n-q \frac{n}{N})$ is identified with the charge\footnote{The  $q$ mod $N$ identification of the 6D discrete charges is resembled by the integer shift symmetry of the KK charges $n$ in 5D.} under $U(1)_E$ which is the shifted $U(1)_{KK}$. The $U(1)_E$ charges can be computed geometrically  via the discrete Shioda map\footnote{
Note that the $U(1)_{KK}$-tower is still integer spaced,
while we will mostly use a $U(1)_E$ charge normalization for the N-section such that KK-charges have $N-$integer spacing.}. Note that states with weight $w \in \mathbf{R}$ and non-trivial $U(1)_E$ charges stay massive at the origin of the Coulomb branch $\xi_I=0$.    

Whenever $\mathfrak{g}$ is an untwisted algebra the codimension one and two singularity structure in either elements of the TS group $\Sha(X)$ are the same, as observed in all known examples
\cite{Klevers:2014bqa,Baume:2017hxm,Buchmuller:2017wpe,Schimannek:2019ijf}, with the only exception that singularities are terminal in the Jacobian. Hence 
for computational purposes it is often times more convenient to use the genus-one fibration to compute states and charges, as the geometry can be fully resolved\footnote{As shown in \cite{Schimannek:2021pau,Katz:2022lyl}, terminal singularities
can be resolved with non-K\"ahler non-commutative resolutions in the type IIA compactifications. Using those resolutions, discrete charged Gopakumar-Vafa invariants have been proposed \cite{Schimannek:2021pau}, that allow to compute charges and states in the Jacobian directly.
}. These states then become massless in the 6D F-theory uplift and therefore must satisfy the 6D (SUGRA) anomaly cancellation conditions. \\
 
When moving away from untwisted algebras to twisted ones however, the simple relationship between states no longer holds.
In the following we discuss in detail that these geometries admit multiple fibers and hence, we expect that the Weil-Ch$\hat{\text{a}}$talet group $WC(X)$ will not reduce to the Tate-Shafarevich group $\Sha(X)$ in these cases.

\section{Twisted Dimensional Reductions in F-theory?}\label{sec:Circle_reduc_spec}
The geometries described in Section \ref{sec:Intro} highlight an intriguing possibility, namely that circle reductions of F-theory to M-theory may \emph{dramatically change the nature of non-Abelian factors in the gauge group} in a way that cannot be explained by standard circle fluxes alone. The natural question then arises: \emph{what is the physical mechanism that accomplishes this change?}

At least a partial answer seems to lie within the study of \emph{twisted circle reductions} of gauge theories which have appeared in a wide range of contexts (see e.g. \cite{Tachikawa:2011ch,Kim:2019dqn,Duan:2021ges,Kim:2021cua,Bhardwaj:2019fzv,Lee:2022uiq}). In brief, if we begin with a theory with gauge symmetry $G$, in $n$-dimensions and consider the dimension reduction of this theory on an $S^1$ to produce an $(n-1)$-dimensional theory, it is natural to consider the possibility of non-trivial boundary conditions on $S^1$ for the gauge theory. Denoting the coordinate along the $S^1$ as $y$, the fields of the theory could experience 'twisting' conditions around the circle. For example the gauge potentials could obey a schematic relationship of the form
\beq\label{sigma}
A^B(y)=\sigma^B(A^C(y+2\pi))
\eeq
where $\sigma$ is some discrete action mixing components of the gauge field. For pure gauge theories, the possible discrete actions $\sigma$ are well understood \cite{Bhardwaj:2019fzv}. In particular, $\sigma$ must be compatible with gauge transformations of the theory but not trivializable under them, leaving the \emph{group of outer automorphisms} of $G$ as a natural choice.

\begin{figure}
    \centering
    \begin{picture}(0,100)
    \put(-100,20){\includegraphics[scale=0.4]{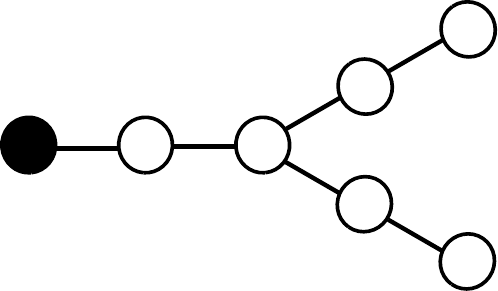}}
    \put(50,40){\includegraphics[scale=0.4]{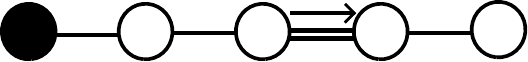}}
\put(-100,30){$\alpha_0$}
\put(-15,10){$\alpha_1$}
\put(-15,55){$\alpha_5$}
\put(-35,45){$\alpha_4$}
\put(-35,20){$\alpha_2$}
\put(-55,30){$\alpha_3$}
\put(-75,30){$\alpha_6$}

\put(50,30){$\alpha_0$}
\put(74,30){$\alpha_4$}
\put(95,30){$\alpha_3$}
\put(118,30){$\widehat{\alpha_2}$} 
\put(140,30){$\widehat{\alpha_1}$} 
    
    \end{picture}
    \caption{\textit{Depiction of the Dynkin diagrams of $\fe_6^{(1)}$ and $\ff_4^{(1)}$, where the affine node is highlighted in black. The later is can be obtained by folding by the $\mathbb{Z}_2$ automorphism of $\fe_6$. Node that $\ff_4^{(1)}$ is Langland dual to $\fe_6^{(2)}$ given in Figure~\ref{fig:E7E62Folding0}.
    }}
    \label{fig:E61-F41Dynkin}
\end{figure}
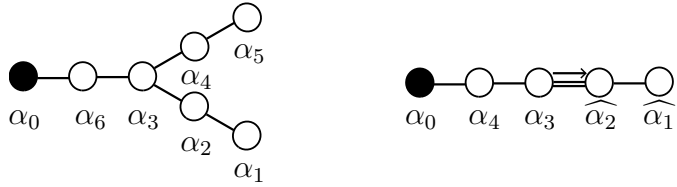

Importantly for our present purposes, for simple Lie algebras the group of outer automorphisms are exactly the group of graph automorphisms of the associated Dynkin diagram. This means that there is a natural identification of nodes of the Dynkin diagrams that can ``fold" the diagram in way a that feels similar in spirit to the relationship between fibers of the twisted genus one fibered manifold $X$ and it's Jacobian, $J(X/B_2)$ sketched in Section \ref{sec:Intro} and explored in detail in Section \ref{sec:main_5D_example}. For example, a natural $\mathbb{Z}_2$ automorphism of $\fe_6$ can fold the group to its $\ff_4$ subgroup (see Figure \ref{fig:E61-F41Dynkin}). Physically, then this symmetry in \eqref{sigma} would lead to a breaking of $\fe_6$ symmetry to $\ff_4$ under a circle reduction.

Within gauge theories, the mechanism above is well understood and frequently utilized. However in theories coupled to supergravity new subtleties arise. The most important of these is that it is believed that discrete symmetries such as those appearing in \eqref{sigma} must be \emph{gauged} in the full supergravity theory. Thus, one must ask how the local discrete actions such as $\sigma$ in \eqref{sigma} are realized explicitly in a given string/M-/F-theory context?

An interesting example of this question arises in the CHL compactifications of the heterotic string \cite{Chaudhuri:1995fk}. Beginning with the $E_8 \times E_8$ heterotic theory in $10$-dimensions it is possible to perform a circle reduction to obtain a $9$-dimensional theory with, for example, $E_8$ symmetry. In this case the discrete action \eqref{sigma} is simply the $\mathbb{Z}_2$ outer automorphism of the product group $E_8 \times E_8$ which interchanges the two $E_8$ factors, leaving a single $E_8$ gauge group in $9$-dimensions. The question of how this apparent global symmetry is gauged might be somewhat opaque within the $10$-dimensional heterotic theory. However the local nature is clearly visible from the viewpoint of $11$-dimensional heterotic M-theory \cite{Horava:1996ma} in which this $\mathbb{Z}_2$ symmetry is realized as a discrete automorphism of the $S^1/\mathbb{Z}_2$ geometry itself which interchanges the two $E_8$-fixed planes (i.e. the 9-dimensional theory is found from compactification on a Mobius strip \cite{Aharony:2007du}). As an important final note on this example, it is useful to observe that this symmetry \emph{only appears at special/tuned (i.e. higher co-dimensional) loci within the moduli space} of the heterotic compactification (or heterotic M-theory). At generic points in the moduli space the two $E_8$ bundles are not symmetric. It is only by moving to a special locus where the gauge connections, etc are identical that the symmetry becomes manifest. We will return to this important point in other contexts below.

With these general comments in hand we come now to the question of how can such twisted dimensional reductions be realized in F-theory? There are several immediate observations that can be made -- which in turn lead to questions about the nature of any twisting symmetries akin to \eqref{sigma} in this context. The first of these is that since we expect this discrete action to be gauged in a compact geometry, it is natural to expect that the symmetry must be realized in some way in the Calabi-Yau geometry of F-theory itself (indeed, in a sense F-theory is just a geometric encoding of field theory information in Calabi-Yau geometry). But that raises the question of \emph{how can a twisting symmetry act on the elliptically fibered geometry?}

To answer this question it is useful to first recall how diagrams such as in Figure~\ref{fig:E61-F41Dynkin} are realized in F-theory. To work directly with the Calabi-Yau geometry it is helpful to consider the resolved limit as guide, despite the fact that the physical theory is determined by (singular) Weierstrass models (i.e. thinking of F-theory as a limit of M-theory). In the resolved CY geometry, the Dynkin diagrams of gauge groups are realized geometrically (via blow-ups of degenerate elliptic fibers) as diagrams of \emph{affine} groups. This affine diagram is reduced to that of the actual gauge group by the inclusion of a section to the elliptic fibration which intersects one component of the fiber. In the F-theory limit, all components of the fiber that do not intersect the section (i.e. all fibral divisors) are taken to zero size.

In this context we might expect that a twisting symmetry should be visible in the F-theory geometry itself as a discrete automorphism. But at this point several immediate questions arise. The first puzzle is that it is the Dynkin diagrams of affine Lie groups rather than their finite counterparts that arise in the explicit (resolved) elliptically fibered Calabi-Yau geometry. Thus, we must ask how is the discrete action in \eqref{sigma} uplifted to the full elliptically fibered geometry? And how does such a symmetry interact with the \emph{section} (or sections) to the elliptic fibration? 

It is clear that any discrete symmetry which ``folds" a Dynkin diagram such as in Figure \ref{fig:E61-F41Dynkin} realized in an elliptic fibration must act compatibly with section(s) of the fibration. In Figure \ref{fig:E61-F41Dynkin} this is straightforward as the affine node "spectates" through the folding, but for some actions on affine diagrams this is not so simple. For example, the $\mathbb{Z}_2$ action which folds $\fe_7^{(1)}$ to $\fe_6^{(2)}$ identifies the affine node (intersected by the zero-section in an F-theory fiber) with another node as in Figure \ref{fig:E7E62Folding0}. Here the presence of a section intersecting the fibers could generically break the symmetry. Such a symmetry could only be restored if the fibration admitted more than one rational section (which could intersect the diagram in an appropriately symmetric way). It is exactly this latter configuration that we find explicitly realized in the Jacobians of CY genus-one fibrations studied here. 

Summarizing the observations above: generic features of the examples we find explicitly realized in (resolved) elliptic Calabi-Yau geometries indicate that
\begin{itemize}
\item The twisting symmetries are expected to ``lift" to discrete automorphims (i.e. foldings) of the affine Dynkin diagrams realized in the elliptic fibers.
\item In order for symmetries to exist which can serve as twisted boundary conditions for the actual (finite) gauge groups, the Mordell-Weil group of the geometry is frequently required to be non-trivial, i.e. there must be more than one rational section. The presence of these sections can "restore" folding symmetries to the fibers that would not otherwise be present with a single rational section.
\end{itemize}
Due to the geometric relationships between CY genus-one fibrations and their Jacobians, we do not find the standard twisted boundary conditions that arise in gauge theory (for example the $\fe_6 \to \ff_4$ folding in Figure \ref{fig:E61-F41Dynkin}, though they may exist). Instead, in the explicit CY geometry we find that only foldings such as those in Figure \ref{fig:E7E62Folding0} can be realized. From the geometric perspective, the appearance of $\fe_7$ instead of $\fe_6$ in the Jacobian is perhaps more expected, as its affine extension is the geometric cover for $\fe_6^{(2)}$ fibers. 
These foldings arise in the context of more complicated geometries with higher rank Mordell-Weil groups and discrete symmetries. As a result, twisted dimensional reductions appear to be part of the rich geometric structure linking genus one fibered Calabi-Yau manifolds and their Jacobian manifolds as described in Section \ref{sec:Intro} and \ref{sec:Circle_review}.

Finally, similarly to the example of the CHL string mentioned above, it is clear that we may need to tune the F-theory geometry to a special locus in complex structure moduli space to make any possible discrete twisting automorphisms visible. Here we expect that these tunings may be more difficult to find/engineer explicitly depending on the interplay between the discrete symmetries in \eqref{sigma} and the full F-theory gauge group (including non-Abelian and Abelian factors as described by reducible elliptic fibers in the resolved geometry and a higher rank Mordell-Weil group \cite{Morrison:1996pp}).

\subsection{A twisted circle reduction of a $U(1) \times U(1)$ theory}\label{sec:u1u1}
In the sections below we will begin by outlining a twisted dimensional reduction in which the twisting symmetry is made completely manifest in the geometry by restricting ourselves to a highly symmetric point in the complex structure moduli space of the elliptic Calabi-Yau variety. In later sections we will postulate that for many examples milder tunings are possible and can still lead to twisting symmetries involving interchange of rational sections (similar to the freedom studied in \cite{Grimm:2015wda}).

\subsubsection{The Calabi-Yau Geometry}
Here we briefly review the geometry of a singular elliptic CY variety which leads to an F-theory background with Abelian gauge symmetry $U(1) \times U(1)$.

To begin we recall the logic taken by Morrison and Taylor in \cite{Morrison:2014era} in characterizing a geometry with two sections to its elliptic fibration (i.e. Mordell-Weil rank 1) and hence a $U(1)$ gauge symmetry in F-theory. As a starting point, one may consider a genus one fibration defined in an ambient $\mathbb{P}^{1,1,2}_{X,Y,Z}$, defined by \eqref{eq:GenericQuartic}. Fixing $s_9$ as a constant, redefining and shifting the coordinates $X,Y,Z$ and classes $s_i$, we can obtain the reduced form of the quartic
\begin{align}
Z^2 = s_1 X^4 + s_2 X^3 Y + s_3 X^2 Y^2 + s_4 X Y^3 + s_5 Y^4 \, .
\end{align} 
The $s_i$ are sections of line bundles in the base read off from Appendix~\ref{app:LBData} as
\begin{align}
[s_1]\sim  2 (c_1-\mathcal{S}_9)\, , \quad [s_2] \sim 2 c_1 - \mathcal{S}_9\, ,\quad  s_3 \sim 2 c_1\, , \quad [s_4] \sim 2 c_1 + \mathcal{S}_9\, , \quad   [s_5] \sim 2 (c_1 + \mathcal{S}_9)\, ,
\end{align}
with $\mathcal{S}_9$ some line bundle class\footnote{The line bundle $\mathcal{S}_9$ is only restricted by the conditions that all $[s_i]$ are all effective sections.} defined in the base $B_2$ of the genus one fibration $\pi: X_{CY} \to B_2$ and $c_1$ the anti-cannonical class of $B_2$. This genus one fibration degenerates over higher codimensional loci in the base (in particular the $I_2$ loci) where the discrete singlets $\mathbf{1}_1$ are located (see \cite{Morrison:2014era} for an explicit discussion).

The Jacobian of this geometry, is a hypersurface in $\mathbb{P}^{2,3,1}_{X,Y,Z}$ given as 
\begin{align}
y^2 = x^3 - 
 s_3 x^2 z^2 + (s_2 s_4 - 4 s_1 s_5) x z^4 - (s_2^2 s_5 + s_1 s_4^2 - 
    4 s_1 s_3 s_5) z^6
\end{align}
from which we can read off the Weierstrass coefficients  
\begin{align}
f=&(s_2 s_4 - \frac13 s_3^2 - 4 s_1 s_5) \\
g=&(-s_1 s_4^2 + \frac13 s_2 s_3 s_4 - \frac{2}{27} s_3^3 + \frac83 s_1 s_3 s_5 - s_2^2 s_5) ~.
\end{align}
It should be noted that at the singlet locus, the discriminant becomes singular (and the genus-one fibration reducible).

The geometry given above can be U(1) enhanced by tuning in a section.
There are actually two inequivalent ways of doing so. We call the enhancement $U(1)_a$ and $U(1)_b$ that are given by setting
\begin{align}
\label{eq:symU1tuning}
U(1)_a: s_1 \rightarrow d_1^2/4 \, \qquad U(1)_b : s_5\rightarrow d_5^2/4 \, .
\end{align}
The above factorizations make sense, as $s_1$ and $s_5$ are both elements of even classes in the base and therefore $d_1$ and $d_5$ are integral. We can directly see that the above factorization yields another section in the genus one geometry e.g. in the first enhancement as
\begin{align}
\frac14 (d_1 X^2 - 2 Z) (d_1 X^2 + 2 Z) =  s_2 X^3 Y + s_3 X^2 Y^2 + s_4 X Y^3 + s_5 Y^4
\end{align}
where we have two sections at $Z= \pm \frac12 d_1 X^2$ \, . So now we have a section and thus, this quartic equation is in birational to the Weierstrass model. The same factorization above happens for the $U(1)_b$.

Finally, a $U(1)_a \times U(1)_b$ model (our goal in this section) is then given as
\begin{align}\label{twou1model}
Z^2= \frac14(d_1^2 X^4) + s_2 X^3 Y + s_3 X^2 Y^2 + s_4 X Y^3 + \frac14(d_5^2 Y^4) \, .
\end{align}
The above geometry admits a symmetry under the exchange 
\begin{align}
d_1 \leftrightarrow d_5 \, \qquad s_2 \leftrightarrow s_4 \qquad X \leftrightarrow Y 
\end{align}
while keeping $s_3$ and $Z$ fixed. This is only possible for the line bundle choices $\mathcal{S}_9=0$ (but of course for many choices of base manifold, $B_2$, in particular for every weak Fano base). As we will see in future sections this type of tuned Weierstrass model (and its symmetries) is similar in spirit to the Jacobians varieties of genus-one fibered manifolds with twisted fibers studied in this work.

\subsubsection{Field Theory: Twisted Dimensional Reduction}\label{sec:u1u1_fieldtheory}
With the geometry above in hand as a background for F-theory, we can consider the compactification of the resulting six-dimensional $U(1) \times U(1)$ gauge theory on a circle. We will consider giving a vev to a scalar field $\phi_{q_1,q_2}$ charged under both $U(1)$ factors and including Wilson lines $\xi_1$ and $\xi_2$ in both gauge factors. The Fourier expansion of the scalar field, and the mass term for the associated modes are then very similar to (\ref{fourier1}) and (\ref{mass1}) above.
\begin{eqnarray} \label{fourierandmass2}
\phi_{q_1,q_2} (x,y) &=& \sum_{n \in \mathbb{Z}} \phi_{q_1,q_2,n}(x) e^{2 \pi i n y \tau} \\ \nonumber
m_n &=& | n \tau +q_1 \xi_1+q_2 \xi_2|
\end{eqnarray}

For simplicity let us consider the special case where one gives an expectation value to a field where $q_1=q_2=q$ which we will denote simply by $\phi$. We will also specialize to the case where $\xi_1=\xi_2 = \xi$. In such an instance, the mass in (\ref{fourierandmass2}) simplifies and we can isolate a special case of backgrounds.

The analysis of the unbroken five-dimensional gauge group in this case mirrors the situation of a single six-dimensional $U(1)$. Taking $\xi= -k/2 q$ we find from (\ref{fourierandmass2}) that the massless mode is $n=k$. This mode transforms under the symmetries as 
\begin{eqnarray} \label{u1u1mode}
\phi_n \to e^{i\left((\theta_1+\theta_2)q + \psi n \right)} \phi_n
\end{eqnarray}
where $\theta_1$ and $\theta_2$ are the parameters of the two six-dimensional $U(1)$'s and $\psi$ is that of the KK gauge factor. The expectation value of $\phi_n$ is therefore invariant if the following condition holds for some $\gamma \in \mathbb{Z}$.
\begin{eqnarray} \label{cond2}
q(\theta_1 + \theta_2) + n \psi = 2 \pi \gamma
\end{eqnarray}
Clearly, we will always have a $U(1) \times U(1)$ symmetry simply by taking $\gamma=0$ and $\psi= -q/n(\theta_1 + \theta_2)$. As before, to complete the analysis we simply need to decide if any non-zero values of $\gamma$ are not identified with the $\gamma=0$ case (and so are not part of the continuous symmetry group already identified).

The gauge parameters are identified under
\begin{eqnarray}
\theta_1 \approx \theta_1 + 2\pi \alpha_1 \;\;,\;\; \theta_2 \approx \theta_2 + 2 \pi \alpha_2 \;\;,\;\; \psi \approx \psi+ 2\pi \beta \;.
\end{eqnarray}
Therefore we have the following identification.
\begin{eqnarray} \label{shift2}
q(\theta_1 + \theta_2) + n \psi \approx q(\theta_1 + \theta_2) + n \psi + 2\pi(q(\alpha_1 + \alpha_2)+n \beta)
\end{eqnarray}
Comparing (\ref{shift2}) and (\ref{cond2}), we see that values of $\gamma$ for which $\gamma=q(\alpha_1 + \alpha_2)+n \beta$ for some $\alpha_1$, $\alpha_2$ and $\beta$ can be identified with zero. Since $\alpha_1 +\alpha_2$ is still just some general integer, as $\alpha$ was in the proceeding analysis, the answer is unchanged. The full gauge group seen in five-dimensions is thus $U(1) \times U(1) \times \mathbb{Z}_{\textnormal{gcd}(q,n)}$.

Where this simple example of a $U(1) \times U(1)$ reduction differs from the analysis with a single abelian factor is in the possibility of there being additional symmetries of the system which can be used as boundary conditions in the circle reduction. In the proposal we present in this paper this symmetry is an emergent one, only appearing on special loci in the moduli space of the Calabi-Yau manifold (i.e. the tuning given in \eqref{eq:symU1tuning} for this example). For those choices of complex structure there exists a  $\mathbb{Z}_2$ symmetry which exchanges the two $U(1)$ factors in six-dimensions as described in the previous subsection. If we use such a symmetry as a boundary condition in traveling once around the $S^1$ upon which we are compactifying, then the analysis of the nature of the five-dimensional theory changes.

A symmetry of the type described in the previous paragraph leaves fields of the form $\phi_{q,q}$ invariant. Therefore the mode expansions (\ref{u1u1mode}) is unchanged in this case. The symmetry does identify $\theta_1$ and $\theta_2$ $\textnormal{mod} \; 2\pi$ at the point where the boundary condition is applied, however, and if we assume that these have no $y$ dependence, then they are set equal $\textnormal{mod}\; 2 \pi$ everywhere. We are thus left with a $U(1) \times \mathbb{Z}_{\textnormal{gcd}(q,n)}$ theory in five-dimensions.

\vspace{0.2cm}

It is useful to also briefly review the fate of the particle spectrum on dimensional reduction. From fields of type $\phi_{Q,Q}$ in six-dimensions which were not those which for which there was an expectation value we obtain degrees of freedom of charge $2Q$ under the low energy $U(1)$. From the field of this type which we gave a vev, we of course obtain an uncharged degree of freedom. In order for the symmetry which we have used as a boundary condition to exist, fields of type $\phi_{q_1,q_2}$ must always be paired with fields of type $\phi_{q_2,q_1}$ in the six-dimensional spectrum. This is automatic when starting with six-dimensional theories of course, but the entire discussion here is essentially unchanged in compactifying from four-dimensions where this requirement becomes non-trivial. The symmetry identifies the fields in these pairs, and as such, each pair gives rise to a single field of charge $q_1+q_2$ in five-dimensions.

\vspace{0.3cm}

As a final comment, it should be noted that we could give an expectation value to a pair of fields of the form $\phi_{q_1,q_2}$ and $\phi_{q_2,q_1}$ instead of the case of equal charges under the two six-dimensional $U(1)$'s described above. In such a case, the mode expansion (\ref{u1u1mode}) is more complicated as the boundary condition transforms one field into the other as the $S^1$ is traversed. We have the following.
\begin{eqnarray}
\phi_{q_1,q_2} &=& \sum_{n \in \mathbb{Z}} \phi_n(x) e^{\pi i n y\tau} \, , \\ \nonumber
\phi_{q_2,q_1} &=& \sum_{n \in \mathbb{Z}} \phi_n(x) e^{\pi i n (y+1)\tau}\, .
\end{eqnarray}
The mass of the KK mode $\phi_n$ is then given by the following expression.
\begin{eqnarray}
m_n = \left|\frac{n}{2} \tau  +q_1 \xi + q_2 \xi \right| \, .
\end{eqnarray}
By choosing $\xi = -k/(2( q_1+q_2))\tau$ we can therefore ensure that the mode $n=k$ is massless. Under the symmetries present in the system, this mode transforms as follows.
\begin{eqnarray}
\phi_n \to \phi_n e^{i(q_1 \theta_1 + q_2 \theta_2 + \frac{n}{2} \psi)} \, .
\end{eqnarray}
Thus the condition that we have an unbroken symmetry is simply,
\begin{eqnarray}
q_1 \theta_1 + q_2 \theta_2 + \frac{n}{2} \psi = 2 \pi \gamma\; \, ,
\end{eqnarray}
for $\gamma \in \mathbb{Z}$. Since the symmetry operation equates $\theta_1=\theta_2 \;\; \textnormal{mod}\;\; 2\pi$ the $\gamma=0$ possibility here leads to a $U(1)$ symmetry in the five-dimensional theory. The gauge parameters are identified under $2\pi$ shifts as in (\ref{shift2}), and as such, while non-vanishing values for $\gamma$ are potential elements of a discrete symmetry group, they will be identified with a vanishing transformation if the following condition holds.
\begin{eqnarray} \label{lastcond}
2 \alpha_1 q_1 + 2\alpha_2 q_2+\beta n = 2 \gamma
\end{eqnarray}
The integer on the left hand side of this expression is always a multiple of
\begin{eqnarray}
\gcd(2q_1+2q_2+n) = \gcd(\gcd(2 q_1+2q_2),n) = \gcd(2\gcd(q_1,q_2),n)\;.
\end{eqnarray}
Therefore (\ref{lastcond}) has a solution iff $\gamma$ is a multiple of $\textnormal{lcm}\left(\gcd(2\gcd(q_1,q_2),n),2\right)/2$.  Finally, then we arrive at a symmetry group for the compactified theory of,
\begin{eqnarray}
U(1) \times \mathbb{Z}_{\textnormal{lcm}\left(\gcd(2\gcd(q_1,q_2),n),2\right)/2}\;.
\end{eqnarray}
The discussion of the $U(1)$ charges of the matter spectrum in this case is identical to that seen in the cases where $\phi_{q,q}$ was given an expectation value.

\subsection{A non-Abelian twisted circle reduction: the $E_7 \times U(1)/\mathbb{Z}_2$ Theory}
\label{ssec:E7U1TwistExampe}

With the observations of the previous sections in hand, we can return now to the geometry mentioned in Section~\ref{sec:Intro}. In the CY geometry defined by \eqref{eq:discrE7U1} we see a smooth, genus one fibered CY 3-fold with $\fe_6^{(2)}$ fibers which yields an elliptically fibered Jacobian geometry (as given by \eqref{eq:discrE7U1}) with $e_7^{(1)}$ fibers. We expect these geometries to be part of a single group of CY torsors and that the 5D M-theory vacua should be related to circle compactifications of the 6D F-theory. We will explore the structure of the genus one geometry and its associated 5D M-theory physics in Section \ref{sec:main_5D_example}, but to begin we will consider briefly the 6D F-theory background and the associated circle reduction (which is dual to M-theory on the genus one geometry in 5D). Keeping the connection between the genus-one geometry and its Jacobian in mind, we can try to understand how the 6D/5D theories are related. In particular, we hope to shed some light on the difference between the 6D and 5D gauge groups.

It is worth briefly taking stock of the ingredients at our disposal. As reviewed in previous sections, for ease of description, we begin with a 6D F-theory compactification defined via a Weierstrass model with $(E_7\times U(1))/\mathbb{Z}_2$ symmetry. How would this theory reduce to a $F_4 \times U(1)$ theory upon circle reduction? As in Section \ref{sec:Circle_review} we expect that the F-theory compactified on $S^1$ vacuum could include

\begin{enumerate}
\item Vevs for matter with a profile along the circle (as in \eqref{fourier1})
\item Gauge fields (i.e. fluxes) around the circle (i.e. \eqref{flux1})
\item Non-trivial boundary conditions (i.e. "twisting" of the circle reduction as in \eqref{twist_action})
\end{enumerate}

Considering the form of the genus one geometry and its Jacobian (with $E_7 \times \mathbb{Z}_2$ symmetry) we expect the first two elements of the list above to be present. The first of these is necessary in that a matter field vev for a $U(1)$ charged field (as we will see in Section \ref{sec:main_5D_example}, $\langle {\bf 1}_2\rangle \neq 0$ for the present example) is required to break the 6D $U(1)$ symmetry to the finite group $\mathbb{Z}_2$ seen in the Jacobian. Likewise, we expect non-trivial gauge flux along the circle in order for the theory to be dual to M-theory on a genus one fibration (see \cite{Mayrhofer:2014opa,Cvetic:2015moa}). 

However it is clear that these two ingredients alone are not enough. Given the 6D $E_7$ symmetry visible in the Jacobian it is clear that only $U(1)$-charged matter (i.e. $E_7$ singlets) can possibly be given a vev. Likewise with the allowed matter vevs, purely Abelian circle fluxes could not change the rank of the gauge group. Thus, these possibilities alone could not lead to the correct 5D gauge group to match M-theory compactified on the twisted genus one geometry. As a result, we must consider the the third item in the list above and the possibility of a ``twisted" reduction of F-theory as described in previous subsections.

As described in the start of this Section, the key question in this context is \emph{what is the F-theory origin of the twisting symmetry} of \eqref{twist_action}? As discussed previously, the resolution of this singular elliptic fibration sketched in \eqref{eq:discrE7U1} displays the affine algebra $\mathfrak{e}_7^{(1)}$. Taken in isolation from the rest of the geometry, this affine Dynkin diagram admits an outer automorphism $\widetilde{\sigma}_2$ of order two. Modding the diagram by this outer automorphism leads to $\mathfrak{e}_6^{(2)}$ (i.e. finite gauge group $F_4$) which is depicted in Figure \ref{fig:E7E62Folding0}.

Importantly, as noted in Section \ref{sec:Intro} these ``folded" Dynkin diagrams (here $\fe_6^{(2)}$) appear to arise in the fibers of a compact, genus one fibered CY threefold which admits an $E_7 \times \mathbb{Z}_2$ Weierstrass model as its Jacobian in the 6-dimensional theory. This is intriguing structure as the diagram for the \emph{affine} $\mathfrak{e}_7^{(1)}$ admits such a $\mathbb{Z}_2$, but of course its finite counterpart $\mathfrak{e}_7$ does not. Thus, the central puzzles to be addressed include the following:
\begin{itemize}
\item Can we perform a twisted dimensional reduction to connect a 6D Weierstrass model with our ``folded fiber" genus one geometry in 5D?
\item Does the finite $\mathbb{Z}_2$ symmetry which "twists" $E_7 \to F_4$ under a dimensional reduction appear as a true symmetry of the elliptic fibers (of the Jacobian Weierstrass model) which produces the folded (i.e. genus one) threefold? Is this symmetry apparent not just in an action on fibers, but on the full 6D theory/geometry? What is the action on the physical theory?
\end{itemize}
In trying to address these questions we will begin by using the full, smooth resolved CY Jacobian geometry to explore the structure of possible twisting symmetries, realizing that we must take a singular limit (by blowing down all reducible curves in the fibers that do not intersect the zero section) in order to discuss the 6D F-theory physics. It should be emphasized that in this context we are only using the symmetries of the resolved Jacobian as a guide to the symmetries of 6-dimensional theory, since in the (singular) Weierstrass limit these symmetries may not be apparent due to some K\"ahler moduli being taken to the zero-volume limit. Note that this is the same logic that allowed us to deduce the $\mathbb{Z}_2$ action on the $U(1) \times U(1)$ theory in Section \ref{sec:u1u1}.

As we will see below, a complete answer to the above questions proves to involve a number of subtle aspects and it is beyond the scope of the present work. However, we will attempt to sketch some of the ideas and the obstacles that arise in fully determining the effective physics of the reduction.

In the resolved geometry, the diagram above makes it clear that the $\mathfrak{e}_7^{(1)}$ fiber could admit an appropriate $\mathbb{Z}_2$ folding to correspond to the desired $\mathfrak{e}_6^{(2)}$ fiber in the 5D genus one geometry. However, as discussed in the previous Section this apparent symmetry is manifestly broken in the full geometry if there exists only a \emph{single} section to the elliptic fibration which would intersect the affine node ($\alpha_0$ above). However, the folding symmetry could possibly exist if there were to exist \emph{two rational sections} $s_0, s_1$ to the elliptic fibration (i.e. Mordell-Weil rank 1) and those sections intersected appropriate (i.e. symmetric) nodes in the affine fiber appropriately as shown below:
\begin{align}
\label{eq:E7weights}
\begin{array}{c}
 (\alpha_7) \\
\underset{s_0}{\underset{|}{(\alpha_0)}}(\alpha_6)(\alpha_5)(\alpha_4)(\alpha_3)(\alpha_2)\underset{s_1}{\underset{|}{(\alpha_1)}}
  \end{array}  \, . 
\end{align}
In this case, the $\mathbb{Z}_2$ automorphism that folds the fiber would only be a true automorphism of the (resolution of the) F-theory Weierstrass model at a higher-codimensional locus in moduli space. For example, it could arise if the geometry were fully symmetric under the interchange of the two rational sections (similar to the geometry in Section \ref{sec:u1u1}). It is natural to ask whether this can be achieved by tuning non-Abelian symmetry into a model such as that given in \eqref{twou1model}? Unfortunately, In this context we find that the geometry becomes too singular for a purely field theoretic analysis. More precisely, tuning the Weierstrass model given in Section \ref{sec:u1u1} to achieve a $E_7 \times U(1)$ symmetry leads to vanishings of $(f,g,\Delta)\geq (4,6,12)$ at points. This implies that the theory contains so-called "SCFT loci" and is not described purely as an ordinary perturbative field theory. Importantly, such geometric limits are likely still be within the realm of ``good" string vacua, but we are not able to write down a purely perturbative field theoretic circle reduction of the form shown in Section \ref{sec:u1u1_fieldtheory} in this setting. 

Alternatively, the higher codimensional locus that makes manifest the $\mathbb{Z}_2$ twisting symmetry should be obtainable by a suitable tuning of the resolution of the Jacobian CY 3-fold given in \eqref{eq:discrE7U1}. Unfortunately, without knowing the particular form of the $\mathbb{Z}_2$ symmetry this is difficult to reverse engineer.

There is a final observation that can be made regarding the possible geometric origin of the $\mathbb{Z}_2$ folding symmetry. In the resolved geometry of the Jacobian CY 3-fold, either of the two rational sections can be chosen as the zero section. The resulting Weierstrass models that arise from blowing down fibral components not intersecting the chosen section are identical. Moreover, the action of this interchange changes the Dynkin weights as expected for the $\mathbb{Z}_2$ folding symmetry of the Affine Dynkin diagram and acts on the $U(1)$ charged matter as charge conjugation. This action (combined with the action of R-symmetry on the hypermultiplets) leads to an apparent symmetry of the 6D theory. Moreover, in the 5D theory, this can be directly seen when constructing the two corresponding Shioda maps that are required to compute the $U(1)$ charges: Their forms, in terms of the weights $\alpha_i$ is given as
\begin{align}
\sigma(s_1) =& [s_1]-[s_0] + \frac12(3 \alpha_1 +4 \alpha_2+5 \alpha_3 +6 \alpha_4+4 \alpha_5 +2 \alpha_6 +3 \alpha_7)\, , \\
\sigma(s_1^\prime) =& [s_0]-[s_1] + \frac12(3 \alpha_0 +4 \alpha_6+5 \alpha_5 +6 \alpha_4+4 \alpha_3 +2 \alpha_2 +3 \alpha_7) \, .
\end{align}
From the above form we find that $U(1)$ charges flip sign under exchange, which as expected, implements a charge conjugation operation. The symmetry of this interchange in the 5D theory provides a hint that this symmetry may survive under the M-theory to F-theory uplift.

We view the observations above as suggestive of the origin of the twisting symmetry in 6D, but due to the higher co-dimensional nature of this symmetry in the F-theory moduli space (and the potential presence of $(4,6,12)$ points), there is no clear path to obtain its explicit action, and as a result it is beyond the scope of the present work to solve. For now, we turn our attention to the 5-dimensional theory and relationships between twisted and untwisted genus one fiberations in the 5D M-theory geometries.

\section{Phases and Lifts of  $\fe_6^{(2)}$ in 5D}\label{sec:main_5D_example}
In this section we discuss the geometry of an $\fe_6^{(2)}$ twisted algebra in a torus fibered threefold. The geometry origin of this geometry is the
fact that affine $\fe_7^{(1)}$ (unlike its finite sub-algebra) admits a $\mathbb{Z}_2$ outer automorphism. Using a genus-one fibration we can therefore engineer a 
monodromy in the base of the fibration that folds by this automorphism, which yields $\fe_6^{(2)}$  as shown in Figure \ref{fig:E7E62Folding0}. The key ingredient to achieve such a monodromy, is to start with a torus fibration that does not have a section but a two-section. We will also show, that this monodromy is missing in the respective Jacobian 
sketched in \eqref{intro_g_one}, precisely due to the presence of the section, such that those types of fibres evade the famous Koraira/Tate classification of singular fibers.   

Below, we will study the intersection matrix of the fibral curves 
of $\mathfrak{e}_6^{(2)}$ type inside the genus one fibered manifold (over a compact two-fold base) in some detail. 
We also discuss the structure of the Jacobian and its type $III^{*}$ fiber structure. By directly studying the geometry, we compute the light 5D BPS states in the M-theory compactification in both theories explicitly for a general base.

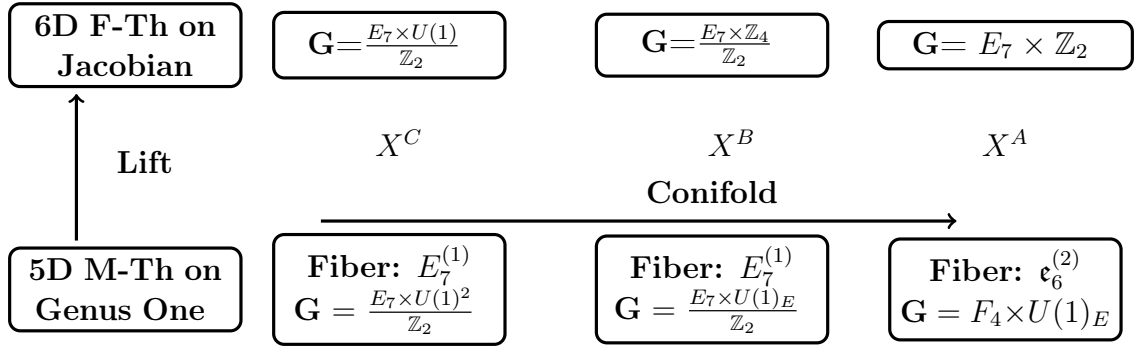
\begin{figure}[t!]
\begin{center} 
	\begin{tikzpicture}[scale=1.3]
	 \draw[very thick, ->] (-4.5,1) -- (-4.5,2.5);  
  \node (A) at (-3.8,1.8) [ text width=3cm,align=center ] {\textbf{Lift}};

 \node (A) at (2,1.5) [ text width=3cm,align=center ] {\textbf{Conifold}};
 \draw[very thick, ->] (-2,1.2) -- (4.5,1.2);  

\node (A) at (2.2,2) [ text width=3cm,align=center ] {$ X^{B} $};
\node (A) at (5,2) [ text width=3cm,align=center ] {$ X^{A} $};
\node (A) at (-1.2,2) [ text width=3cm,align=center ] {$ X^C $};
 
	\node (A) at (-4,0.5) [draw,rounded corners,very thick,text width=2.8cm,align=center ] {{\bf \textcolor{black}{5D M-Th on \\ Genus One }}};
	
	\node (A) at (-4,3) [draw,rounded corners,very thick,text width=2.8cm,align=center ] {{\bf \textcolor{black}{6D F-Th on\\ Jacobian  \\}}};

  	\node (A) at (-1.3,3) [draw,rounded corners,very thick,text width=2.8cm,align=center ] {{\bf \textcolor{black}{ G=$\frac{E_7 \times U(1)}{\mathbb{Z}_2}$  }}};

  \node (A) at (-1.3,0.5) [draw,rounded corners,very thick,text width=2.8cm,align=center ] {{\bf \textcolor{black}{Fiber: $  E_7^{(1)} $ \\ $\mathbf{G}= \frac{E_7\times U(1)^2}{\mathbb{Z}_2}$  }}};
 
  \node (A) at (2,3) [draw,rounded corners,very thick,text width=2.8cm,align=center ] {{\bf \textcolor{black}{ G=$   \frac{ E_7 \times \mathbb{Z}_4}{\mathbb{Z}_2}$  }}};

  \node (A) at (2,0.5) [draw,rounded corners,very thick,text width=2.8cm,align=center ] {{\bf \textcolor{black}{Fiber: $  E_7^{(1)} $ \\ $\mathbf{G}=\frac{E_7\times U(1)_E}{\mathbb{Z}_2}$  }}};

 \node (A) at (5,0.5) [draw,rounded corners,very thick,text width=2.8cm,align=center ] {{\bf \textcolor{black}{Fiber: $\fe_6^{(2)} $ \\ $\mathbf{G}=F_4\times U(1)_E$}}};

 \node (A) at (5,3) [draw,rounded corners,very thick,text width=3.1cm,align=center ] {{\bf \textcolor{black}{G= $  E_7 \times \mathbb{Z}_2 $  }}};
	\end{tikzpicture}
\end{center} 
\caption{\textit{M-and F-theory compactifications of three types of geometries $X^I$, connected via geometric transitions. For $I\neq C$, the threefolds have a non-trivial TS group $\Sha$, which results in disconnected 5D vacua. 
}}
\label{fig:E7E62chain}
\end{figure}

Finally, it is possible to explore other genus one fibered geometries, related to 
$\mathfrak{e}_6^{(2)}$ by geometric transitions as sketched in Figure~\ref{fig:E7E62chain}. Such geometric transitions usually have an interpretation in terms of Higgs transitions across special loci in the Coulomb branch moduli space. Identifying the massless matter multiplets is more intricate in the present case, which we comment on below.

\subsection{The geometry of $\mathfrak{e}_6^{(2)}$}
\label{ssec:E62}  
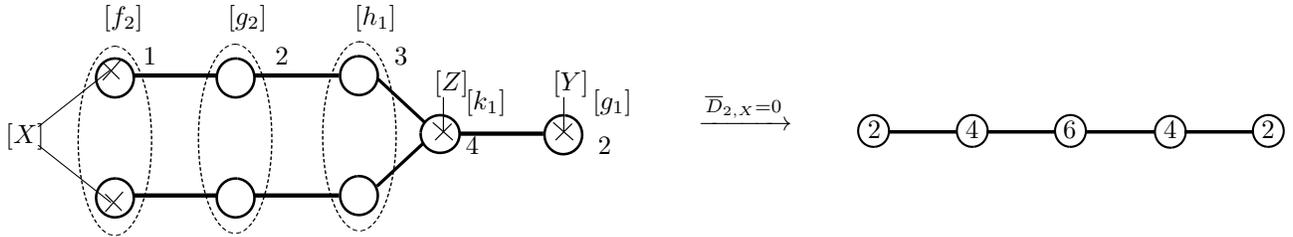
\begin{figure}[t!]
 \begin{center}
 {\footnotesize
\begin{picture}(160, 100)
 \put(-150,10){\includegraphics[scale=0.55]{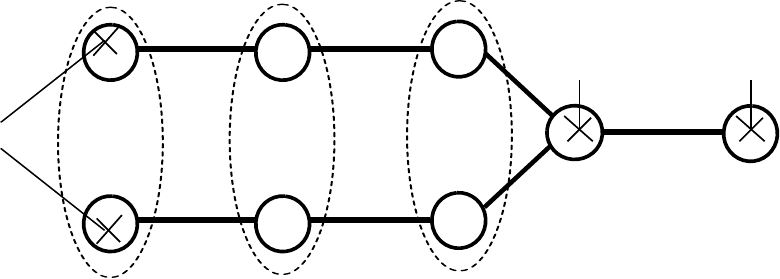} }
 
        \put(-162,45){$[X]$}
 \put(-125,90){$[f_2]$}
 \put(-78,90){$[g_2]$}
  \put(-30,90){$[h_1]$}

   \put(12,58){$[k_1]$} 
   \put(0,65){$[Z]$}  
   
      \put(45,65){$[Y]$}  
    
       \put(12,41){$4$}

            \put(-110,75){$1$}
            
            \put(-60,75){$2$}

            \put(-15,75){$3$}
      
                \put(100,50){$\xrightarrow{  \overline{D}_{2,X}=0}$     }
      
      \put(160,43){\includegraphics[scale=0.45]{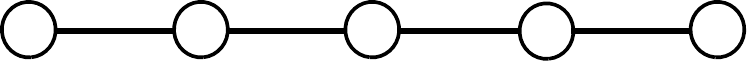} }
   
  \put(60,58){$[g_1]$}  
   \put(62,41){$2$}  
  
    \put(164,47){$2$}  
      \put(201,47){$4$}  
     \put(238,47){$6$}  
    \put(276,47){$4$}  
     \put(313,47){$2$}  
 \end{picture} 
 }
 \caption{\label{fig:E62a}{\it Depiction of the $\fe_6^{(2)}$ fiber. On the left, fibral divisors, curves, intersections and Kac labels are depicted.  Crossing the 2-section monodromy divisor $\overline{D}_{2,X}$, interchanges the two fibral curves in $[f_2]$, $[g_2]$ and $[h_1]$. At $\overline{D}_{2,X}=0$ the fiber degenerates into a multiplicity two fiber, depicted on the right. 
  }} 
 \end{center}
 \end{figure}

To engineer the $\mathfrak{e}_6^{(2)}$ fibration we require a 2-section model to realize the folded fiber type. To obtain such a model we employ a quartic fiber model \cite{Morrison:2014era} that was already introduced in the Section~\ref{sec:Intro}. We repeat the respective hypersurface equation here for convenience, as  
\begin{align}
\label{eq:Quartic}
p=s_1   X^4 + s_2  X^3 Y + s_3  X^2 Y^2 + s_4   X Y^3 + 
 s_5  Y^4 + s_6   X^2 Z + s_7 X Y Z + s_8  Y^2 Z + s_9 Z^2 \, ,
\end{align}
where we can fix $s_9=1$ globally and 
with the monodromy divisor for the 2-section $X=0$
\begin{align}
\label{eq:monoX}
    \overline{D}_{2,X}=  s_8^2 - 4 s_5 \, .
\end{align} 
  In order to engineer an $\mathfrak{e}_6^{(2)}$ type of fiber over the base divisor $\mathcal{Z}:= z=0$, we treat the  $s_i$ of \eqref{eq:Quartic} as the generalization of the Tate coefficients in the quartic genus-one model. All of those sections are associated to certain line-bundles\footnote{Sometimes care has to be taken, that the $s_i$ do not become constants. In such situations the multi-section may split into multiple one-sections and Abelian gauge enhancement may occur \cite{Klevers:2014bqa}.} that are given in Appendix~\ref{app:LBData}. Similarly to the Tate-model, we engineer singularity by imposing certain vanishing orders of the $s_i$ in codimension one. In order to do so, we factor 
  powers of the line bundle $z$ out of the $s_i$ as 
\begin{align}
\label{eq:SectionShift}
s_i \rightarrow d_i z^{n_i} \, , \text{ such that } [d_i] \sim [s_i - n_i \mathcal{Z}] \, ,
\end{align}
which characterises the singularity in the genus-one fiber at the vanishing lovus of $z$. We package those vanishing orders into the generalized Tate-vector
vector $\Vec{n}$. For an elliptic fibration those singularities are classified, but for a genus-one model this remains an open problem\footnote{At present even a classification of genus-one fiber models and the range of possible N-sections is not known. The state of the art is the 5-section model presented in \cite{Knapp:2021vkm}.}.
A toric classification  has been given in \cite{Bouchard:2003bu} in terms of tops, introduced in \cite{Candelas:1996su} which we are using throughout this work. In this section it is our goal to engineer what we call a 
  $III^*$ non-split singularity that corresponds to an $\mathfrak{e}_6^{(2)}$ resolution, specified by the Tate vector 
\begin{align}
\label{eq:TateE62}
    \vec{n} = \left\{3,3,2,1,0,2,1,0 \right\} \, .
\end{align}
Note that this factorization is specifically chosen such that the monodromy divisor \eqref{eq:monoX} intersects $\mathcal{Z}$ generically. 
The resolution (following \cite{Bouchard:2003bu}) is given via the following four additional coordinates  
$g_1,g_2,k_1,l_1$ given in the toric hypersurface\footnote{See \cite{Buchmuller:2017wpe,Dierigl:2018nlv} for a discussion of related techniques used here.} 
\begin{align}
p=   &\phantom{+} d_1 f_2^3 g_2^2 h_1 X^4 + d_2 f_2^3 g_1^2 g_2^3 h_1^3 k_1^3 X^3 Y + 
 d_3 f_2^2 g_1^2 g_2^2 h_1^2 k_1^2 X^2 Y^2 + d_4 f_2 g_1^2 g_2 h_1 k_1 X Y^3 \nonumber \\ & + 
 d_5 g_1^2 Y^4 + d_6 f_2^2 g_1 g_2^2 h_1^2 k_1^2 X^2 Z + 
 d_7 f_2 g_1 g_2 h_1 k_1 X Y Z + d_8 g_1 Y^2 Z + Z^2
\end{align}
 which admits the following Stanley-Reisner ideal\footnote{Note that we have resolved the ambient space, such that the singular point $X=Y=0$ is absent.} 
\begin{align}
\mathcal{SRI}: \{ Y X, Y f_2, Y g_2, Y h_1, Y k_1, Z g_1,   X g_1, X g_2, X h_1, X k_1, f_2 h_1, f_2 k_1, g_2 k_1 \}\, .
\end{align}
In our toric description, the base divisor $\mathcal{Z}$ is then given via the projection
\begin{align}
    \pi: X_3 : (f_2, g_1, g_2, h_1 k_1) \rightarrow \mathcal{Z}=  f_2 g_2^2  h_1^3 k_1^4 g_2^2\, . 
\end{align}
The intersections of the fibral curves is depicted in 
 Figure~\ref{fig:E62a}. Note that we have highlighted when multiple curves are part of a fibral divisor, e.g. when the self-intersection is $-4$ instead of $-2$. E.g. the fibral intersection $D^{i,j}=D_i \cdot D_j$ and the Cartan matrix $\mathfrak{C}^{i,j}= - 2  \frac{D^{i,j}}{D^{ii}}$ is given as
 \begin{align}
     D=\left(
\begin{array}{ccccc}
-4 & 2 & 0 & 0 & 0 \\
2 & -4 & 2 & 0 & 0 \\
0 & 2 & -4  & 2 & 0\\
0 & 0 & 2 & -2 & 1 \\
0 & 0 & 0 & 1 & -2 
     \end{array}
     \right) \quad \text{ and } \quad     \mathfrak{C}=\left(
\begin{array}{ccccc}
2 & -1 & 0 & 0 & 0 \\
-1 & 2 & -1 & 0 & 0 \\
0 & -1 & 2  & -2 & 0\\
0 & 0 & -1 & 2 & -1 \\
0 & 0 & 0 & -1 & 2 
     \end{array}
     \right)
 \end{align}

 Those fibral curves are given as  the expressions
\begin{align}
\begin{split}
\mathbb{P}^1_{\pm,0}: \{ p =f_2 = 0\} :&\quad d_5   g_1^2+d_8 g_1 Z+Z^2 \\
\mathbb{P}^1_{\pm,1}: \{p =g_2 = 0\}:&\quad d_5   g_1^2+d_8   g_1 Z+  Z^2 \\
\mathbb{P}^1_{\pm,2}: \{p =h_1 = 0 \} :&\quad d_5 g_1^2 + d_8 g_1 Z +  Z^2 \\
\mathbb{P}^1_3: \{p =k_1 = 0\}:&\quad  d_5 g_1^2+d_1 h_1+d_8   g_1 Z+ Z^2\\
\mathbb{P}^1_4: \{p =g_1 = 0\}:&\quad 1 + d_1 f_2^3 g_2^2 h_1 \, .
\end{split}
\end{align}
Importantly inside of $\mathbb{P}^1_{\pm,j}$ with $j=0,1,2$
we find two curves that are interchanged along the monodromy divisor \eqref{eq:monoX} of the 2-section $X=0$. This is also clear by inspection of the intersection picture given in 
Figure~\ref{fig:E62a}.
There we find that the 2-section $X$ intersects each of the two curves of $\mathbb{P}^1_{\pm,0}$, inducing a monodromy action on the respective curves as well. 
From the intersection picture we find the affine Cartan matrix of $\mathfrak{e}_6^{(2)}$.

An important consequence for this type of geometries is the appearance of \textit{multiple fibers} \cite{Anderson:2018heq,Anderson:2019kmx,Kohl:2021rxy,Grassi:2023aks}:
These appear exactly at the smooth codimension two degeneration loci where the monodromy divisor $\overline{D}_{2,X}=0$ intersects 
the base curve of the type $III^*$ fiber $\mathcal{Z}$. In order to see this, recall that the generic fiber $\mathcal{E}$ over $\mathcal{Z}$ becomes reducible and degenerates into \begin{align}
    [\mathcal{E}] \rightarrow [f_2]+ 2 [g_2] + 3 [h_1]+ 4 [k_1] + 2 [g_1] \, ,
\end{align}
where the multiplicity factors are the respective Kac labels. Now at the codimension two intersection locus with $\overline{D}_{2,X}=0$
the fibral curves in $[f_2], [g_2]$ and $[h_1]$ degenerate into a single curve of multiplicity two as depicted on the right side of Figure~\ref{fig:E62a}.
Hence the whole fiber becomes 
\begin{align}
    [\mathcal{E}] \xrightarrow{\overline{D}_{2,X}=0} \, \, 2\cdot ( [f_2]+ 2 [g_2] + 3 [h_1]+ 2[k_1] + [g_1]) \, ,
\end{align}
that is a non-reduced fiber of multiplicity two. 
Note again, that those multiplicity two fibers appear over smooth points in $B_2$ and hence are of very different kind than those discussed in \cite{Anderson:2018heq,Anderson:2019kmx,Kohl:2021rxy,Grassi:2023aks}. More generally, the presence of those multiple fibers implies that the group of CY torsors does not reduce to the TS group $\Sha(X^A/B_2)$ but the Weil-Ch$\hat{\text{a}}$talet group $WC(X^A/B_2)$ \cite{dolgachev1993elliptic}.
  
Similar degeneration structures also appear for non-simply laced groups, at the loci 
where two fibral curves are interchanged. The important difference in the present case is that this monodromy action can never involve a section, as these are just a single point. A section, in turn, must intersect a multiplicity one fibral curve, typically the affine node. Hence an elliptic fibration can never have a monodromy action on the affine node of some ADE fiber and therefore can only lead to untwisted algebras. 

Having discussed the geometry of a non-trivial $WC(X^A/B_2)$ element we can discuss the geometry of the Jacobian $J(X^A/B_2)$. The singular Jacobian is obtained, by employing the tuning  as given by the generalized Tate-Vector in \eqref{eq:TateE62} and substituting it into the Jacobian map given in Appendix~B of \cite{Klevers:2014bqa}. To compare the fiber structures over $\mathcal{Z}$, we only give the relevant leading order Weierstrass coefficients as  
\begin{align}
\begin{split}
\label{eq:E62Jac}
    f= z^3 d_1  \overline{D}_{2,X} + \mathcal{O}(z^4)\, , \quad   
    g= z^5 d_1 P_{(4d)}  + \mathcal{O}(z^6)  \, ,  \quad
    \Delta= 4 z^9 d_1^3 \overline{D}_{2,X}^3 + d_1^2 \mathcal{O}(z^{10})\, .
    \end{split}
\end{align}
Note that we find a type $III^*$ fiber, just as in the genus-one geometry, i.e. an $\fe_7$ singularity. This may come as a surprise as one could have expected a $IV^*$ split singularity here, i.e. an $\fe_6$ instead, as this is the field theory lift of an $\fe_6^{(2)}$ reduction. Instead the Jacobian yields the geometric cover of the twisted algebra. Interestingly the monodromy divisor of the genus one fibration appears at leading order of the discriminant in \eqref{eq:E62Jac} and leads to local matter fields\footnote{For non-simply laced groups, the monodromy divisor also appears at leading order in $\Delta$ but does not directly imply localized matter there.
}. Unlike in the genus-one fibration, we can not expect the monodromy divisor to lead to a non-simply laced type of group in the Jacobian, simply because $E_7$ does not possess an outer automorphism, only its affine extension does. 
  Consequently we find the genus-one fibration and its Jacobian do not have the same number of K\"ahler moduli similar to the cases discussed in \cite{Oehlmann:2016wsb,Baume:2017hxm,Oehlmann:2019ohh} which had multiple fibers as well. It appears that this may be a common feature in the presence of non-trivial Weil-Ch$\hat{\text{a}}$telet groups. 

Note that we use the typical polynomial map from the genus-one fibration to the Jacobian, which might not see potential non-polynomial deformations. Indeed, such deformations are often times present, which we will discuss in Section~\ref{sec:TVSUT} in more detail. 

\subsection{Geometric state counting}
Having discussed the difference in codimension one structure between the genus-one threefold and its Jacobian, we now compute the number of charged states for each geometry. We will focus in particular on those states that carry non-trivial representations under the maximal 5D finite sub-group $\ff_4 \subset \fe_6^{(2)}$ and ignore massive singlet states in the rest of this work. We wish to return for a complete discussion of all such states in future work.

\subsubsection*{State Counting in the $\mathfrak{e}_6^{(2)}$ Genus-one Fibration}
Having discussed the general features of the genus-one geometry, we turn now to the generic matter spectrum of the $\mathfrak{e}_6^{(2)}$ fibration. To do so, we first need to understand which line bundles the $s_i$, and upon tuning the $d_i$, are sections of. These are determined by the line bundles/divisor classes summarized in Appendix~\ref{app:LBData} and tuning by the generalized Tate-Vectors in \eqref{eq:SectionShift}. From the resolved geometry as well as its Jacobian we find a non-minimal singularity over $z=d_1=0$ that exhibits a vanishing order  ord$_{\text{van}}(f,g,\Delta)=(4,6,12)$. Those points are generally associated to superconformal matter, i.e. E-string theories associated with tensionless strings in the spectrum. To avoid further complications we use the line bundle class $\mathcal{S}_7$ dependency of $[d_1]$ to restrict to fibrations where such loci are absent by demanding
 \begin{align}
 \label{eq:E62tuning46}
[z]\cdot [d_1]= 4 c_1 \cdot \mathcal{Z} - 2 \mathcal{S}_7 \cdot \mathcal{Z} - 3 \mathcal{Z}^2 = 0 \, .
 \end{align} 
 In this way we can solve for the intersection $\mathcal{S}_7 \cdot \mathcal{Z}$, which yields the intersections 
\begin{align}
\label{eq:d8Intersection}
[d_8] \cdot \mathcal{Z}=  \mathcal{S}_7 \cdot \mathcal{Z} = 2 c_1 \cdot \mathcal{Z}-\frac32 \mathcal{Z}^2 = 4(1-g)+ \frac12 \mathcal{Z}^2  \, , 
\end{align}
with $g$ being the genus of the curve $\mathcal{Z} \subset B_2$. Throughout this work we implicitly assume the base $B_2$ to be smooth and hence all base divisors to be Cartier \cite{DelZotto:2014fia} resulting in integral intersection numbers. It may therefore seem puzzling to find a factor $\frac12$  factor in the intersection formula \eqref{eq:d8Intersection}. However as we explicitly assumed a smooth base, we should view this ocurance rather as a geometric condition that forces $\mathcal{Z}^2$ to be an even self-intersection.
This condition has also been found in \cite{Lee:2022uiq}. In Section~\ref{ssec:E62Transitions} we give a physical interpretation of this phenomenon
which is related to $\mathbf{56}$-plets that are not pseudo-real and must therefore appear in pairs of half-hypers.

To compute the multiplicity of BPS particles, charged under the physical gauge group inside $\mathfrak{e}_6^{(2)}$, we need to compute the moduli spaces of divisors $D$
that collapse to curves $\mathcal{Z}$. The collapsing 
$\mathbb{P}^1$'s are wrapped by M2 branes in M-theory which yield states in the 5D theory. In order to do so we adopt the geometric framework pioneered in \cite{Witten:1996qb,Katz:1996fh,Aspinwall:2000kf} and \cite{Braun:2014oya}. The main complication to consider is the monodromy under which the fibral 
$\mathbb{P}^1$'s are branched in the base. Before computing the exact moduli spaces we want to fight the respective weights of the states.
To do so we start with the covering algebra and then implement the monodromy in a second step. In our case, we have an $\mathfrak{e}_7$ cover and the M2 states that wrap those shrinkable curves form the adjoint representation. 
 We write those curves  schematically as $\mathcal{C}= \sum_i a_i \mathcal{C}_i$ for $i=1 \ldots 8$. The curves satisfy the $\mathcal{C}_i \cdot \mathcal{C}_j = -\mathfrak{C}_{i,j}(\mathfrak{e}_7^{(1)})$ being the affine Cartan matrix. 
In the following we write those curves graphical as 
 \begin{align}
 \mathcal{C}= a_1 -a_2 <\begin{array}{c}a_3-a_4-a_5 \\ a_6-a_7-a_8       \end{array}
  \end{align}
  We can then generate the 126 curves with
 $a_i \geq 0$ and $\mathcal{C}^2=-2$ that generate the roots of $\fe_7$ following \cite{Braun:2014oya}.

In the next step we implement the monodromy on the 126 curves. In order
to do so, we need to
decompose all curves into orbits under the $\mathbb{Z}_2$ action 
\begin{align}
 \mathbb{Z}_2:\,  \mathcal{C}_5 \leftrightarrow \mathcal{C}_8 \, , \qquad  \mathcal{C}_4 \leftrightarrow \mathcal{C}_7 \, ,  \quad \mathcal{C}_3 \leftrightarrow \mathcal{C}_6  \, , 
 \end{align}
and group them into the invariant curves such as
\begin{align}
 \mathcal{C} = (\mathcal{C}_1, \mathcal{C}_2, \mathcal{C}_3+\mathcal{C}_6, \mathcal{C}_4+\mathcal{C}_7)_{(\mathcal{C}_5+\mathcal{C}_8)} \, .
 \end{align}
with the combination $\mathcal{C}_5 + \mathcal{C}_8$ denoting the  affine curve. 
We have summarized this split in Table~\ref{tab:E62} for several example curves. 
Shrinking all but 
the affine curve yields a maximal $\mathfrak{f}_4 \times \mathfrak{u}_{1,E}\in \mathfrak{e}_6^{(2)}$ sub-algebra.
\begin{table}[t!] 
\begin{center}
\begin{tabular}{|l|c|c|c|c|c}\hline
Mono & $\#$ Curves  &  Example Curve &  $\mathfrak{f}_4$    weight &  $\chi$   \\ \hline   
  Inv & 24 & $  1-1<\begin{array}{c}0-0-0 \\ 0-0-0   \end{array}$    &  $(-1,-1,2,0)_0 $  &  g  \\ \hline
  $\mathbb{Z}_2$  & 24   &  
  $1-1<\begin{array}{c}1-0-0 \\ 0-0-0    \end{array}   \quad + \quad  1-1<\begin{array}{c}0-0-0 \\ 1-0-0        \end{array}$  & $(-1, 0, 0, 1)_0  $  & $\begin{array}{c}  \overline{g}  \end{array}$     \\ \hline 
   $\mathbb{Z}_2$ &        24   &  
  $0-1<\begin{array}{c}1-0-0 \\ 1-1-1        \end{array} \quad + \quad   0-1<\begin{array}{c}1-1-1 \\ 1-0-0       \end{array} $ & $(1, 0, -1, 1)_{-1}$   &   $\overline{g}  $  \\ \hline    
  $\mathbb{Z}_2$  &  3   & $ 
  1-2<\begin{array}{c}2-1-1 \\ 1-1-0        \end{array} \quad + \quad   1-2<\begin{array}{c}1-1-0 \\ 2-1-1       \end{array}$  & $(0, 0, 0, 0)_{1}$  &  $\overline{g}  $       \\ \hline  
   \end{tabular}  
   \caption{
\label{tab:E62}\textit{
Schematic reduction of the 126 curves of $\mathfrak{e}_7$ roots into
$\mathbb{Z}_2$ invariant linear combinations,
their $\mathfrak{f}_4 \subset \mathfrak{e}_6^{(2)}$ charges and respective moduli spaces dimension $\chi$ computed in \eqref{eq:F4branched}. 
   }}
   \end{center}
\end{table}
We can verify this explicitly when computing the representations and multiplicities of states obtained by M2 branes that wrap the collapsing fibral $\mathbb{P}^1$'s. 
First note that the 
  moduli space of a degree $d$ branched curve $\overline{g} $ can be computed via the Riemann-Hurwitz (RH) theorem    
\begin{align} 
\label{eq:RHT}
\overline{g}-g= (d-1)(g-1)+\frac12 \text{deg}(R) \, .
\end{align} 
Here $g$ denotes the genus of the curve $\mathcal{Z} \subset B_2$, 
$d$ denotes the order of the cover and $\text{deg}(R)$ the degree of the ramification. The degree $\text{deg}(R)$ can be computed from the intersection of the monodromy divisor with $\mathcal{Z}$ as 
\begin{align}
    \text{deg}(R) = [\overline{D}_{2,X}]\cdot [\mathcal{Z}]=2 [d_8] \cdot [\mathcal{Z}] \, ,
\end{align}
given in \eqref{eq:d8Intersection} and in our case we have $d=2$.
Putting it all together we get
\begin{align}
\label{eq:F4branched}
  \overline{g}-g=  3(1-g) + \frac12 \mathcal{Z}^2 \, .
\end{align}
 Now we can read off the associated 5D states, when the respective curves collapse.
We first consider the shrinkable curves, given by those that admit trivial affine weight, depicted by the $24+24$
curves in the first and second row of Table~\ref{tab:E62}. 
Upon including the four invariant Cartan generators those make up the $\mathbf{52}$-dimensional adjoint representation of $\ff_4$. In total those curves contribute a 5D vectormultiplet and $g$ hypermultiplets in the adjoint representation \cite{Witten:1996qb}.

Note that the moduli space of the 
of the $24$ $\mathbb{Z}_2$ branched states, computed in \eqref{eq:F4branched} is $\overline{g}$-dimensional. Hence there are still $\overline{g}-g$ massless hypermultiplets that did not get paired up to the adjoint representation. Those 
left-over states are completed to
$\mathbf{26}_0$-plet charged hypermultiplets, upon adding two uncharged fields. The later ones are counted as complex structure deformations on the CY geometry (see Section~\ref{sec:TVSUT} for more details).

The other representations work similarly: There are 24 non-shrinkable curves that admit an affine charge and weights of the $\mathbf{26}_1$ with multiplicity $\overline{g}$. 
Note that two singlets from the last row in Table~\ref{tab:E62} are needed to complete the $24$ states to a full  $\mathbf{26}_1$-plet.  
 Summarizing the hypermultiplet sector in terms of geometric data we then have 
\begin{align}
\label{eq:e62Spec}
\begin{split}
    n_{\mathbf{52}_0} &=g \, , \\ 
    n_{\mathbf{26}_0} &= \overline{g}-g=  3(1-g) + \frac12 \mathcal{Z}^2 \, , \\
    n_{\mathbf{26}_1} &= \overline{g}= 3(1-g) + \frac12 \mathcal{Z}^2 + g \, .
    \end{split}
\end{align} 
The residual charged singlets are given as 
\begin{align}
    n_{\mathbf{1}_1}=   \overline{g}= 3(1-g) + \frac12 \mathcal{Z}^2 + g \, .
\end{align}
Note again that all fields, that have a non-trivial $\mathfrak{u}_{1,E}$ charge should be regarded as having a shifted KK-momenta and thus, stay massive at the origin of the $\ff_4$ Coulomb branch. Evidently, 
the above formula does not satisfy the 6D anomaly cancellation condition for an $\mathfrak{f}_4$ algebra 
by exactly one 
\begin{align}
    n_{\mathbf{26}_0}+ n_{\mathbf{26}_1} = 5(1-g) + \mathcal{Z}^2+1 \, .
\end{align}
As the 6D lift of the above theory is not an $\mathfrak{f}_4$ however there is no direct reason to expect such a cancellation to occur. In fact, in our construction we argue that the correct lift is given by the Jacobian, i.e. an $\fe_7$ theory with a matter spectrum that we will discuss momentarily. It is important to remark, that our results are different than the interpretation provided in \cite{Braun:2014oya,Kohl:2021rxy},
where it was proposed that the maximal finite subgroup, $\fe_4$ in our case,  would lift to 6D.
It turns out however, that there do exist distinct but related geometries which exhibit $\mathfrak{f}_4^{(1)}$ fibers instead of $\mathfrak{e}_6^{(2)}$ that do lift to Jacobians with $F_4$ gauge symmetry in 6D, as we discuss in Section~\ref{sec:TVSUT}.

\subsubsection*{Comparison with $\fe_6 \rightarrow \mathfrak{e}_6^{(2)}$ reduction}
The typical way in which a 5D twisted theory with $\mathfrak{f}_4$ symmetry is obtained in the literature \cite{Bhardwaj:2019fzv,Kim:2020hhh,Lee:2022uiq}, is by starting from a 6D $\mathfrak{e}_6$ gauge theory and performing a twist by an outer automoprhism (as in Figure \ref{fig:E61-F41Dynkin}) when compactifying on a circle. In contrast to this, we find an $\fe_7$ theory in the Jacobian, instead of $\fe_6$. Despite this difference however, we find the $\fe_6 \subset \fe_7$ subgroup to be acted upon as in the ``usual" twisted reduction. In the following we show that the spectrum agrees between these two viewpoints.   

To begin we will consider a 6D $\mathfrak{e}_6$ gauge theory.
In order to make contact with our geometric formulas, we will express the number of 6D hypermultiplets in terms of the standard formulas that include the self-intersection and the genus $g$ of the curve $\mathcal{Z}$ that a stack of 7-branes with $\fe_6$ world-volume gauge theory wraps.
Assuming a geometric engineering of such a symmetry in F-theory the massless hypermultiplet spectrum can be expressed as
\begin{align}
    n_{\mathbf{27}} = 6(1-g) + \mathcal{Z}^2 \, ,\quad n_{\mathbf{52}} =g \,.  
\end{align} 
This 6D theory is the starting point for a twisted compactification. 
The $\mathbb{Z}_2$ twist $\sigma$ appearing in \eqref{twist_action} acts as the outer automorphism of $\mathfrak{e}_6$ by
exchanging the two legs of the $\fe_6$ Dynkin diagram as shown in Figure \ref{fig:E61-F41Dynkin}. This acts as complex conjugation on the representations  
\begin{align}
\sigma: \, \, \mathbf{27} \leftrightarrow \overline{\mathbf{27}}\, .
\end{align} 
Upon the $S^1$ reduction we impose
$\sigma$ twisted boundary conditions when going around the circle. This effectively shifts the KK-tower by a half-integer multiple and leaves a $\mathfrak{f}_4$ finite sub-algebra in 5D. For an integer-spaced KK tower we obtain the decomposition of the adjoint \cite{Bhardwaj:2019fzv,Lee:2022uiq} as 
\begin{align}
\mathbf{78} \rightarrow \mathbf{52}_0 \oplus \mathbf{26}_{\frac12}\, ,
\end{align}
where the subscript denotes the shifted KK-charge. 
Since a 6D hyper contains a fundamental and its conjugate, we can take a pair of them and mod by the $\sigma$ upon the circle reduction. This condition requires an even amount of 6D hypermultiplets which in terms of the F-theory geometry requires $\mathcal{Z}^2$ to be even. Note that this is exactly the geometric condition we found in the section before\footnote{Cases with unpaired matter have been considered in \cite{Bhardwaj:2020kim} and are not described by our geometric setup.}. 
 Decomposing a pair upon the $\sigma$-quotient one obtains 
\begin{align}
    (\mathbf{27} \oplus \overline{\mathbf{27}}) \rightarrow (\mathbf{26}_0 \oplus \mathbf{26}_\frac12 \oplus \mathbf{1}_\frac12 \oplus \mathbf{1}_0  ) \, .
\end{align}
The subscript denotes the shifted KK-mass of the fields. 
As discussed before, we expect those fields to stay massive, as e.g. the  $\mathbf{26}_\frac12$ vectormultiplets.  Collecting the multiplicities of 5D hypermultiplets in terms of the geometric data of the $\mathfrak{e}_6$ theory we find
\begin{align}
\label{eq:e62spectrum}
\begin{split}
\begin{array}{ll}
     n_{\mathbf{26}_0}= 3(1-g)+\frac12 \mathcal{Z}^2  \, ,\quad & 
     n_{\mathbf{26}_\frac12}= 3(1-g)+\frac12 \mathcal{Z}^2 +g \, ,\\   n_{\mathbf{1}_\frac12}= n_{\mathbf{1}_0}= 3(1-g)+\frac12 \mathcal{Z}^2\, , \qquad & n_{\mathbf{52}_0}= g \,,
    \end{array}
    \end{split}
\end{align}
where we added the contribution of fundamentals from 6D adjoint hypermultiplets. 

Upon rescaling the shifted KK-charges by two, to obtain the same charge normalization as used in our geometric derivation, we find full agreement of the multiplicities
 \eqref{eq:e62Spec}. Moreover, we also find a prediction for the amount of additional neutral singlets $n_{\mathbf{1}_0}$ in the geometry
 that we were not able to obtain from
 in the geometric computations. Those later ones we identify as complex structure moduli specific non-polynomial realized complex structure deformations, that we will discuss in more detail in 
 Section~\ref{sec:TVSUT}.

While the state counting coincides with those of a typical $\fe_6 \rightarrow \fe_6^{(2)}$ twisted reduction, the $\fe_7$ in the Jacobian suggests a potential different 6D uplift in our case. Note however, that the 6D $\fe_7$ is always Higgsable to the desired $\fe_6$. Hence, it might be that the Jacobian map that we used in this work might somehow implicitly require those complex structure deformations to acquire non-trivial values and hence breaks $\fe_7 \rightarrow \fe_6$ in the Jacobian. As discussed in Section \ref{sec:Circle_reduc_spec} we expect the 6D $\mathbb{Z}_2$ twisting symmetry to be of more intricate origin in the elliptic fibered geometry. But as seen above, its action on the $\fe_6$ subgroup of $\fe_7$ must be consistent with that given above.

Another interesting distinction between the reduction proposed in this work and the standard $\fe_6$ construction is their relationships to the IIB axio-dilaton $\tau$: As the genus-one fibration admits the same $\tau$-profile as the Jacobian, the type $III^*$ singularity fixes this value to
$\tau=i$. An $\fe_6$ singularity on the other hand, fixes the coupling to 
$\tau=e^{-2\pi i/3}$ locally. In this regard, our construction with $\fe_6^{(2)}$ fibers seems to have a different 6D origin to the usual twisted reductions considered in the literature. 

\subsubsection*{State counting in the $\mathfrak{e}_7^{(1)}$ Jacobian}
Having analyzed the $\mathfrak{e}_6^{(2)}$ fibration, we turn now to its Jacobian given by the Weierstrass model in 
 \eqref{eq:E62Jac}. For this we consider the codimension two enhancements of the Weierstrass model of the $\mathfrak{e}_7$ divisor $\mathcal{Z}: z=0$ with the monodromy divisor $\overline{D}_{2,X}$ of the genus-one model.
At those loci the vanishing orders enhance as
\begin{align}
z=0:\quad  \text{ord}_{|\text{van}}(f,g,\Delta)=(3,5,9) \xrightarrow{\overline{D}_{2,X}=0}
\text{ord}_{|\text{van}}(f,g,\Delta)=(4,5,10) \, ,
\end{align}
  i.e. from a type $III^*$ to a type $II^*$ singularity which corresponds to an $E_8$. Those enhancement loci we associate with $\mathbf{56}$-plet representations.
  Note that the multiplicity of those states is precisely given by the number of ramification points in the genus-one fibration
\begin{align}
n_{56}= 2 \cdot [d_8] \cdot \mathcal{Z}=deg(R)=8(1-g) + \mathcal{Z}^2 \, ,
\end{align}
consistent with the 6D anomalies.
Next we should infer the $\mathbb{Z}_2$ charges of the $\fe_7$ fundamentals. 
In order to do so, we would typically use the genus-one fibration and uplift the 5D shifted $U(1)_E$ charges to discrete ones in 6D. As the genus-one fibration admits a very different 5D gauge group however, we can not use this approach here.  In fact, the following two charge assignments are consistent with the 
pseudo-reality of the states
\begin{align}
    \mathbf{56}_q \text{ with } q=0,1 \text{ mod } 2 \, .
\end{align} 
This is still a non-trivial charge assignment, as there could have been a non-trivial mixing of the $\mathbb{Z}_2$ discrete gauge symmetry, with the center of $\fe_7$, which have lead to fractional charges that are not compatible with the pseudo reality, as we demonstrate in the next sections. It would be interesting however to deduce those charges discrete charges from first principles. 

\subsubsection*{Toric examples}
To be fully concrete, we give explicit toric examples. For the first one, we specify the toric rays that make up the 4D ambient variety in Appendix~\ref{app:Example3folds}: The example is a genus-one fibration on an $\mathbb{F}_0$ base, with an 
$\mathfrak{e}_6^{(2)}$ on a $\mathcal{Z}^2=0$ curve of genus $g=0$. The Hodge numbers are computed as 
\begin{align}
    (h^{1,1},\, h^{2,1}\, (h^{2,1}_{np})\, )(X^C)=(7,\, 95(9)) \, .
\end{align}
We have also added the $h^{2,1}_{np}$ non-polynomial contribution of the complex structure deformations, that we will discuss in more detail in Section~\ref{sec:TVSUT}. The resulting 5D theory admits $3\times \mathbf{26}_0\oplus 3\times \mathbf{26}_1$-plets in its spectrum. The spectrum also contains several $I_2$ fibers and $95$ uncharged singlets. We will not discuss those charged singlets but comment on the $95$ complex structures via a Higgs transition in the next section. 
As expected, we find three K\"ahler parameters from generic fiber and Hirzebruch base as well as four more from the $\mathfrak{e}_6^{(2)}$. 

Consulting the general expressions for the spectrum in \eqref{eq:e62spectrum}, we find the 
possibility for a 5D theory with a trivial Higgs branch, i.e. a trivial hypermultiplet spectrum for
for $g=0$ and $\mathcal{Z}^2=-6$. Indeed, we can engineer a simple toric example, where the genus-one fibration has an
$\mathbb{F}_6$ base as given in Table~\ref{tab:E62F6Rays}.
\begin{table}[t!]
\begin{center}
\begin{tabular}{ccc}
$
\begin{array}{|c|c|}
\multicolumn{2}{c}{\text{ Generic Fiber }} \\ \hline 
X&(-1,1,0,0)\\
Y&(-1,-1,0,0)\\
Z&(1,0,0,0)\\ \hline  
\end{array}$  & $ 
\begin{array}{|c|c|}
\multicolumn{2}{c}{\text{ $\mathbb{F}_6$ Base }} \\ \hline  
x_1 & (1, 0, 0, 1) \\
y_0 & (-4, 5, -1, -6) \\
y_1 & (-10, -10, 1, 0)\\ 
f_2&(-3,0,0,-1) \\ \hline
\end{array}$ &$
\begin{array}{|c|c|}
\multicolumn{2}{c}{\text{$\mathfrak{e}_6^{(2)}$ Fiber }} \\ \hline    
%f_3^*& (-2,-1,1,0)\\ 
g_1& (-5, -2, 0, -2)\\
g_2  & (-5,-1,0,-2)\\ 
h_1 &(-7,-2,0,-3)\\ 
 k_1 & (-9,-3,0,-4) \\ \hline
\end{array} $  
\end{tabular}
\caption{\label{tab:E62F6Rays}\textit{
The toric rays, of an $\mathfrak{e}_6^{(2)}$ genus-one fibration $\hat{X}^C$ over a $\mathbb{F}_6$ base with $f_2$ being the affine node.
}}
\end{center}
\end{table}
The associated threefold $\hat{X}^C$ admits Hodge numbers 
\begin{align}
    (\, h^{1,1}\, ,\, h^{2,1}\, (h^{2,1}_{np})\,  )(\hat{X}^C)=(\, 7\, ,\, 151 (0)) \, .
\end{align}
The structure of the fibration can be readily be read off from the structure of the toric vertices: The $\mathbb{F}_6$ base is given via the projection onto the last two coordinates of each ray. The identification of base rays are
\begin{align}
  \left\{ x_1; y_0; y_1; f_2, g_1, g_2, h_1 , k_1  \right\} \xrightarrow{\pi}  \left\{ \hat{x}_1; \hat{y}_0; \hat{y}_1; \hat{x}_0  \right\} \text{ with } \,  \hat{x}_0= f_2 g_1^2 g_2^2 h_1^3 k_1^4 \, .
\end{align}
The base curves $D_{\hat{y}_i}$ are the respective $0$-curves while $D_{\hat{x}_1}$ is the $+6$ curve
and we identify $\mathcal{Z}=D_{\hat{x}_0}: \hat{x}_0 = 0$ as the $-6$ curve. This example is significant for various reasons: First, it has a trivial $\ff_4$ charged hypermultiplet spectrum, i.e. a trivial Higgs branch, analogously to the Non-Higgsable cluster theories. Second, this 5D theory makes it clear, that this theory can not
be uplifted to 6D with an $\ff_4$ gauge theory,  as the $-6$ self-intersection would make it impossible to satisfy the 6D gauge anomalies.

\subsection{Transitions to $\mathfrak{e}_6^{(2)}$} 
\label{ssec:E62Transitions}

Another way to understand various 6D and 5D theories is to connect them via geometric transitions. In the following we will carry this out for the $\mathfrak{e}_6^{(2)}$ fibration and its Jacobian.  
These geometric transitions typically map to chains of Un/Higgsings in the effective theories. We can therefore learn a lot about the (change of) degrees of freedom in the various theories by moving in a larger moduli space. In particular we want to discuss the relationship between the twisted affine fibration and an untwisted one. Hence we need to consider tunings in the complex structure moduli space that result in an unfolding of the fibers and match the geometric quantities to the associated 5D states. Our starting point is therefore a tuning of the geometry that introduces a section to the genus-one fibration 
such that it becomes birationally equivalent to the Jacobian fibration. This gives us a useful starting point from which we can follow the branching structure in 5D towards the phase with the twisted $\fe_6^{(2)}$.

For this we recall the critical role of the monodromy divisor $\overline{D}_{2,X}$ and its intersection with the curve $\mathcal{Z}$ in the base to produce the affine folding. To unfold the geometry we thus have two options:
\begin{enumerate}
\item Tuning the monodromy divisor, such that it does not intersect $\mathcal{Z}$ anymore, resulting in an un-folded genus one fibration with $\mathfrak{e}_7^{(1)}$ fiber. This model might correspond to the generalized Tate-tuning
\begin{align}
\label{eq:E7NoSec}
   \Vec{ n}=  \left\{3,3,2,1,1,2,1,0 \right\} \,. 
\end{align}
\item Tuning the geometry to an elliptic fibration results in a complete removal of the monodromy divisor and adding a two sections via the generalized Tate-vector
\begin{align}  
\label{eq:E7Sec}
     \Vec{n}=  \left\{3,3,2,1,-,2,1,0 \right\} \, \text{ for } \mathfrak{e}_7^{(1)} \, .
\end{align}
Here the entry $``-"$ refers to setting the coefficient $d_5$ to zero globally. Doing so splits the monodromy divisor into a perfect square
\begin{align}
    \overline{D}_{2,X} \xrightarrow{d_5 \rightarrow 0} d_8^2 \, .
\end{align} 
\end{enumerate} 
In the following we will discuss a chain of transitions, that will go through both phases. For ease of exposition we start from the tuning to an elliptic fibration, i.e. the second type of tuning. 
 \subsubsection*{The $\mathfrak{e}_7^{(1)} \times \mathfrak{u}_1$ model}
\begin{figure}[t!]
 \begin{center}
 {\footnotesize
 \begin{picture}(00, 90)
 \put(-150,30){\includegraphics[scale=0.6]{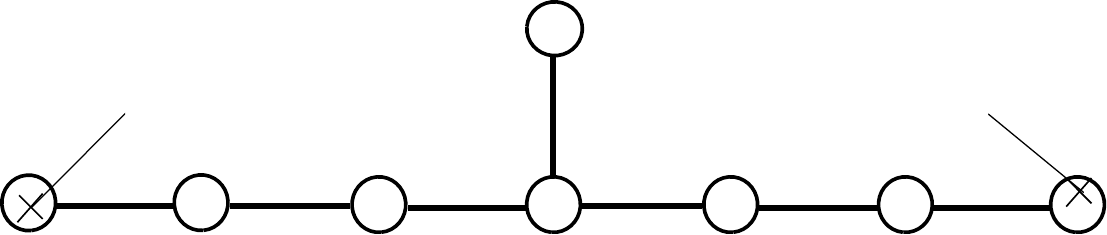} } 
   \put(20,90){$[g_1]$}
  \put(20,78){$2$}
     \put(-7,50){$[k_1]$}
       \put(20,40){$4$}
     
        \put(-150,50){$[f_2]$}
        \put(-120,70){$[X]$}
          \put(-100,50){$[g_2]$}
           \put(-80,40){$2$}
    \put(155,50){$[f_4]$}
         \put(120,70){$[e_1]$}  
         \put(173,40){$1$}
        
  \put(-47,50){$[h_1]$}                
     \put(-30,40){$3$}              
                                
            \put(-130,40){$1$}
      
  \put(57,50){$[h_2]$} 
  \put(74,40){$3$} 
  
   \put(107,50){$[g_3]$} 
  \put(124,40){$2$} 
  
 \end{picture} 
 }
 \caption{\label{fig:E7U1}\textit{Depiction of the $\mathfrak{e}_7^{(1)}$ fiber and their corresponding 
 fibral divisors. Depicted are also the two section $X=0$ and $e_1=0$ that intersect the affine and fundamental node respectively.}} 
 \end{center}
 \end{figure}
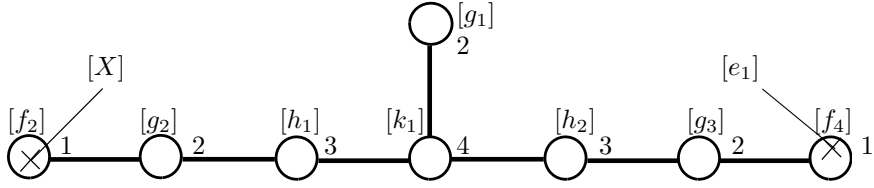

Using the tuning in Eqn.~\eqref{eq:E7Sec} we obtain a Morrison-Park type of $U(1)$ model. The singular model can  directly be mapped into the birational Weierstrass model where we find at leading orders
\begin{align}
\begin{split}
\label{eq:discriE7U1}
f=z^3 d_1 d_8  + \mathcal{O}(z^4) \, , \quad
g=z^5 d_1  p_4(d_i) + \mathcal{O}(z^6) \,, \quad
\Delta=z^9 d_1^3 d_8^6 + \mathcal{O}(d_1 ^2 z^{10}) \, .
\end{split}
\end{align}We can directly infer the gauge group and matter structure from the above model.  We find two discriminant components. The first component is given by $z=d_1=0$. This is a $(4,6)$ locus which we have tuned to zero via the specialization given in \eqref{eq:E62tuning46}. The second component is a $(4,5,10)$ locus at $z=d_8=0$ where we find matter in the $\mathbf{56}$ representation which admits a non-trivial charge under the additional $U(1)_{6D}$.
By resolving the full geometry, we can compute its charge directly.
 The $U(1)_{6D}$ symmetry is associated to the higher rank Mordell-Weil group of the geometry whcich is generated by the two 1-sections
 \begin{align}
     s_0: X=0\, \quad \text{ and } \, \quad  s_1: e_1=0 \, ,
 \end{align}
 as reviewed in Section~\ref{sec:Circle_review}. The fully resolved model is then given as the hypersurface
\begin{align}
    p=&d_1 f_2^3 f_4^2 g_2^2 g_3 h_1 X^4 e_1 + 
 d_2 f_2^3 f_4^2 g_1^2 g_2^3 g_3^2 h_1^3 h_2^2 l_1^3 X^3 Y e_1^2 + 
 d_3 f_2^2 f_4 g_1^2 g_2^2 g_3 h_1^2 h_2 l_1^2 X^2 Y^2 e_1 \nonumber \\& + 
 d_4 f_2 g_1^2 g_2 h_1 l_1 X Y^3  + 
 %d_5 f_2 g_1^4 g_2^2 g3 k1^3 k2^2 l1^4 Y^4  + 
 d_6 f_2^2 f_4^2 g_1 g_2^2 g_3^2 h_1^2 h_2^2 l_1^2 X^2 Z e_1^2+ 
 d_7 f_2 f_4 g_1 g_2 g_3 h_1 h_2 l_1 X Y Z e_1 \nonumber \\  & + d_8 g_1 Y^2 Z  + f_4 g_3 k_2 Z^2 e_1 \, ,
\end{align}
with the Stanley-Reisner ideal
 \begin{align}
     \mathcal{SRI}= \{&
   X Y, e_1 Y, f_2 Y, f_4 Y, g_2 Y, g_3 Y, k_1 Y, h_1 Y, h_2 Y, X Z, f_2 Z, g_1 Z, \
g_2 Z, k_1 Z, h_1 Z, g_1 X,\nonumber \\ & e_1 g_1, f_4 g_1, g_1 g_3,  g_1 h_2, f_4 X, g_2 X, g_3 X, 
k_1 X, h_1 X, h_2 X, e_1 g_2, e_1 g_3, e_1 k_1, e_1 h_1, \nonumber \\ & e_1 h_2, f_2 g_3, f_2 k_1, f_2 
h_1, f_2 h_2, f_4 k_1, f_4 h_2, g_2 h_2, h_1 h_2, g_2 g_3, g_2 k_1
      \} \, .
 \end{align}
 The fibral divisors $f_2,f_4,g_1,g_2,g_3,h_1,h_2,l_1$ resolve the fibral singularity fully and exhibit an intersection picture as depicted in Figure~\ref{fig:E7U1}. 
 The fibral divisors intersect like the affine $\mathfrak{e}_7^{(1)}$-Cartan matrix as expected, with ordering as
\begin{align}
\label{eq:E7weights2}
\begin{array}{c}
 (g_1) \\
 (f_2) (g_2)(h_1)(k_1)(h_2)(g_3) (f_4) 
  \end{array} =\begin{array}{c}
 (\alpha_7) \\
 (\alpha_0) (\alpha_6)(\alpha_5)(\alpha_4)(\alpha_3)(\alpha_2) (\alpha_1)  
  \end{array} \, . 
\end{align}

The intersection picture is relevant as the additional section $s_1$ intersects the fundamental node $\alpha_1$ of the $\mathfrak{e}_7^{(1)}$ fiber. This signals a non-trivial mixing of the $U(1)_{6D}$ with the center  of $\mathfrak{e}_7$ which yields the global gauge group 
\begin{align*}
    G_{6D}= \frac{E_7 \times U(1)_{6D}}{\mathbb{Z}_2} \, .
\end{align*}
This is clear, when writing the $U(1)$ generator in terms of the orthogonal divisors. First we note that the naive $U(1)$ generator is generated by the difference of the sections $[e_1]-[X]$. However, this combination is not properly orthogonalized with respect to $f_4$, that is the fundamental node $\alpha_1$. 
We thus must therefore orthogonalize the $U(1)_{6D}$ properly by taking the linear combination of fibral divisors
\begin{align}
\label{eq:E7Shift}
    \sigma(s_1)= [e_1]-[X] + \frac12 \underbrace{\sum  b_i \alpha_i}_{W_C} \, , \quad \text{ with } b_i=\{3, 4, 5, 6, 4, 2, 3\} \, ,
\end{align}
to obtain the actual $U(1)_{6D}$ that is orthogonal to the $\fe_7$.
The important point is that the linear combination $W_C$ corresponds to the $\mathbb{Z}_2$ center generator of $E_7$. Hence its appearance in the $U(1)_{6D}$ generator implies a global shift by a $\mathbb{Z}_2$ center values \cite{Grimm:2015zea,Cvetic:2017epq} of $E_7$.  
Note that the above configuration admits also a  $\mathbb{Z}_2$ symmetry 
when interchanging the two sections\footnote{
Similarly, one can obtain the dual Shioda map 
\begin{align}
\omega:   \sigma(\tilde{s_1}) =  \sigma(\tilde{s_1}) + \Delta \, ,
\end{align}
upon relabeling roots and zero-section. The difference term   
 $\Delta=2 (s_0-s_1)+
\frac32 (\alpha_1-\alpha_0) + (\alpha_6-\alpha_2)  + \frac12 (\alpha_5 -\alpha_3) $ vanishes upon the identification of the four classes that are related via the folding procedure.   
} and reordering the fibral divisors:
\begin{align}
    \omega: \quad \alpha_{0} \leftrightarrow {\alpha_{1}}\,, \alpha_{6} \leftrightarrow {\alpha_{2}}\, ,\alpha_{3} \leftrightarrow {\alpha_{4}} \quad \text{ and }\quad s_0 \leftrightarrow s_1  \, .
\end{align}
From this perspective we directly see that a fundamental $\mathbf{56}$ must have a $\frac12$-fractional charge under the $U(1)_{6D}$. Clearly, the $U(1)_{6D}$ get conjugated upon exchanging the Shioda map generator. Hence we have constructed a concrete geometric realization of the section-interchange symmetry, discussed in Section~\ref{ssec:E7U1TwistExampe}.
This can be verified by investigating the explicit degeneration locus of the $E_7$ fiber over 
  $d_8 =0$, where $\mathbf{56}$-plet states reside. Their charges and multiplicities can then be computed as
\begin{align}
\label{eq:56hypers}
    n_{\mathbf{56}_\frac12} = 4 (1-g) + \frac12 \mathcal{Z}^2 \, ,
\end{align}
which we have expressed in terms of geometric intersection numbers. 
Note that this  might look puzzling at first: As $\mathbf{56}$-plets are pseudo-real, one typically obtains half-hypermultiplts in 6D. 
In our case the $\mathbf{56}$-plets carry $U(1)_{6D}$ charge and hence renders the states not pseudo-real anymore. We therefore expect to find full hypermultiplets, which is exactly accounted for by the multiplicity formula \eqref{eq:56hypers}.  

In an example model over an $\mathbb{F}_0$ base given in Appendix~\ref{app:E62Example} we find the Hodge numbers
\begin{align}
\label{eq:E7U1Euler}
    (h^{1,1},h^{2,1})(X^A)=(11,59) \,.
\end{align}
We identify the K\"ahler moduli as the eight from the gauge group and three from the fiber and base respectively. We then identify
\begin{align}
\mathcal{Z} \sim H_1 \, , \qquad \mathcal{S}_9 \sim 2H_2  \, ,
\end{align}
and $c_1\sim 2H_1 + 2 H_2$ where $H_{1,2}$ are the hyperplane classes in $\mathbb{F}_0$. Using the divisor classes given in Appendix~\ref{app:LBData} and \eqref{eq:E7Sec} we can then identify $[d_4]\sim 3H_1+6H_2$ and $[d_8]\sim 2H_1 + 4H_2 $. This allows us to deduce the 6D/5D F/M-theory spectrum which includes the following matter 
\begin{align}
\label{eq:E7U1Spec}
\begin{split}
    H= 4 \cdot \mathbf{56}_\frac12 \oplus 24 \cdot \mathbf{1}_2 \oplus   70 \cdot \mathbf{1}_1 \oplus  60 \cdot \mathbf{1}_0\,,  \qquad T=1\,, \qquad V=\mathbf{133} \oplus \mathbf{1} \, .
   \end{split}
\end{align}
\subsubsection*{The $\fe_7^{(1)}$ genus-one model}
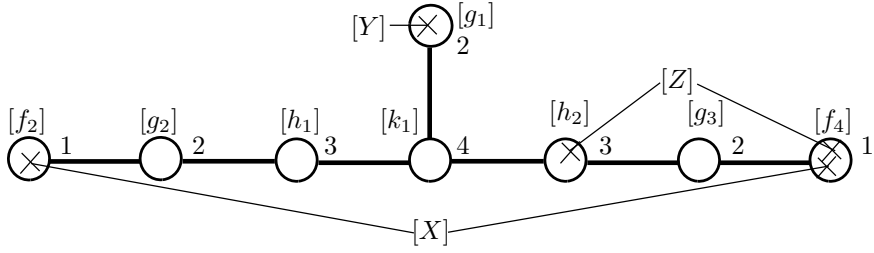
\begin{figure}[t!]
 \begin{center}
 {\footnotesize 
 \begin{picture}(00, 90)
 \put(-150,10){\includegraphics[scale=0.6]{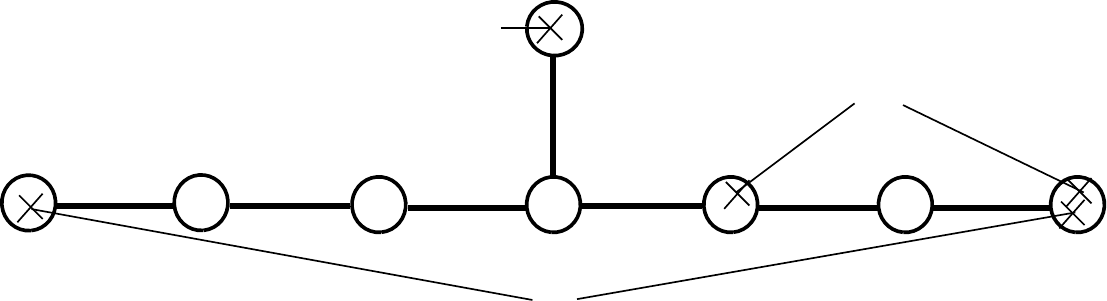} } 
   \put(20,90){$[g_1]$}
      \put(-20,85){$[Y]$}
  \put(20,78){$2$}
     \put(-9,50){$[k_1]$}
       \put(3,8){$[X]$}
       \put(20,40){$4$}
     
        \put(-150,50){$[f_2]$}
          \put(-100,50){$[g_2]$}
           \put(-80,40){$2$}
    \put(155,50){$[f_4]$}
         \put(97,65){$[Z]$}  
         \put(173,40){$1$}
        
  \put(-47,50){$[h_1]$}                
     \put(-30,40){$3$}              
                                
            \put(-130,40){$1$}
      
  \put(55,53){$[h_2]$} 
  \put(74,40){$3$} 
  
   \put(107,53){$[g_3]$} 
  \put(124,40){$2$} 
  
 \end{picture} 
 }
 \caption{\label{fig:E7Unhiggs}\textit{Depiction of the $(E_7 \times \mathbb{Z}_4)/\mathbb{Z}_2$ Dynkin diagram obtained by unfolding $\fe_6^{(2)}$. We depict the divisor classes of the respective $\mathbb{P}^1$'s as well as their Dynkin multiplicity and intersections with the two sections $[X]$ and $[Y]$ as well as the 4-section $[Z]$. The affine node is given by the fibral curve at $f_2=0$.}}
 \end{center}
 \end{figure}

The next step in the chain towards the $\mathfrak{e}_6^{(2)}$ theory
is to remove the sections. This can be done via a conifold transition which does not affect the $\fe_7^{(1)}$ fiber but only the additional $U(1)_{6D}$ part as reviewed  in Section~\ref{ssec:Genus1Review}.
From a geometric point of view, we simply change the generic fiber structure from an $Bl^1 \mathbb{P}^2_{112}$ to $\mathbb{P}^2_{112}$ which does not affect the
$E_7$ fiber itself. This is associated to a Higgsing on a field which is a hypermultiplet with charges $\mathbf{1}_2$ and a non-trivial KK charge.
The resulting resolved genus-one fibration $X^B$ is given via the hypersurface
 \begin{align}
 p= & d_1   f_2^3 f_4^2 g_2^2 g_3 h_1 X^4 + 
 d_2   f_2^3 f_3 f_4^2 g_1^2 g_2^3 g_3^2 h_1^3 h_2^2 k_1^3 X^3 Y + 
 d_3  f_2^2 f_3 f_4 g_1^2 g_2^2 g_3 h_1^2 h_2 k_1^2 X^2 Y^2 \nonumber \\ &+ 
 d_4   f_2   g_1^2 g_2 h_1 k_1 X Y^3 + 
 d_5   f_2   g_1^4 g_2^2 g_3 h_1^3 h_2^2 k_1^4 Y^4  + 
 d_6  f_2^2   f_4^2 g_1 g_2^2 g_3^2 h_1^2 h_2^2 k1^2 X^2 Z  \nonumber \\ &  + 
 d_7   f_2  f_4 g_1 g_2 g_3 h_1 h_2 k_1 X Y Z + d_8    g_1 Y^2 Z + 
 d_9   f_4 g_3 h_2 Z^2 \, ,
 \end{align}
  with the following Stanley-Reisner ideal  
  \begin{align}
\mathcal{SRI}: \{ &Y X, Y f_2, Y f_4, Y g_2, Y g_3, Y h_1, Y h_2, Y k_1,    Z f_2, Z g_1, Z g_2, Z h_1, Z k_1,  k_1 h_1, \nonumber \\ &  X g_1, f_4 g_1, g_1 g_3, g_1 h_2, X g_2, X g_3, X h_1, X h_2, X k_1, f_4 h_1,f_4 h_2, f_4 k_1, f_2 g_3, f_2 h_1,\nonumber \\ &  f_2 h_2, f_2 k_1, g_2 h_2, g_3 k_1, g_2 k_1 \} \, .
\end{align}
Note that the fiber model is the quartic  model which we obtain by adding the monomials $d_5 Y^4$ after removing the MW generator given by the divisor $e_1=0$. The intersection picture is summarized in Figure~\ref{fig:E7Unhiggs} which yields the expected $\mathfrak{e}_7^{(1)}$ shape.
 As a result of the conifold, the two 1-sections $s_0$ and $s_1$ of the Morrison-Park model have merged into a single 2-section $s_0^{(2)}: X=0$ that maintains their intersections with the fiber as shown in Figure \ref{fig:E7Unhiggs}. However as we identify 
the class of the 2-section with the generator of generic fiber class which  contains the $\mathfrak{u}_{1,KK}$ we find that it is not orthogonalized to the $\fe_7$ fibral divisors. This is not surprising when recalling that the 5D $U(1)_E$ is a linear combination of KK U(1) and the massive $\mathfrak{u}_{1,6D}$ which mixed with the center of $E_7$. 
Hence the orthogonalized 2-section analog in the genus-one fibration is given as  
\begin{align}
    \sigma(s^{(2)}_0)= [X]+ \frac12 W_C    \, ,  
\end{align}
shifted by the same center Wilson line combination $W_C$ as in \eqref{eq:E7Shift}. 
In the above linear combination we have fixed the $\mathfrak{u}_{1,E}$ charges and its mixture
with the KK-tower. E.g. since $X=0$ is a two-section we have 
\begin{align}
   \sigma(s^{(2)}_0) [T^2]=2 \, .
\end{align} Thus an M2 brane state that wraps the a fibral curve and the generic torus $n$ times yields 5D particle with $\mathfrak{u}_{1,E}$ charges $q+2n$. The above structure is in fact useful to deduce the discrete charges $q$ in the 6D (and Jacobian) compactification as we discuss next.

Shrinking all nodes but the affine $f_2$ we can map the configuration into the Jacobian, i.e. the singular Weierstrass form. At leading order in $z$ we still find the form to be given as \eqref{eq:discriE7U1}. 

\begin{figure}[t!]
 \begin{center}
 {\footnotesize
 \begin{picture}(00, 90)
 \put(-150,10){\includegraphics[scale=0.6]{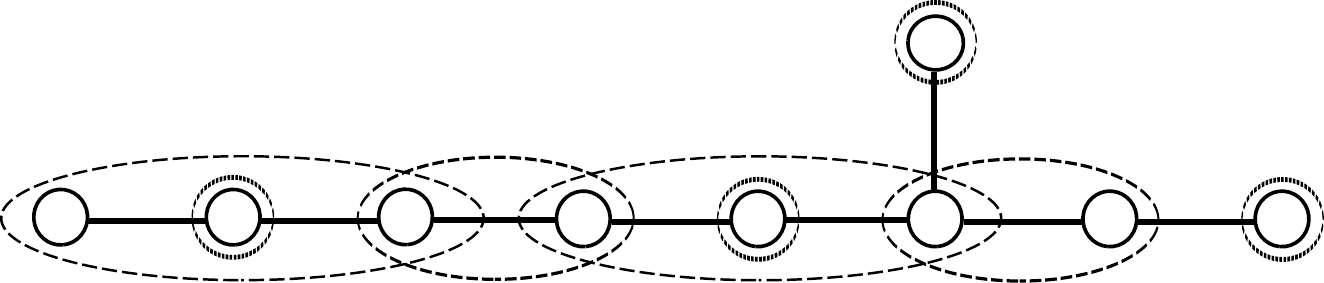} } 

      \put(90,80){$[h_2]$}
        \put(-90,55){$[f_4]$}
         \put(-10,55){$[g_2]$}
           \put(60,55){$[g_3]$}
          \put(140,55){$[k_1]$}
         \put(200,45){$[g_1]$}
          \put(42,15){$[h_1]$}
          \put(-110,15){$[f_2]$}

  \put(-143,27){  {\tiny$ \mathbb{P}^1_{1,2 } $}}
   \put(-93,27){  {\tiny$ \mathbb{P}^1_{\alpha_0 } $}}
   \put(-43,27){  {\tiny$ \mathbb{P}^1_{1,1 } $}} 
       \put(9,27){  {\tiny$ \mathbb{P}^1_{2,1 } $}}
           \put(59,27){  {\tiny$ \mathbb{P}^1_{2,2} $}}
        \put(110,27){  {\tiny$ \mathbb{P}^1_{2,3} $}}
               \put(161,27){  {\tiny$ \mathbb{P}^1_{4,1} $}}
                \put(211,27){  {\tiny$ \mathbb{P}^1_{\alpha_7} $}}
                \put(111,78){  {\tiny$ \mathbb{P}^1_{\alpha_3} $}}

 \end{picture}
 }
 \caption{\label{fig:E8Degen}\textit{The fiber structure over $\mathcal{Z}$ and $d_8=0$ where $\mathbf{56}_{\frac12}$ matter resides. The fiber admits an $E_8$ structure consistent with the expectation of the Jacobian.}} 
 \end{center}
 \end{figure}
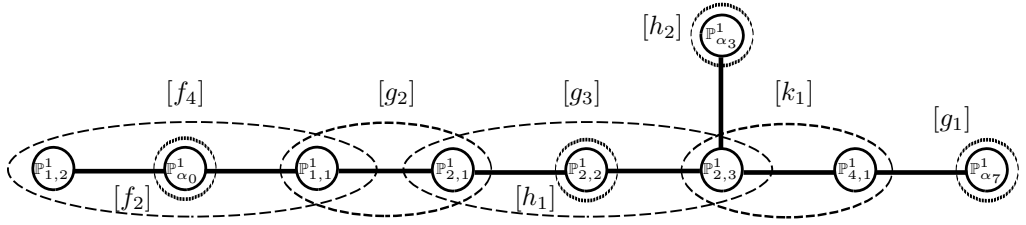

In the following we investigate the locus of the $\fe_7$ fundamentals (inside $d_8=0$) in a bit more detail. In order to discuss the degeneration at this codimension two locus we list all eight fibral curves of $\mathfrak{e}_7^{(1)}$ in detail as 
\begin{align}
\begin{split}
 \mathbb{P}_{\alpha_0}: f_2=&0 =  d_9 f_4+d_8 g_1 \, , 
\\
\mathbb{P}_{\alpha_1}: f_4=&0=d_4 f_2 g_2+d_5 f_2 g_2^2 g_3+d_8 Z \, ,\\
\mathbb{P}_{\alpha_2}: g_3 =&0 = d_4 g_2 h_1 k_1+d_8 Z \, , \\
\mathbb{P}_{\alpha_3}: h_2=&0=d_1 g_3+d_4 k_1+d_8 Z \, ,\\
\mathbb{P}_{\alpha_4}: k_1=&0=d_8 g_1+d_1 g_3 h_1+d_9 g_3 h_2 \, ,\\  
\mathbb{P}_{\alpha_5}: h_1 = & 0  =  d_8 g_1+d_9 g_3 \, , \\
\mathbb{P}_{\alpha_6}: g_2=&0= d_8 g_1+d_9 f_4 g_3 \, ,\\
\mathbb{P}_{\alpha_7}: g_1 =&0= 1+d_1 f_2^3 g_2^2 h_1 \, .
\end{split}
\end{align}
Some curves above become reducible when imposing  $d_8=0$ where curves split as
\begin{align}
    \mathbb{P}^1_{\alpha_i} \xrightarrow{d_8=0}  \sum_j \mathbb{P}^1_{i,j}  \, .
\end{align} 
Indeed, we find the curves to degenerate into $9$ distinct  curves with an intersection structure which resembles the affine $E_8$ Dynkin diagram as given in Figure~\ref{fig:E8Degen}. This is in fact also what we expect from the Jacobian, which highlights again the close connection between the two (non-birational equivalent) geometries.  

We can use the resolved geometry to deduce the properties of the 5D theory on the Coulomb branch. We do so by considering M2 branes that wrap the reducible fibral curves and compute their charges under the Cartan subalgebra. 
This allows us to compute $E_7$ weights and $U(1)_E$ charges by considering the following set of intersections
\begin{align}
    (  f_4, g_4,h_2, k_1, h_1 , g_2, g_1 | f_2)_{\sigma(s_0^2)} \cdot \mathbb{P}^1_{i,j} \, .
\end{align}
We pick two of the reducible curves, given in Figure~\ref{fig:E8Degen} and compute the intersections
\begin{align}
     \mathbb{P}^1_{1,1} :\qquad&  ( -1,1,0,0,0,-1,0|1)_{(\frac12)}\, , \\
     \mathbb{P}^1_{2,1} :\qquad& (1,-1,0,0,1,-1,0|1)_{(\frac12)} \, .
\end{align}
Note that we have also highlighted non-trivial charges under the affine node $f_2$, which is non-trivial for both curves. 
The above charges are weights of an $\mathbf{56}$ representation with $\mathfrak{u}_{1,E}$ charges $q=\frac12$ as expected.

When taking the singular limit, the above genus-one fibration 
admits the 5D gauge group $G_{5D}=(E_7 \times U(1)_E)/\mathbb{Z}_2$. The global shift by $W_C$ effectively mixes the $E_7$ center with $U(1)_E$ similar to the case of the Shioda map \cite{Cvetic:2018bni}.
This non-trivial global structure lifts in a non-trivial way to the 6D theory as the $\mathfrak{u}_{1,E}$ contains the massive $\mathfrak{u}_{1,6D}$ and in particular its discrete remnant.
Recall that the 6D theory is obtained from the Higgsing on a $\mathbf{1}_2$ state, which breaks the $\mathfrak{u}_{1,6D}$ to a $\mathbb{Z}_2$ subgroup. The non-trivial center mixing however changes this group to  
\begin{align}
G_{5D}= (E_7 \times U(1)_E)/\mathbb{Z}_2 \xrightarrow{\text{ F-theory lift}} G_{6D}= \frac{E_7 \times \mathbb{Z}_4}{\mathbb{Z}_2} \, ,
\end{align}
 (see \cite{Garcia-Etxebarria:2014qua} for related effects).
This global structure is compatible with the 6D charge assignment
\begin{align}
    \mathbf{56}_{\frac12} \, , \quad \mathbf{1}_1  \, ,
\end{align}
given that they are taken mod 2, even though that the $\fe_7$ fundamentals have  a fractional charge. 
Furthermore these discrete charges make clear that the $\fe_7$ fundamentals are not pseudo-real and thus not half hypermultiplets, consistent with the count of their multiplicities 
\eqref{eq:56hypers}.

When performing the transition from $X^A$ to $X^B$ as shown in Figure \ref{fig:TwistingChain}, the $U(1)$ generator gets eaten by the $\bf{1}_2$ Higgs fields. The remaining degrees of freedom of the same charge become neutral singlets. This can be seen in the example geometry, which admits Hodge numbers
\begin{align}
(h^{1,1},h^{2,1})(X^B) = (10,82) \, .
\end{align}
Note the decrease in K\"ahler parameters in $X^B$ compared to $X^A$ (see \eqref{eq:E7U1Euler}) agrees with the breaking of $U(1)_{6D}$ and the enhancement of 23 complex structure parameters can be explained by the $24$ $\mathbf{1}_2$ fields. 

\subsubsection*{Transition to $\mathfrak{e}_6^{(2)}$} 
The geometry $X^B$ provides a key step in our goal of engineering a transition to the $\fe_6^{(2)}$ fibration. In terms of the singular geometries this is due to the fact that the order two monodromy divisor does not intersect the $\fe_7$ singularity, while it does in the $\fe_6^{(2)}$. 
This can also be observed by comparing the vanishing orders in the two relative singular genus-one geometries (i.e. one can compare the Tate-vectors in \eqref{eq:TateE62} with \eqref{eq:E7NoSec}). The $\fe_6^{(2)}$ leaves the monodromy divisor invariant, while the $\fe_7$ tuning does not. 

In the following we want to discuss the geometric transition and match it to the 5D field theory. 
The two types of states of relevance, are the scalars in the 5D vector multiplets, that break the theory to a (partial) Cartan subgroup and parametrize the Coulomb branch, 
and the $\mathbf{56}_{\frac12}$-plets whose VEV parametrize the Higgs branch of the $\fe_7$ theory.
A straight forward way to obtain the $\ff_4$ theory from $\fe_7$, is via the non-Levy type branching as given as 
\begin{align}
    \fe_7 \times \mathfrak{u}_{1,E} \rightarrow \ff_4 \times \fsu_2 \times \mathfrak{u}_{1,E}\, \rightarrow \ff_4 \times  \widehat{\mathfrak{u}}_{1,E} \, ,
\end{align}
which can be attributed to a VEV in the $(\mathbf{1},\mathbf{4})_\frac12 \in \mathbf{56}_{\frac12}$. 
This however raises an immediate puzzle: In order to trigger the Higgsing on the $\mathbf{56}_\frac12$-plets, we require those fields to be massless.

Those states however admit a $\mathfrak{u}_{1,E}$ charge, and hence one expects them to admit a non-trivial mass, proportional to the $\mathfrak{u}_{1,E}$ Coulomb branch parameter. As the $\mathfrak{u}_{1,E}$ Coulomb branch parameter is proportional to the inverse of the 6D circle radius. It cannot be taken to zero size at finite distance in the moduli space\footnote{Recall that Higgsing from the $\fe_7 \times \mathfrak{u}_{1,6D}$ theory has fixed the $U(1)_{6D}$ Coulomb branch parameter to $\xi_{6D}=\tau/2$. This also implies that $\mathbf{56}_\frac12$ must be massive at the origin of the $\fe_7$ Coulomb branch as well.}.  
However, the fact that we can perform an infinitesimal deformation in complex structure moduli space to go from the singular $X^B$ geometry to $X^C$ suggests that those states should be massless. 
%%%%%%%%%%%%%%%%

%%%%%%%%%%%%%%%%%%%%%%%%%%%%%%%%%%%%%%%%%
We do not resolve the above puzzle here, instead we bypass it by a different mechanism to obatin a massless Higgs field. Instead of a simple complex structure deformation in the singular limit, we first move on the 5D Coulomb branch by performing a particular resolution of the singular fiber. Thus we break the gauge algebra to $\fe_7 \rightarrow \fe_6 \times \mathfrak{u}_1$ which then will allow us to obtain massless states, that can be given a vev. 

In geometry this branch corresponds to only shrinking the $\fe_6$ sub-diagram inside of the affine $\fe_7^{(1)}$ given in Figure~\ref{fig:E7Unhiggs}, keeping the affine and fundamental nodes at finite sizes.  The resulting 5D theory admits  an $\fe_6 \times \mathfrak{u}_{1,A} \times \mathfrak{u}_{1,E}$  gauge algebra and the $\fe_7$ representations decompose as \begin{align}
\begin{split}
    \mathbf{133}_0& \rightarrow \mathbf{78}_{0,0} \oplus 
\mathbf{27}_{\frac23,0} \oplus  \overline{\mathbf{27}}_{-\frac23,0} \oplus \mathbf{1}_{0,0}\ , \\ 
    \mathbf{56}_\frac12& \rightarrow \mathbf{27}_{-\frac13,\frac12} \oplus  \overline{\mathbf{27}}_{\frac13,\frac12} \oplus \mathbf{1}_{-1,\frac12} \oplus \mathbf{1}_{1,\frac12} \, .
    \end{split}
\end{align}
As expected only  5D states with trivial $\mathfrak{u}_{1,A} \times \mathfrak{u}_{1,E}$ charges are massless which again is not possible at a generic point in moduli space. The masses of the resulting particles becomes
\begin{align}
    m_{\mathbf{27}_{-\frac13,\frac12}}=| -\frac13 \xi_A + \frac12 \tau | \, .
\end{align}
Hence when taking the linear combination 
\begin{align}
\hat{\mathfrak{u}}_{1,E} = \frac32 \mathfrak{u}_{1,A} - \mathfrak{u}_{1,E} \,,
\end{align}
there is a combined CB parameter
\begin{align}
    \hat{\xi}=\frac32\xi_A-\xi_E\, .
\end{align}
Setting $\hat{\xi}=0$ makes the $\mathbf{27}_{-\frac13,\frac12}$ plets massless. Now we are in the position to give those states a vev and break the gauge symmetry. The resulting gauge theory breaking is then given as  $\fe_6 \mathfrak{u}_{1,A} \times \mathfrak{u}_{1,E} \rightarrow  \ff_4 \times \hat{\mathfrak{u}}_{1,E}$. Note that the $\fe_6 \rightarrow \ff_4$ folding inside of the $\fe_7$ resembles nicely the geometry action of the $\fe_7^{(1)} \rightarrow \fe_6^{(2)}$ fiber. 
After collecting the charges under the massless $\mathfrak{u}_1$ generator, we can obtain the multiplicities of the massless $\ff_4$ BPS states in terms of the $\fe_7$ ones as 
\begin{align}
\label{eq:260Transition}
n_{\mathbf{26}_0}= n_{\mathbf{56}_\mathbf{1/2}}+n_{\mathbf{133}_0}-1 \, ,
\end{align}
which matches the geometric computation.   Similarly, the massive $\mathbf{26}_1$ plets are given as
\begin{align}
n_{\mathbf{26}_1}= n_{\mathbf{56}_\mathbf{1/2}}+2 n_{\mathbf{133}_0} \, .
\end{align}
The above expression however, deviates from the geometric computation of $\mathbf{26}_1$ multiplicities, performed in the $\fe_6^{(2)}$ theory by over counting by one. 
This suggests that one more massive field should decouple in the transition. It would be interesting to investigate this miss-match in future work. 

We can proceed similarly with the $\hat{\mathfrak{u}}_{1,E}$-charged singlets.
However we want to focus in particular on the change of neutral singlets $\mathbf{1}_0$, as those contribute to the change in complex structure moduli $h^{2,1}$ which is easy to match across the transition. E.g. there are neutral fields $\mathbf{27}_{-\frac13,\frac12}\rightarrow \mathbf{26}_0 \oplus \mathbf{1}_0$ upon the breaking to $\ff_4 \times \hat{\mathfrak{u}}_{1,E}$ that contribute uncharged singlets to the geometry. However further singlet contributions are more subtle and deserve a more detailed discussion: Recall that the shifted KK symmetry is given as 
$\mathfrak{u}_{1,E}=\mathfrak{u}_{1,6D}-2 \mathfrak{u}_{1,KK}$. Thus before Higgsing, a singlet state with charges 
$\mathbf{1}_{-1,\frac12} \in \mathbf{56}_{\frac12}$ under $\mathfrak{u}_{1,A}\times \mathfrak{u}_{1,E}$
is accompanied with a KK-tower $\mathbf{1}_{-1,\frac12+2n}$ having KK momenta $n$. Upon the Higgs transition such singlet towers therefore change into  of $\mathbf{1}_{-2+2n}$ under the unbroken $\hat{\mathfrak{u}}_{1,E}$.
This is important as there is a mode in the tower with $n=-1$ that is neutral and massless and therefore contributing to $h^{2,1}$ in the geometry $X^A$ of the $\fe_6^{(2)}$ theory. 

   The second source of neutral singlets are the massless representations $\mathbf{26}_{0}$. Recall that those hypermultiplets admit only 24 non-trivial weights and two more trivial ones that contribute to $h^{2,1}(X^A)$ each. (see Section~\ref{sec:TVSUT} for more details).
   Hence, each $\mathbf{56}_{\frac12}$ KK-tower of states, contributes four neutral singlet components upon the Higgsing. 
Finally we need to take the $7$ neutral states in any $\mathbf{133}$ hypermultiplet that reduce to just $4$ in $\mathbf{52}$ and therefore need to be subtracted. Also recall, that there are the Goldstone modes $\mathbf{26}_0 \oplus \mathbf{1}_0$ that need to be removed and thus, subtract three more neutral fields. 

 Summing up all contributions, we obtain the following change in neutral charged fields 
\begin{align}
\Delta \mathbf{1}_{0} = -3 \cdot n_{\mathbf{133}_{0}} -3 + 4 \cdot \mathbf{56_{1/2}} =  13 (1 - g) + 2 \mathcal{Z}^2 - 6g \, ,  
\end{align} 
which is identified with the change in complex structure moduli upon the conifold transition.

The above considerations are confirmed in the example geometries. 
Recall the Hodge numbers of the two geometries are
\begin{align}
  (h^{1,1},h^{2,1}):  (10,82) \rightarrow (7,95) \, .
\end{align}
As the respective gauge algebras are over 
$\mathcal{Z}^2=0$ of genus $g=0$, we find exactly the difference of $13$ complex structure moduli that was expected.

Another puzzle remains, the Higgs transition in the 6D theory associated to the Jacobian geometry appears to be more subtle than that realized in the 5D genus one geometries. When making the transition from the Jacobian of $X^B$ to that of $X^C$ we find the codimension one structure unchanged i.e. the $E_7$ singularity is preserved. At codimension two however, the locus of $\mathbf{56}$-plets has changed to the intersection locus with the two section monodromy divisor.
In particular we find that the $\mathbf{56}_q$-plets should now be counted as half-hypers which implies a change in the discrete $\mathbb{Z}_2$ charge to $q=0$ or $q=1$ but not $q=\frac12$. This furthermore implies a subtle modification of the global gauge group structure in the 6D theory as 
\begin{align}
G_{6D}= \frac{E_7 \times \mathbb{Z}_4}{\mathbb{Z}_2} \rightarrow G_{6D}=E_7 \times \mathbb{Z}_2 \, .
\end{align}
It would be interesting to explore the origin of the corresponding transition and its relation to the 5D twisted reduction in more detail in the future.

%%%%%%%%%%%%%%%%%%%%%%%%%%%%%%%%%%%%%%%%%%%%%%%%%%%%%%
\section{More Twisted Affine Fibrations}
\label{sec:MoreTwistedTheories}

\begin{figure}[t!]
\begin{center} 
	\begin{tikzpicture}[scale=1.3]
	 \draw[very thick, ->] (-4.5,1) -- (-4.5,2.5);  
  \node (A) at (-3.8,1.8) [ text width=3cm,align=center ] {\textbf{Lift}};

 \node (A) at (2,1.5) [ text width=3cm,align=center ] {\textbf{Conifold}};
 \draw[very thick, ->] (-2,1.2) -- (5,1.2);  

\node (A) at (1.2,2) [ text width=3cm,align=center ] {$ X^{B} $};
\node (A) at (3.8,2) [ text width=3cm,align=center ] {$ X^{A} $};

\node (A) at (-1.2,2) [ text width=3cm,align=center ] {$ X^C $};
 
	\node (A) at (-4,0.5) [draw,rounded corners,very thick,text width=2.8cm,align=center] {{\bf \textcolor{black}{5D M-Th on \\ Genus One }}};
	
	\node (A) at (-4,3) [draw,rounded corners,very thick,text width=2.8cm,align=center] {{\bf \textcolor{black}{6D F-Th on\\ Jacobian  \\}}};
  
  	\node (A) at (-1.3,3) [draw,rounded corners,very thick,text width=2.3cm,align=center ] {{\bf \textcolor{black}{ G=$ \overline{\mathfrak{g}} \times U(1)$}}};

  \node (A) at (-1.3,0.5) [draw,rounded corners,very thick,text width=2.3cm,align=center ] {{\bf \textcolor{black}{Fiber: $  \overline{\mathfrak{g}}^{(1)} $
  \\
 $ \mathbf{G}= \overline{\fg} \times \mathfrak{u}^2_1 $
  }}};
 
  \node (A) at (1.2,3) [draw,rounded corners,very thick,text width=2.5cm,align=center ] {{\bf \textcolor{black}{ G=$   \overline{\mathfrak{g}} \times \mathbb{Z}_n $  }}};

  \node (A) at (1.2,0.5) [draw,rounded corners,very thick,text width=2.5cm,align=center ] {{\bf \textcolor{black}{Fiber: $  \overline{\mathfrak{g}}^{(1)} $ \\
   $ \mathbf{G}= \overline{\fg} \times \mathfrak{u}_{1,E} $
} }};

 \node (A) at (3.8,0.5) [draw,rounded corners,very thick,text width=2.8cm,align=center ] {{\bf \textcolor{black}{Fiber: $\mathfrak{g}^{(n)} $ \\ $\mathbf{G}=\mathfrak{g}^\circ \times \mathfrak{u}_{1,E} $  }}};

 \node (A) at (3.8,3) [draw,rounded corners,very thick,text width=2.8cm,align=center ] {{\bf \textcolor{black}{G= $ \overline{\mathfrak{g}}  \times \mathbb{Z}_n $  }}};

	\end{tikzpicture}
\end{center} 
\caption{\textit{The chain F/M-theory vacua on torus fibered threefolds $X^I$, highlighting the 6D gauge symmetry $G$ and 5D fiber structures. 
}}
\label{fig:TwistingChain}
\end{figure}
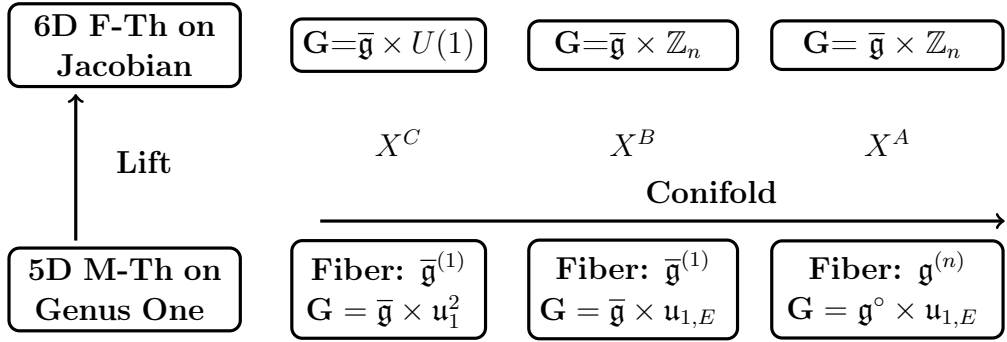

In this section we will follow the same approach used in Section \ref{sec:main_5D_example} to study the  $\mathfrak{e}_6^{(2)}$ geometry and discuss further examples of twisted fibers. We will consider the twisted algebras $\mathfrak{su}_3^{(2)}$ and $\mathfrak{so}_8^{(3)}$. For each we will discuss the structure of their Jacobians and the general conifold relations as given in Figure~\ref{fig:TwistingChain}.

\subsection{The geometry of $\mathfrak{su}_3^{(2)}$ }
\label{ssec:su32Geometry}
\begin{figure}[t!]
 \begin{center}
 {\footnotesize
 \begin{picture}(00, 120)
 \put(-150,10){\includegraphics[scale=0.6]{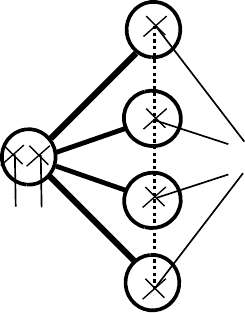} }
 \put(-150,70){$[g_1]$}
 \put(-95,15){$1$}
  \put(-160,53){$2$}
  \put(-155,30){$[X]$}
  	  \put(-140,30){$[Y]$}
 \put(-110,110){$[f_1]$}
 \put(-78,51){$[Z]$} 
 \put(-40,55){$\xrightarrow{\overline{D}_{4,Z}=0}$     }
 \put(50,50){\includegraphics[scale=0.6]{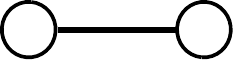} }
 
  \put(56,55){$2$} 
    \put(106,55){$4$} 
 \end{picture} 
 }
 \caption{{\it \label{fig:A2Folding}Depiction of an $\mathfrak{su}_3^{(2)}$ fiber. The two two-sections $X, Y$ both intersect the  middle node. All four fibral curves in $[f_1]$ are permuted by the four-section $Z$. At the intersection with the monodromy divisor the fiber degenerates to a multiplicity two fiber.}}
 \end{center}
 \end{figure}
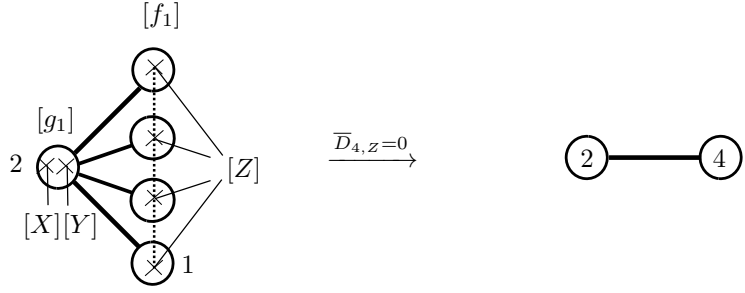

In this section we describe the geometry of the $\mathfrak{su}_3^{(2)}$ fiber. Its fiber structure is given on the left of Figure~\ref{fig:A2Folding}. When interpreted as a fiber graph, we see that its geometric cover is that of an $\fso_8^{(1)}$ with a degree 4 folding acting on the four outer nodes. More precisely, the folding 
proceeds by its  $\mathbb{Z}_2 \times \mathbb{Z}_2$ center.
In order to perform this geometric action, we require a four-section with the appropriate monodromy. One such example can be found in a manifold with CICY genus-one fiber given by $\mathbb{P}^3[2,2]$ \cite{Schimannek:2019ijf}. Luckily, there is a simpler model which allows us to resolve the singularity in a single hypersurface: The model is the quartic fibration given in \eqref{eq:Quartic}. Besides the two 2-sections $X,Y$ there is also a 4-section $Z=0$ that brings with it the right monodromy. 

 \begin{table}[t!]
 \begin{center} 
$\begin{array}{|c|c|c|c|c|c|c|c|c||c|c|c||c|c|}\hline
     s_1   & s_2 & s_3 & s_ 4 & s_5 & s_6 & s_7 & s_8 & s_9 &f & g & \Delta & \text{Genus-1 Fiber} & \text{Jac Fiber} \\ \hline
    0 & 0 & 0 & 0 &1 & 1 & 1 & 1 & 1 & 2 & 3 & 6 & \mathfrak{g}_2^{(1)} & \mathfrak{g}_2^{(1)} \\ \hline
    0 & 0 & 0 & 0 & 0 & 1 & 1 & 1 & 1 & 2 & 3 & 6 & \mathfrak{su}_3^{(2)} & \mathfrak{g}_2^{(1)} \\ \hline 
      0 & 0 & 0 & 0 & 0 & 0 & 0 & 0 & 1 & 0 & 0 & 2 &\mathfrak{su}_2^{(1)} & \mathfrak{su}_2^{(1)}  \\ \hline 
\end{array}   $
\caption{\label{tab:QuarticTuninga22}\textit{Tunings of sections in the quartic genus one model \eqref{eq:Quartic} for $\fsu_3^{(2)}$ fibers and related models. The fiber structure in genus-one and Jacobian is given in the two rightmost columns. }}
\end{center} 
\end{table}

To proceed, we employ the same algorithm used to analyze the geometry in Section \ref{sec:main_5D_example}: We need to tune an $I_0^*$ type of singularity over $Z=0$ in the quartic model and let it intersect the order four monodromy divisor. This  monodromy divisor can be found by first considering the four-section $Z=0$ which yields
\begin{align}
\sigma_4: \{ Z=p=0\}: s_1 X^4 + s_2 X^3 Y + s_3 X^2 Y^2 + s_4 X Y^3 + s_5 Y^4
\end{align}
and then finding its discriminant locus 
%\begin{align}
%\overline{D}_{4,Z}= -4 (s_3^2 - 3 s_2 s_4 + 12 s_1 s_5)^3  + (2 s_3^3 - 9 s_3 (s_2 s_4 + 8 s_1 s_5) + 
%   27 (s_1 s_4^2 + s_2^2 s_5))^2 \, .
%\end{align} 
\begin{align}
\begin{split}
 \overline{D}_{4,Z}=&    2 s_ 2^3 (2 s_ 4^3 - 9 s_ 3 s_ 4 s_ 5) + 
 2 s_ 1 s_ 2 s_ 4 (-9 s_ 3 s_ 4^2 + 40 s_ 3^2 s_ 5 + 96 s_ 1 s_ 5^2) \\ & + 27 s_ 2^4 s_ 5^2 +
  s_ 2^2 (-s_ 3^2 s_ 4^2 + 4 s_ 3^3 s_ 5 + 6 s_ 1 s_ 4^2 s_ 5 - 
    144 s_ 1 s_ 3 s_ 5^2) \\ & + 
 s_ 1 (4 s_ 3^3 s_ 4^2 + 27 s_ 1 s_ 4^4 - 
    16 s_ 3 (s_ 3^3 + 9 s_ 1 s_ 4^2) s_ 5 + 128 s_ 1 s_ 3^2 s_ 5^2 - 
    256 s_ 1^2 s_ 5^3) \, .
    \end{split}
\end{align}
Proceeding further, in Table~\ref{tab:QuarticTuninga22} we have summarized the vanishing of the generalized Tate-coefficients and the resulting fiber structures in the quartic threefold and its Jacobian fibration.
Hence the singular genus-one $\fsu_3^{(2)}$ is given via the tuning
\begin{align}
s_i \rightarrow z^{n_i} d_i\, , \quad  \text{ with } \quad 
    \Vec{n}=(
    0 , 0 , 0 , 0 , 0 , 1 , 1 , 1 , 1) \, .
\end{align}
Note that we are required to keep the $s_9$ section explicitly at this point to be able to perform the tuning. 

The twisted affine fiber is given by the second row of Table~\ref{tab:QuarticTuninga22}. The tuning is chosen so that it respects the generic four-monodromy locus. The resolved quartic hypersurface is then given as 
\begin{align}
p= & d_1 X^4 + d_2 X^3 Y + d_3 X^2 Y^2 + d_4 X Y^3 + d_5 Y^4 \nonumber  \\ & + d_6 f_1 g_1 X^2 Z + 
 d_7 f_1 g_1 X Y Z + d_8 f_1 g_1 Y^2 Z + d_9 f_1 Z^2\, .
\end{align}
The geometry can be fully analyzed given the Stanley-Reisner ideal
\begin{align}
\mathcal{SRI}: \{ Y X Z,
   Z g_1 \} \, ,
\end{align} 
and the two fibral divisors are explicitly given as  
\begin{align}
\begin{split}
\mathbb{P}^1_{0,i}: \{ p =f_1 = 0\} : & \, d_1 X^4 + d_2 X^3 Y + d_3 X^2 Y^2 + d_4 X Y^3 + d_5 Y^4   \, , \qquad i=1\ldots 4 \, , \\
\mathbb{P}^1_{1}: \{ p =g_1 = 0\} : & \,  d_1 X^4 + d_2 X^3 Y + d_3 X^2 Y^2 + d_4 X Y^3 + d_5 Y^4 +  d_9 f_1 \,  .
\end{split}
\end{align}  
Note that the four $\mathbb{P}^1$s inside of $f_1=0$ admit an $S_4$ symmetry and take the very same form as the four-section $Z=0$ and hence, experience the same mondromy effect. 

In order to obtain the correct gauge algebra factor of the quartic model we need to choose the fibration so that upon factorizing the divisor $\mathcal{Z}: z=0$, the residual $d_9$ polynomials becomes a constant. For this we set 
\begin{align}
    [z\, \, d_9] \sim c_1 - \mathcal{S}_7 + \mathcal{S}_9 \, ,
\end{align}
and solving for $\mathcal{S}_7\sim c_1   + \mathcal{S}_9 - \mathcal{Z}$. In what follows we will then set $d_9=1$.

The fibral graphs and intersections with the 2-sections and 4-section is given in Figure~\ref{fig:A2Folding}. Note again, that we obtain an order two multiple fiber, over each intersection locus of $\overline{D}_{4,Z}$ with the fibral curve $\mathcal{Z}$ as depicted in Figure~\ref{fig:A2Folding}. 
 
\subsubsection{Counting states in the $\fsu_3^{(2)}$ genus-One model}
As in the previous section, we turn our attention now to the BPS states. To count the states, within the twisted affine fiber we split up all curves of the  $\mathfrak{so}_8$ cover into orbits of the Eigenspaces of the monodromy action. The starting point are curves $\mathcal{C}$ of self-intersection $-2$ that are linear combinations $\mathcal{C}= \sum_i a_i \mathcal{C}_i$ for $i=0\ldots 4$ where the basis curves $\mathcal{C}_i$ intersect as nodes of the affine Dynkin diagram $\fso_8^{(1)}$. Graphically we write those curves as 
\begin{align}
\mathcal{C}_i= a_0 - \begin{array}{c} a_1 \\ a_2  \\ a_3  \end{array} - a_4  \, , 
 \qquad \text{ with } a_i \in \mathbb{N} \text{ and } i=0\ldots 4 .
\end{align}
  In this way we setup the 24 states that make up the W bosons of the adjoint of the covering $\mathfrak{so}_8$. In the second step, we incorporate the $\mathbb{Z}_4$ action that fixes $\mathcal{C}_2$ but rotates the other $C_i$. The invariant combination of curves is then given by the collection $(\mathcal{C}_2, \mathcal{C}_1 + \mathcal{C}_3 + \mathcal{C}_4 + \mathcal{C}_5)$ which we will also use in order to compute the weights of states via the intersections of the covering algebra. Then within the 24 states, we need to collect together the Eigenspaces under the  action and split them up into shrinkable and non-shrinkable curves. The result is given in Table~\ref{tab:A22states}.
  
\begin{table}[h!]
\begin{align*}
\begin{array}{|l | c|c|c|c| }\hline
 \text{Mono}& \# Curves  & \text{Example Curve  } &  \fsu_2 \subset \mathfrak{su}_3^{(2)} \text{  weight } & \chi  \\ \hline 
\text{Inv}  & 2 &
   \begin{array}{c}0 \\ 0-1-0 \\ 0  \end{array}
   &  (-2)  & g    \\ \hline
  \mathbb{Z}_4  & 2 & {\scriptsize \begin{array}{c} \begin{array}{c}0
 \\ 1-0-0 \\ 0  \end{array}+ \begin{array}{c}1
 \\ 0-0-0 \\ 0  \end{array} \\ +
 \begin{array}{c}0
 \\ 0-0-0 \\ 1  \end{array}+ \begin{array}{c}0
 \\ 0-0-1 \\ 0  \end{array} \end{array}}
  & (-1)  & g\prime  \\ \hline 
   \mathbb{Z}_4 & 2 &  {\scriptsize \begin{array}{c} \begin{array}{c}0
 \\ 1-1-0 \\ 0  \end{array}+ \begin{array}{c}1
 \\ 0-1-0 \\ 0  \end{array} \\ +
 \begin{array}{c}0
 \\ 0-1-0 \\ 1  \end{array}+ \begin{array}{c}0
 \\ 0-1-1 \\ 0  \end{array} \end{array}} & (-1)  & g\prime  \\ \hline 
    \mathbb{Z}_4 & 1 &  {\scriptsize \begin{array}{c} \begin{array}{c}1
 \\ 0-1-0 \\ 1  \end{array}+ \begin{array}{c}1
 \\ 0-1-0 \\ 1  \end{array} \\ +
 \begin{array}{c}0
 \\ 1-1-0 \\ 1  \end{array}+ \begin{array}{c}0
 \\ 1-1-1 \\ 0  \end{array} \end{array}} & (0)  & g\prime  \\ \hline
   \mathbb{Z}_2 & 1 &  {\scriptsize \begin{array}{c} \begin{array}{c}1
 \\ 1-1-0 \\ 0  \end{array}+ \begin{array}{c}0
 \\ 0-1-1 \\ 1  \end{array}   \end{array}} & (0)  & \overline{g}\prime  \\ \hline 
   \end{array}
   \end{align*}
    \caption{
\label{tab:A22states}\textit{
Schematic splitting of the 24 curves of $\mathfrak{so}_8$ roots into $\mathbb{Z}_4$ invariant curves,
their $\mathfrak{su}_2 \subset \mathfrak{su}_3^{(2)}$ charges and respective moduli space dimensions $\chi$ computed in \eqref{eq:modspacea22}.}}
\end{table} 
The permutation action $\sigma_4$ acts on the outer roots as
\begin{align}
   \sigma_4: \quad \begin{array}{c}
2. \\
4.-0-1. \\ 
3.
  \end{array}  \, ,
\end{align}
where we have chosen the $i-th$ root to be permuted to $i+1$
under the $\mathbb{Z}_4$ we quotient by\footnote{Note that there might be other potential subgroups of the $S_4$ symmetry to quotient by and an interesting point for future research (see. e.g. \cite{Esole:2017qeh} for $S_3$ example of $\fso_8$).
}.
We find two $\mathbf{2}_{\frac12}$ doublets as well as two types of singlets as shown in Table~\ref{tab:A22states}. Note that one of the singlets is only an order two ramified curve under the order four action. 

In order to compute the moduli space dimensions, we need the divisor class of the monodromy divisor. In order to deduce it, we first note that the ramification divisor is associated to the base line bundle class
\begin{align}
[ \overline{D}_{4,Z}] \sim   2[d_1] + 4 [d_4] \, ,
\end{align}
and we further observe
\begin{align}
    [d_1] \sim 2 c_1 - 2 \mathcal{S}_9 + \mathcal{Z}\,  \qquad 
    [d_4] \sim 2 c_1 + \mathcal{S}_9 - 2 \mathcal{Z}\, .
\end{align}
 Plugging those in, we find that $\mathcal{S}_9$ cancels out and we obtain  
 \begin{align}
 [ \overline{D}_{4,Z}]  \sim 12 c_1 -6 \mathcal{Z} \, .
\end{align}
Intersecting the monodromy divisor $\overline{D}_{Z,4}$ with $\mathcal{Z}$ yields the points of ramification
that we need to compute the moduli spaces of fibered $\mathbb{P}^1$'s. There we count the moduli spaces $g\prime$ and $\overline{g}\prime$ for degree 4 and 2 covered curves respectively which are given as 
\begin{align}
\label{eq:modspacea22}
    g\prime   = 9(1-g)+3 \mathcal{Z}^2 + g \, , \qquad 
     \overline{g}\prime   = 11 (1-g)+3 \mathcal{Z}^2 + g \, , 
\end{align}
When adding up, the non-trivial representations, we obtain
\begin{align}
\label{eq:A22Spectrum}
n_{3_0} = g \, , \qquad n_{2_{\frac12}}=& 18(1-g)+ 6 \mathcal{Z}^2 + 2g \,, \\
n_{\mathbf{1}}=& 20(1-g)+ 6 \mathcal{Z}^2 + 2 g \, .  
\end{align}
Note that all states carry a non-trivial $\mathfrak{u}_{1,E}$ charge and are therefore generically massive. 
\subsubsection*{Comparison with $\fsu_3 \rightarrow \fsu_3^{(2)}$ reduction}
In pure field theories, the 5D $\mathfrak{su}_3^{(2)}$ theories have been observed to arise from an outer automorphism twisted 6D theory with a $\fsu_3$ gauge algebra. The autormorphism acts as complex conjugation among the fundamental representations, and leaves an $\fsu_2$ maximal sub-algebra. Similarly to the examples given in Section \ref{sec:main_5D_example} we will observe below that the Jacobian admits an $\fg_2$ gauge algebra instead of $\fsu_3$. Thus, we expect that the genus one geometry we outline here will arise from a circle reduction of a 6D $\fg_2$ F-theory background, but that the twisting symmetry will act on the $\fsu_3 \in \fg_2$ sub-algebra as in the usual gauge-theoretic twisted reductions.

The decomposition of the fields under the twisted reduction\footnote{Note that our shifted KK charge for the fundamentals differ from those, given in \cite{Lee:2022uiq}.} results in  
 \begin{align}
    \mathbf{8} \rightarrow \mathbf{3}_0 \oplus \mathbf{2}_\frac14 \oplus \mathbf{2}_\frac34 \oplus \mathbf{1}_\frac12  \quad \text{ and } \quad (\mathbf{3} \oplus \overline{\mathbf{3}}) \rightarrow    \mathbf{2}_\frac14 \oplus \mathbf{2}_\frac34 \oplus \mathbf{1}_0 \oplus \mathbf{1}_\frac12 \, .
\end{align}
The subscript $\mathbf{R}_s$ denotes the shifted KK charge $q=s+n$ and KK-momentum $n$ typically used in the literature. 
  
Recollecting the 5D multiplicities of states, we write the 6D spectrum in geometric terms as  
\begin{align}
n_{\mathbf{3}_0}=18(1-g) + 6 \mathcal{Z}^3  \, , \qquad  n_\mathbf{8} = g \, .
\end{align}  
 Since $\mathbf{2}_\frac{1}{4}$ states are in the same tower as $\mathbf{2}_\frac34$ we can simply sum them up to obtain the multiplicity of
\begin{align}
    n_{\mathbf{2}_\frac14}=18(1-g)+ 6 \mathcal{Z}^2 +2g \, .
\end{align}
Finally we need to match the normalization of the shifted KK-charges
with those used for our intersections. In order to do so, we need to multiply the above shifted KK-charges by two to obtain a match of both multiplicities.

Similar as in the $\fe_6^{(2)}$ case, we also obtain a prediction for the neutral singlets that are inherited from the fundamentals, given as
\begin{align}\label{singlet_np}
    n_{\mathbf{1}_0}=  \frac{n_\mathbf{3}}{2} = 9(1-g) + 3\mathcal{Z}^2   \, .
\end{align}
The states above in \eqref{singlet_np} we will interpret as non-polynomial deformations of the toric hypersurface in Section~\ref{sec:TVSUT}. 
\subsubsection*{State counting in the $\mathfrak{g}_2^{(1)}$ Jacobian} 
Lets compare those computations with that of the Jacobian which admits an unaltered 6D uplift. Similar as for the $\fe_6^{(2)}$ geometry, we expect to find a generic $I_0^*$ fiber i.e. a $\fg_2$. Indeed at leading order of the Weierstrass form along the singularity at $z=0$ we find
\begin{align}
\begin{split}
f=&-z^2  \frac13 (d_3^2  + d_2 d_4- 4 d_1 d_5)  + \mathcal{O}(z^3)\, , \\
g=&-z^3
\frac{2}{27}(2 d_3^3 - 9 d_2 d_3 d_4 + 27 d_1 d_4^2 + 27 d_2^2 d_5 - 72 d_1 d_3 d_5) + \mathcal{O}(z^4) \, , \\
\Delta=&z^6 \overline{D}_{4,Z} + \mathcal{O}(z^7) \, .
\end{split}
\end{align}
with $\overline{D}_{4,Z}$ the monodromy divisor. In the Jacobian we find over $z=0$ simply an $I_0^{*}$ non-split fiber, i.e. an $\mathfrak{g}_2$ singularity. The monodromy divisor takes on a similar role as in the genus-one fibration but in this case, it only rotates three of $\fso_8$ outer roots, instead of four. 
Hence the fibral $\mathbb{P}^1$'s undergo a degree $d=3$ monodromy. 
This is important to compute the moduli spaces of the curves that yields $n_\mathbf{7}=(\overline{g}-g)$ fundamental hypermultiplets of $\fg_2$. Using the same amount of ramification points as for $\fsu_3^{(2)}$ but this time $d=3$ in the RH theorem yields the multiplicities 
\begin{align}
n_{\mathbf{14}}=g\, , \qquad n_{\mathbf{7}} = 10 (1-g) + 3 \mathcal{Z}^2 \, ,
\end{align}
which satisfies the $\fg_2$ anomalies in 6D.

\subsubsection*{Toric examples}
In the following we discuss two concrete toric examples .
The first threefold, admits a $\mathbb{P}^2$ base 
with an $\fsu_3^{(2)}$ over a $\mathcal{Z}^2=+1$ curve of genus $g=0$. The toric rays are given in Appendix~\ref{app:Example3folds} and the respective Hodge numbers are computed as 
\begin{align}
\label{eq:a22onP2}
    (h^{1,1}, h^{2,1}(h^{2,1}_{np}))=(3,107(12)) \, .
\end{align}
Note that the number of K\"ahler parameters matches the expectation. The charged spectrum comes with   multiplicities  
\begin{align}
    n_{\mathbf{3}_0}=0 \, , \qquad n_{\mathbf{2}_{\frac12}} = 24 \, .
\end{align} 
We will return to the model above, when discussing another transition to an actual $\fsu_2^{(1)}$ fiber, which shares the very same Hodge numbers as the model above in Section~\ref{sec:TVSUT}.

Finally we also present the geometry that correspond to the twisted reduction of a non-Higgsable $\fsu_3$ theory. From a 6D perspective, this comes when $g=0$ and $\mathcal{Z}^2=-3$. From the general considerations we find, that we should also find no $\fsu_3^{(2)}$ particles. Such an example can easily be constructed, over an $\mathbb{F}_3$ base.  
\begin{table}[t!]
\begin{center}
\begin{tabular}{ccc}
$
\begin{array}{|c|c|}
\multicolumn{2}{c}{\text{ Generic Fiber }} \\ \hline 
X&(-1,1,0,0)\\
Y&(-1,-1,0,0)\\
Z&(1,0,0,0)\\ \hline  
\end{array}$  & $ 
\begin{array}{|c|c|}
\multicolumn{2}{c}{\text{ $\mathbb{F}_3$ Base }} \\ \hline  
x_1 & (-2, -1, 0, 1) \\
y_0 & (10, 10, -1, -3) \\
y_1 & (-9, -10, 1, 0)\\ 
f_1&(1,0,0,-1) \\ \hline 
\end{array}$ &$
\begin{array}{|c|c|}
\multicolumn{2}{c}{\text{$\mathfrak{su}_3^{(2)}$ Fiber }} \\ \hline    
g_1& (1, 0, 0, -2)\\ \hline
\end{array} $  
\end{tabular}
\caption{\label{tab:A22F3Rays}\textit{
Toric rays, of a genus-one fibration over 
$\mathbb{F}_3$ base, with $\mathfrak{su}_3^{(2)}$ 
where the ray $f_1$ yields the affine node.
}}
\end{center}
\end{table}
The toric rays for this model are given in Table~\ref{tab:A22F3Rays} and
the Hodge numbers can be computed as 
    \begin{align} 
    (h^{1,1}, h^{2,1}(h^{2,1}_{np}))=(4,130(0)) \, ,
\end{align}  
admitting the four expected K\"ahler parameters and notably only polynomial complex structure deformations. 

\subsubsection{Transitions to $\fsu_3^{(2)}$ }
In the following we explore various transitions towards the the 5D 
  $\mathfrak{su}_3^{(2)}$ theory. We have observed already, that this theory admits a  
  $\mathfrak{g}_2 \times \mathbb{Z}_2$ theory in its Jacobian. We therefore tune the theory to an $\mathfrak{g}_2 \times \mathfrak{u}_{1,6D}$ in 6D via various transitions. 
  
Geometrically, the way this fibration is constructed follows the same approach as in the $\fe_6^{(2)}$ theory, by having a monodromy that acts non-trivially an the affine node. 

 \subsubsection*{The $\mathfrak{g} _2 \times [ U(1)_{6D} \rightarrow  \mathbb{Z}_2]$ transition}
The basis for the $\mathfrak{su}_3^{(2)}$ model was the existence of the 4-section monodromy divisor that folds the $\fso_8^{(1)}$ covering fiber.  
We will not present a complete such unfolding, but only a deformation 
which grants us a section and hence an elliptic model. As the generic fiber description is the same as in the $\mathfrak{e}_6^{(2)}$ model. 
  
For this we can start with an $\mathfrak{g}_2\times \mathfrak{u}_{1,6D}$ 6D theory
by using the tuning given by general Tate-vector\footnote{Note that we have used $f_1$ as the affine coordinate to make contact with the $\mathfrak{su}_3^{(2)}$ geometry. The affine coordinate however is given as $f_2$, which can equally well be given by the tuning
\begin{align}
    \vec{n}=\{ 3,2,1,0,-,2,1,0,0\}
\end{align}
in the quartic model. Consistently, this tuning also leaves $\overline{D}_{3,Z}$ invariant.
} specified in Table~\ref{tab:QuarticTuninga22} and at the same time tuning $d_5$ to zero globally.

 The resolved hypersurface is given as 
 \begin{align}
p=& d_1 f_2^3 U^3 X^4 + d_2 f_2^2 U^2 X^3 Y + d_3 f_2 U X^2 Y^2 + d_4 X Y^3  \nonumber \\ &+ d_6 f_1 f_2^2 g_1 U^2 X^2 Z + d_7 f_1 f_2 g_1 U X Y Z + 
 d_8 f_1 g_1 Y^2 Z + d_9 f_1 U Z^2 \, .
 \end{align}
 
 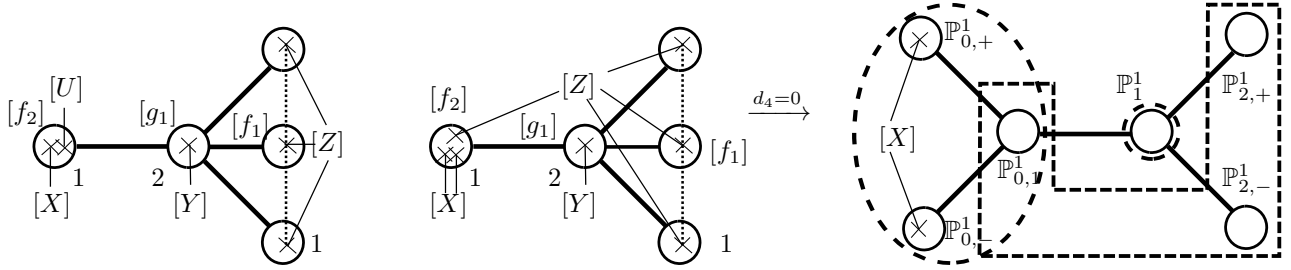
\begin{figure}[t!]
 \begin{center}
 {\footnotesize
 \begin{picture}(00, 100)
 \put(-230,10){\includegraphics[scale=0.6]{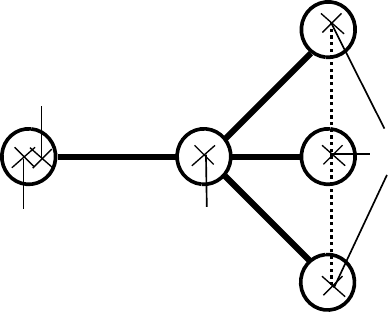} }
   \put(-230,30){$[X]$}
   \put(-177,30){$[Y]$}
     \put(-224,74){$[U]$}
    \put(-240,65){$[f_2]$}
  \put(-190,65){$[g_1]$}
   \put(-155,60){$[f_1]$}
   \put(-125,53){$[Z]$}

 \put(-215,40){$1$}
 \put(-185,40){$2$}
 \put(-125,15){$1$}

  \put(-65,40){$1$}
 \put(-35,40){$2$}
 \put(30,15){$1$}
 
  \put(-80,30){$[X]$}
  	  \put(-30,30){$[Y]$}
  	   \put(-30,75){$[Z]$}
  	  
 \put(-80,70){$[f_2]$} 
  \put(-46,60){$[g_1]$} 
    \put(26,50){$[f_1]$} 
 
  \put(-80,10){\includegraphics[scale=0.6]{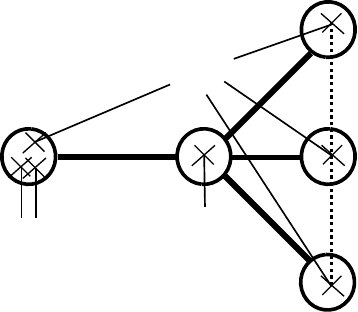} }
 \put(80,10){\includegraphics[scale=0.6]{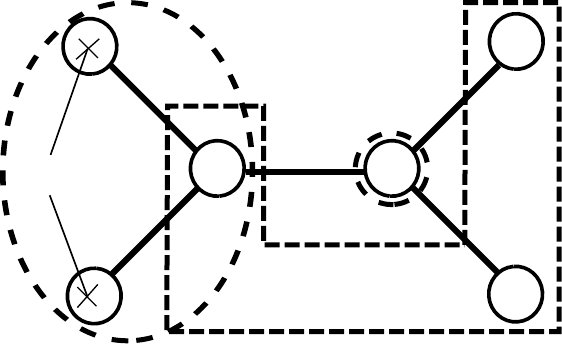} }
   \put(180,75){$\mathbb{P}^1_1$} 
     \put(135,43){$\mathbb{P}^1_{0,1}$} 
        \put(115,95){$\mathbb{P}^1_{0,+}$} 
         \put(115,20){$\mathbb{P}^1_{0,-}$} 
                  \put(220,40){$\mathbb{P}^1_{2,-}$}
                  \put(220,75){$\mathbb{P}^1_{2,+}$} 
   \put(90,55){$[X]$}
 \put(40,65){$\xrightarrow{d_4=0}$}
 \end{picture} 
 }
 \caption{\label{fig:G2Genus1model}\textit{Summary of three fiber structures. On the left is the $\mathbf{g}_2$ fiber in the elliptic model and in the middle in the genus-one model and their intersections. Shown are the two-sections $X,Y$ and the four-section $Z=0$. On the right is the $\mathfrak{so}_{10}$ degeneration shown where matter is localized. }}
 \end{center}
 \end{figure}   
The zero-section is again denoted by $U=0$  and the second section by $X=0$. The above fiber is of $\mathfrak{g}_2$ type as can be seen from the fibral curves and their intersections using the
\begin{align}
\mathcal{SRI}: \{  & Y X, Y U, Y f_2, e_1 Z, e_1 U, e_1 f_1, e_1 g_1, e_1 f_2, 
 X Z, Z g_1, Z f_2, X f_1, f_1 f_2, X g_1      \} \, .
\end{align}
 This leads to the following form of fibral curves:
 \begin{align}
 \begin{split}
 \mathbb{P}^1_{i}: [f_1]: \quad & d_1 U^3 + d_2 U^2 Y + d_3 U Y^2 + d_4 Y^3 \, , \qquad i=1,2,3\, , \\
 \mathbb{P}^1_2: [g_1]:\quad & d_9 f_1 U + d_1 f_2^3 U^3 + d_2 f_2^2 U^2 Y + d_3 f_2 U Y^2 + d_4 Y^3 \, , \\
  \mathbb{P}^1_0: [f_2]:\quad & d_8 g_1 + d_9 U + d_4 X \, .
  \end{split}
 \end{align}
The divisor $f_1=0$ itself is a cubic polynomial that decomposes into into three irreducible curves $\mathbb{P}^1_i$, that are interchanged by the monodromies. Note that $Z=0$ is a 3-section in the MP model but will become a 4-section in the quartic genus-one fibration.

The intersections in the fiber are depicted in Figure~\ref{fig:G2Genus1model}. To compute the spectrum it is convenient to map the geometry into the Weierstrass form which we sketch to leading orders here
\begin{align}
\label{eq:Jacobiang2}
\begin{split}
f=&-\frac13 (d_3^2 - 3 d_2 d_4) z^2 + \mathcal{O}(z^3)\, , \\
g=&-\frac{1}{27} (2 d_3^3 - 9 d_2 d_3 d_4 + 27 d_1 d_4^2)z^3 + \mathcal{O}(z^3) \, ,\\
\Delta=& z^6 d_4^2 \overline{D}_{3,Z}+\mathcal{O}(z^7)\, ,
\end{split}
\end{align}
with $\overline{D}_{3,Z}$ the $\mathbb{Z}_3$ monodromy divisor of the three-section $Z=0$, that also permutes the three $\mathbb{P}^1$'s in $f_1=0$. Its polynomial equation is  explicitly given as
\begin{align}
\label{eq:Z3Mono}
\overline{D}_{3,Z}=(-d_2^2 d_3^2 + 4 d_1 d_3^3 + 4 d_2^3 d_4 - 18 d_1 d_2 d_3 d_4 + 27 d_1^2 d_4^2) \, .
\end{align} 
Precisely due to the reason that $\overline{D}_{3,Z}$ is not a perfect cube leads to the non-split type $I_0^*$ fiber.
    From the intersections with the additional section, we find the $U(1)$ Shioda map to be simply given as 
\begin{align}
\sigma(s_1)= [X]-[U] \, .
\end{align}
Unlike to the other examples there is no geometric mixing of the U(1) with the $\mathfrak{g}_2$ center at this point as this is simply trivial. The presence of the extra $U(1)$ leads to the expectation of non-trivial charged matter. The first source is non-localized matter induced by the monodromy locus $\overline{D}_{3,Z}$ but we also find a  $(2,3,7)$ enhancement locus over $z=d_4=0$. In the resolved model we find the curves $\mathbb{P}^1_2$ to split into two curves which thus gives the fiber the topology of $\mathfrak{so}_{10}^{(1)}$ as depicted in Figure~\ref{fig:A2Folding}. One of these components, e.g. $g_1 = U=0$ contains the zero-section and admits the following intersections with the Cartan generators $(f_1, g_1, f_2)_{\sigma(s_1)}$
 \begin{align}
 (1,-1;1)_{1} \ni \mathbf{7}_1 \, ,
 \end{align} 
 i.e. a weight from the fundamental of $\fg_2$ with non-trivial $U(1)$ charge.  
 
When Higgsing to the genus-one model, the $\fg_2$ stays basically as it is, so we will not discuss matter  multiplicities up to this point.

  When performing the conifold transition to the genus-one geometry we blow down $U$ and add the deformation term $d_5  f_1 g_1^2 Y^4$  to the hypersurface. This results in changing
  $Z=0$ and $X=0$ into a four- and a two-section respectively. 
   This is reflected by the change in the intersection picture depicted in the middle of Figure~\ref{fig:G2Genus1model} for which we used the fibral Stanley-Reisner ideal
\begin{align}
\mathcal{SRI}= \{ Y X, Y f_2,   Z g_1, X f_1 \} \, .
\end{align}
In analogy to the procedure before we use the two-section $X=0$ as the analog of the zero-section. As all of its strains intersect the affine node $\mathbb{P}^1_{\alpha_0}$ only, we do not have to orthogonalize it with respect to other $\mathfrak{g}_2$ Cartan generators, just as in the case before. Hence we use $X$ as the discrete Shioda map generator itself which in order to compute the 6D discrete $\mathbb{Z}_2$ charges or 5D KK charges.
Lets investigate the reducible fibers below in codimension one, for which we find  
 \begin{align}
 \begin{split}
 \mathbb{P}^1_{\alpha_1,i}& : \quad [f_1] \cap d_1 f_2^3 + d_2 f_2^2 Y + d_3 f_2 Y^2 + d_4 Y^3 \, , \qquad i=1,2,3\, ,  \\ 
 \mathbb{P}^1_{\alpha_2}&: \quad [g_1] \cap  d_9 f_1 + d_1 f_2^3 X^4 + d_2 f_2^2 X^3 Y + d_3 f_2 X^2 Y^2 + d_4 X Y^3\, , \\
 \mathbb{P}^1_{\alpha_0}&: \quad [f_2] \cap d_5 f_1 g_1^2 + d_4 X + d_8 f_1 g_1 Z + d_9 f_1 Z^2 \, .
 \end{split}
 \end{align}
 From the point of view of the four-section $Z=0$ it is notable, that the three fibral curve  $\mathbb{P}^1_i$ undergo only a $\mathbb{Z}_3$ monodromy. Indeed one of the four-section strains of $Z=0$ attaches to the affine node in $f_2=0$. This is also reflected in the Jacobian of the genus-one model which at leading order is exactly the same as given in \eqref{eq:Jacobiang2}. Hence we also find the same degeneration locus over $d_4=0$ again. This is the important locus that triggers the 5D Higgs transition to the $\mathfrak{su}_3^{(2)}$ theory so we will discuss it in a bit more detail.  
 Moving onto the locus $d_4=0$, the fibral curves split as summarized in Table~\ref{tab:G2G1Matterlocus}.
 \begin{table}[t!]
 \begin{align*}
\begin{array}{|c|l |}\hline
\text{node} &  \multicolumn{1}{c|}{\text{Irreducible  components over }    d_4=0 } \\ \hline
\mathbb{P}^1_{1,i}:[f_1] &  \mathbb{P}^1_{0,1} : f_2 \cap f_1 \, ,\qquad \mathbb{P}_{0,\pm}: f_1 \cap (f_2 \pm p Y)   \\ \hline 
\mathbb{P}^1_{2}: [g_1]& \mathbb{P}^1_{2}: g_1 \cap (d_9 f_1 + d_1 f_2^3 + d_2 f_2^2 Y + d_3 f_2 Y^2)    \\ \hline
\mathbb{P}^1_{0}: [f_2]&  \mathbb{P}^1_{0,1}: f_2 \cap  f_1  \qquad \mathbb{P}^1_{0,\pm}: f_2 \cap ( f_0 g_1 \pm q  Z)  \\ \hline
\end{array}
 \end{align*}
 \caption{\label{tab:G2G1Matterlocus}
 {\it The codimension two irreducible components of the fiber at $d_4=0$, where $\fg_2$ fundamentals $\mathbf{7}_1$ are located. Their intersection picture is given on the right of Figure~\ref{fig:G2Genus1model}.   
 }}
 \end{table}
 When investigating the curve
 $\mathbb{P}^1_{0,\pm} $, we find the M2 brane states to support the following weights
%These components admit the following weights with respect to $(f_1, g_1; f_2)_X $:
\begin{align}
\label{eq:g21comp}
\begin{array}{ll}
%\mathbb{P}^1_{0,1} &:(-2,0;1)_0 \, , \\
\mathbb{P}^1_{0,\pm} &:(1,0;-1)_1 \ni \mathbf{7}_1 \, , \\
%\mathbb{P}^1_{2,\pm} &:(-2,0;1)_0 \, , \\
%\mathbb{P}^1_{1} &:(3,1;-2)_0 \, .
\end{array}
\end{align}
Similar as in $U(1)_{6D}$ phase, the new fibral component gives 
rise to a charged fundamental. When lifting to 6D, the charge under $U(1)_E$ becomes a $\mathbb{Z}_2$ charge, while in the genus-one geometry it combines with the $U(1)_{KK}$ charge. 

Up to this point we have not yet computed the multiplicities of the $\mathfrak{g}_2$ charged states in either geometry.
 In order to do so, we take $f_1$ as the affine coordinate that factored out of $d_i$. The full factorization in terms of the quartic model is given   in Table~\ref{tab:QuarticTuninga22}. 

Using the formulas for the classes first, we need to set $d_9$ to be a constant in order to avoid additional $\mathfrak{su}_2$ gauge factors. In order to achieve this, we can fix the line bundle classes of the base as 
\begin{align}
\mathcal{S}_7 \sim c_1 + \mathcal{S}_9 - \mathcal{Z} \, .
\end{align}
Important for what follows are the  classes of the two sections $d_1$ and $d_4$. Those can be 
deduced to transform in the base classes
 \begin{align}
 [d_1] \sim 2 c_1 - 2\mathcal{S}_9 + \mathcal{Z}\, , \quad [d_4] \sim 2 c_1 + \mathcal{S}_9 - \mathcal{Z} \, .
 \end{align}
From this data we can read off the class of the $\mathbb{Z}_3$ monodromy divisor \eqref{eq:Z3Mono} to  
 \begin{align}
[\overline{D}_{Z,3}] \cdot \mathcal{Z} = [2d_1 + 2 d_4] \cdot \mathcal{Z} =  8 c_1 \cdot \mathcal{Z}- 2\mathcal{Z}^2-2 \mathcal{S}_9 \cdot \mathcal{Z} =  16 (1-g) + 6 \mathcal{Z}^2 -2 \mathcal{S}_9 \cdot \mathcal{Z}\, .
\end{align}  
With this topological information we can compute the multiplicities of the charged matter. First there is as usual the adjoint valued fields counted by the genus of $\mathcal{Z}$. Then there are the two sources of $\mathbf{7}$-plets  
   The first source is counted  by the intersections of $z=0$ with $d_4=0$, while the second uses the moduli spaces of the  degree three branched curve $z=0$ via Riemann-Hurwitz. The result is
\begin{align}
\begin{split}
n_{\mathbf{7}_1}=& [d_4] \cdot \mathcal{Z} = 4(1-g) + \mathcal{S}_9 \cdot \mathcal{Z} \, , \qquad n_{\mathbf{14}_0} = g  \, , \\
n_{\mathbf{7}_0} =& 2(g-1)+\frac12 R = 6(1-g)+ 3 \mathcal{Z}^2 - \mathcal{Z} \cdot  \mathcal{S}_9 \, . 
\end{split}
\end{align}
Note that there is the additional free line bundle class $\mathcal{S}_9$ which is not fixed. The freedom of tuning this classes and the intersection with $\mathcal{Z}$ allows to control the relative difference of  charged and uncharged $\mathbf{7}$-plets \footnote{The class $\mathcal{S}_9$ is constrained by the existence of sections $d_i$ to be effective.}.
Consistent with gauge anomalies we find that the sum of the representation does not depend on the additional parameter $\mathcal{S}_9$ as it must be 
\begin{align}
n_{\mathbf{7}_1}+ n_{\mathbf{7}_0} = 10 (1-g)+3 \mathcal{Z}^3 \, .
\end{align} 
To illustrate those general considerations, we discuss a concrete toric example in the following

For this we fix the base to be $\mathbb{P}^2$ and the full toric data is given in Appendix~\ref{app:Example3folds}. 
The model admits an 
$\fg_2$ over a genus 0 curve with the additional data $\mathcal{Z} \cdot \mathcal{S}_9=\mathcal{Z}^2=1$ and Hodge numbers
\begin{align}
(h^{1,1},h^{2,1})=(4,98) \, .
\end{align}
Indeed, besides the two classes of fiber and base, there are two more K\"ahler classes that parametrize the $\fg_2$ Coulomb branch.
Focusing on the $\fg_2$ charged spectrum, we find
\begin{align}
    n_{\mathbf{7}_1}=5 \, , \qquad n_{\mathbf{7}_0}=8 \, .
\end{align} 
\subsubsection*{Transition to $\fsu_3^{(2)}$}
With all the details at hand, we can perform the conifold type of transitions. In order to do so, we first need to go to a partial Coulomb branch, as in the $\fe_7 \rightarrow \fe_6^{(2)}$ case. The reason is again, that the $\mathbf{7}_1$ states admit a non-trivial $\mathfrak{u}_{1,E}$ charge and therefore do  not become massless at the Coulomb branch origin of $\fg_2$. While we can can break the $\fg_2 \rightarrow \fsu_2 \times \fsu_2$ and then move to the Coulomb branch, followed by a Higgsing. 

Instead we break the $\fg_2 \rightarrow \fsu_2 \times \mathfrak{u}_{1,A}$ such that there is a second CB modulus to make states massless. In geometry, the respective curve in
$g_1$ may be shrunk while $f_1$ and $f_2$ stays finite. This realizes an $\fsu_2 \times \mathfrak{u}_{1,A} \times \mathfrak{u}_{1,E}$ gauge algebra
and the $\mathbf{7}$ plets decompose to
\begin{align}
\begin{split}
\mathbf{7}_1& \rightarrow \mathbf{2}_{\frac12,1} \oplus \mathbf{2}_{-\frac12,1}  \oplus \mathbf{1}_{-1,1} \oplus \mathbf{1}_{1,1} \oplus \mathbf{1}_{0,1}\, , \\
\mathbf{7}_0& \rightarrow \mathbf{2}_{\frac12,0} \oplus \mathbf{2}_{-\frac12,0}  \oplus \mathbf{1}_{-1,0} \oplus \mathbf{1}_{1,0} \oplus \mathbf{1}_{0,0} \, ,\\
\mathbf{14}_0& \rightarrow \mathbf{3}_{0,0} \oplus 2\times \mathbf{2}_{-\frac12,0}  \oplus 2 \times \mathbf{2}_{-\frac12,0} \oplus \mathbf{1}_{1} \oplus \mathbf{1}_{-1,0} \oplus \mathbf{1}_{0,0}  \, .
\end{split}
\end{align}
In the $\mathbf{1}_{-1,1}\in \mathbf{7}_1$, we find a good Higgs candidate, that leaves the $\fsu_2$ subgroup invariant. This state can become massless when the  $\mathfrak{u}_{1,A}$  and $\mathfrak{u}_{1,E} $ CB moduli become the same. Now, we can Higgs on the $\mathbf{1}_{-1,1}$ by giving it a VEV that corresponds to a complex structure deformation of the geometry. Field theoretically, the Higgsing leaves a  $\hat{\mathfrak{u}}_{1,E}=\mathfrak{u}_{1,A}+\mathfrak{u}_{1,E}$ symmetry unbroken.

Collecting the $\mathbf{2}_\frac12$-plets \footnote{Note that $\mathbf{2}_\frac32$ are part of the the same KK-tower and are therefore counted as the same representation.} yields a collection of massive states of multiplicity
\begin{align}
    n_{\mathbf{2}_\frac12}= 2 n_{\mathbf{7}_1}+2 n_{\mathbf{7}_0}+4n_{\mathbf{14}_0} \, ,
\end{align}
which matches the geometric computation but over counts by two.  

In a similar way, also the charged and neutral singlets can be counted. 
 Especially the later ones correspond to complex structure moduli, which can be used to predict the change in $h^{2,1}$ upon the conifold transition, given as
\begin{align}
    \Delta h^{2,1} = 2 n_{\mathbf{7}_1} -1 \, .
\end{align}
Note that the $\mathbf{7}_0$ and $\mathbf{14}_0$ representations already contain one and two neutral singlets respectively before and after the transition that are included in the overall count already.  Importantly, we must also count the $\mathbf{1}_2$-plets that originiated from the $\mathbf{7}_1$-plets, as neutral states that contribute to $h^{2,1}$. This is again due to the fact, that the whole tower of KK states, is given as $\mathbf{1}_{2+2n}$ and hence, for $n=-1$, there is a zero mode that contributes a neutral singlet. 
 
When comparing the amount of complex structures upon the $\fg_2^{(1)} \rightarrow \fsu_3^{(2)}$ deformation, we find the Hodge numbers
\begin{align}
    (h^{1,1},h^{2,1}) =(4,98) \rightarrow (3,107) \, ,
\end{align}
where we have $\Delta h^{2,1}=2 n_{\mathbf{7}_1}-1=9$ as expected from the above field theory arguments. 
\subsection{The geometry of $\mathfrak{so}_8^{(3)}$}
\label{ssec:so83Geometry}
In this section we describe the geometry of the twisted  $\mathfrak{so}_8^{(3)}$ algebra. This geometry can be interpreted as the twisted compactification of an $\fso_8$ by its order three automorphism, which admits a $\mathfrak{g}_2$ invariant sub-algebra.
 \begin{table}[t!]
 \begin{center} 
$\begin{array}{|c|c|c|c|c|c|c|c|c|c|c||c|c||c|c|}\hline
     d_1   & d_2 & d_3 & d_ 4 & d_5 & d_6 & d_7 & d_8 & d_9 &d_{10}&f & g & \Delta & \text{Genus-1 Fiber} & \text{Jac Fiber} \\ \hline
        2 & 2 & 1 & 0 & 2 & 1 & 0 & 1 & 0 & 1 & 3 & 4&8 & \mathfrak{f}_4^{(1)} & \mathfrak{f}_4^{(1)} \\ \hline
    2 & 2 & 1 & 0 & 2 & 1 & 0 & 1 & 0 & 0 & 3 & 4&8 & \mathfrak{so}_8^{(3)} & \mathfrak{f}_4^{(1)} \\ \hline 
      2 & 1 & 1 & 0 & 1 & 0 & 0 & 1 & 0 & 0 & 3 & 4&8 & \mathfrak{g}_2^{(1)} & \mathfrak{g}_2^{(1)} \\ \hline 
\end{array}   $
\caption{\label{tab:CubicTuningso83}\textit{Tunings of sections in the cubic genus one model and the resulting fiberes.}} 
\end{center} 
\end{table}
However, when viewed as a fiber graph, the $\fso_8^{(3)}$ Cartan matrix admits reversed arrows when compared to $\fg_2^{(1)}$. Hence its geometric cover is rather that of an $\fe_6^{(1)}$ affine Dynkin diagram folded by its order three automorphism. To engineer a geometry with such an order three (affine) automorphism, we are therefore required to consider genus-one fibrations with a three-section. We must therefore depart from the quartic fiber model, that worked well in the models before, but not in this case. 

 The simplest choice for a three-section genus-one model is the cubic fibration, where the fiber is represented as a generic degree three polynomial, given as 
\begin{align} 
 \label{eq:singcubic}
p= s_1    u^3 + s_2 z^2    u^2 v + 
 s_3    u v^2 + s_4   v^3 + 
 s_5    u^2 w   +s_6        u v w +
  s_7   v^2 w + d_8       u w^2 + s_9   v w^2 + 
 s_{10}   w^3 \, ,
\end{align} 
in $\mathbb{P}^2_{u,v,w}$. To promote the genus-one curve to a fibration, the $s_i$ must become sections of line bundle of the base  $B_2$ which we consider as the generalized Tate-sections of the cubic. 

The specific dependencies of those line bundle classes that are in agreement with the CY conditions can be found in Appendix~\ref{app:LBData} and follow the convention used in \cite{Klevers:2014bqa}. The above model admits a 3-sections, given by the vanishing of any of the toric coordinates which makes the  geometry into a genus-one fibration. 
Following the logic in the sections before, we define the monodromy divisor as the discriminant locus of a three-section. In the following we use $u=0$ as the reference 3-section, which results in the solutions of the polynomial
\begin{align}
s_0^{(3)}=s_4 + s_7 w + s_9 w^2 + s_{10} w^3 \, ,
\end{align}
with discriminant locus 
\begin{align}
\label{eq:Z3Monou}
\overline{D}_{3,u}=27 s_{10}^2 s_{4}^2 + 4 s_{10} s_7^3 - 18 s_{10} s_4 s_7 s_9 - s_7^2 s_9^2 + 4 s_4 s_9^3    \, .
\end{align}
The line bundle classes of the monodromy divisor can then be read off from Appendix~\ref{app:LBData} as
\begin{align}
    [\overline{D}_{3,u}] \sim 2 \mathcal{S}_9+2 \mathcal{S}_7=2 \mathcal{R}\, .
\end{align}
The respective 3-section class 
 $\sigma(s_0^{(3)})=[u]$ is also used to obtain the respective charges under  
$\mathfrak{u}_{1,E}$ of the 5D states. Due to the usual mixing, KK-tower states admits charges   $q_E=q + 3n$ for the $n$-th KK-mode.

Similar as in the quartic, this can be tracked field theoretically by a $\mathfrak{u}_{1,6D}$ Higgsing of a $q=3$ state, that carries 5D KK-momentum \cite{Klevers:2014bqa,Cvetic:2015moa}.

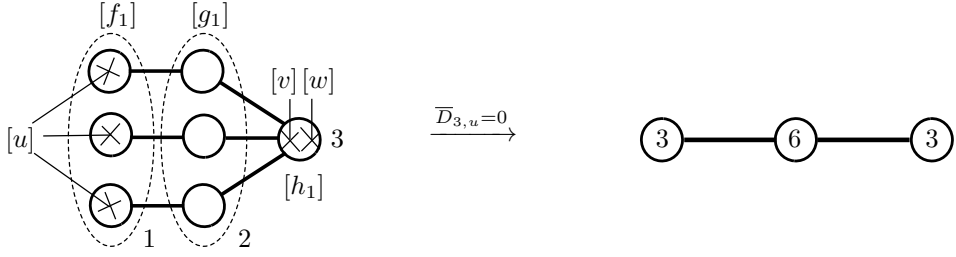
\begin{figure}[t!]
 \begin{center}
 {\footnotesize
 \begin{picture}(250, 100)
 \put(-50,10){\includegraphics[scale=0.6]{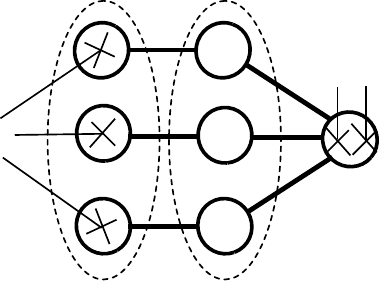} } 
        \put(-60,48){$[u]$} 
  \put(-25,95){$[f_1]$}
   \put(10,95){$[g_1]$}
   \put(45,30){$[h_1]$}  
      %   \put(97,65){$[Z]$}   
                \put(63,48){$3$}  
            %   \put(-12,43){$1$}
           % \put(-13,68){$1$} 
   \put(-8,10){$1$}  
   %  \put(1,43){$2$}
        %    \put(1,68){$2$} 
   \put(28,10){$2$}  
  \put(40,70){$[v]$}  
    \put(52,70){$[w]$}  
    \put(100,50){$\xrightarrow{\overline{D}_{3,u}=0 }$}
     \put(180,42){\includegraphics[scale=0.6]{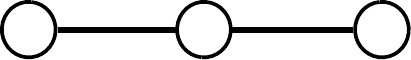} } 
      \put(186,48){$3$}  
        \put(236,48){$6$}  
          \put(288,48){$3$}  
    
 \end{picture}
 }
 \caption{\label{fig:G2t}{\it Depiction of the intersection structure of the $\mathfrak{so}_8^{(3)}$ fiber.  Numbers denote the multiplicity of the divisor, that coincide with the Kac labels.
 The three-section $u=0$ interchanges the three curves in $[f_1]$ and $[g_1]$ along its monodromy divisor.  Along its  intersection the fiber degenerates into an order three multiple fiber depicted on the right.}} 
 \end{center}
 \end{figure}

Using the Arten-Tate algorithm, we can map the cubic into the Weierstrass form (see \cite{Klevers:2014bqa}). Following the usual logics, we need to tune a type $IV^{*}$ singularity that intersects the monodromy divisor $\overline{D}_{3,u}$ non-trivially. This ansatz is motivated as the geometric cover of $\fso_8^{(3)}$ is $\fe_6^{(1)}$.
 We again employ a tuning of the
 $s_i \rightarrow d_i z^{n_i}$ in \eqref{eq:singcubic} and emply the following generalized Tate-vector 
\begin{align}
    \vec{n}=\{  2,2,1,0,2,1,0,1,0,0 \} \, ,
\end{align} 
which yields an $\fso_8^{(3)}$ singularity.
The respective toric resolution requires two more resolution divisors, which yield the toric hypersurface  
      \begin{align}
p= &d_1 f_1^2 g_1 u^3 + d_2  f_1^2 g_1^2 h_1^2 u^2 v + d_3  f_1 g_1 h_1 u v^2 + 
 d_4  v^3 + d_5   f_1^2 g_1^2 h_1^2 u^2 w \nonumber \\ &+ d_6   f_1 g_1 h_1 u v w  + 
 d_7  v^2 w + d_8   f_1 g_1 h_1 u w^2 + d_9   v w^2 + d_{10}   w^3 \, .
\end{align} 
Here $f_1$ replaces the singularity $z$ and  $g_1,h_1$ yield the exceptional divisors. 
Using the Stanley-Reisner ideal
\begin{align}
\mathcal{SRI}: \{ 
%u f_0, f_0 f_1, f_0 g_1,
u g_1, u h_1, f_1 h_1, w v u, w v f_1, w v g_1, w v h_1 \}\,. 
\end{align}
The explicit form of the fibral curves is given as 
\begin{align}
\label{eq:G2tcurves}
\begin{split}
\mathbb{P}^1_{\alpha_0,i}&: \quad  f_1 \cap d_4 v^3 + d_7 v^2 w + d_9 v w^2 + d_{10} w^3 \, , \\
\mathbb{P}^1_{\alpha_1,i}&: \quad  g_1 \cap d_4 v^3 + d_7 v^2 w + d_9 v w^2 + d_{10} w^3 \, ,\\
\mathbb{P}^1_{\alpha_2}&: \quad  h_1 \cap d_1 g_1 + d_4  v^3 + d_7  v^2 w + d_9  v w^2 + d_{10}  w^3 \, .
\end{split}
\end{align}
Note that the first two fibral divisors contain three irreducible curves each that are permuted along the $\mathbb{Z}_3$ monodromy of the 3-section $\overline{D}_{3,u}=0$ given in \eqref{eq:Z3Monou}. This allows us to compute the intersections as given in Figure~\ref{fig:G2t}.  
Similar to the other twisted algebras before the degeneration picture of the fibral curves, with the monodromy divisor is of central importance:
As can be seen from the fibral curves in Eqn.~\eqref{eq:G2tcurves}, at the intersection point of the curve $\mathcal{Z}$ with the monodromy divisor $\overline{D}_{3,u}$, we find the fiber to degenerate into a degree three multiple fiber. This phenomenon is indicative, that the group of CY-torsors, the Weil-Ch$\hat{\text{a}}$talet group does not simply reduce to the Tate-Shafarevich group. 
 
When compared to the singular Weierstrass model, this structure is absent due to the presence of a section. Here we find at leading orders the Weierstrass coefficients 
\begin{align}
\label{eq:g2Jacfiber}
\begin{split}
f=&\frac12 z^3 d_1 (9 d_{10} d_4 d_6 - 6 d_{10} d_3 d_7 + 2 d_7^2 d_8 - d_6 d_7 d_9 - 
   6 d_4 d_8 d_9 + 2 d_3 d_9^2)+ \mathcal{O}(z^4)\, ,  \\
g=&-\frac14 z^4 d_1^2 \overline{D}_{3,u}+d_1 \mathcal{O}(z^5)\, ,  \\ 
\Delta=& \frac{27}{16}d_1^4 z^8 \overline{D}_{3,u}^2+ d_1^3 \mathcal{O}( z^9)\, ,
\end{split}
\end{align}
which yields the expected type $IV^{*,ns}$, i.e. $\ff_4$ singularity. Just as in the other twisted cases, we find the monodromy divisor $\overline{D}_{3,u}$ of the genus-one fibration, to also appear as a codimension two component in the discriminant in the Weierstrass model. However, note that there is also an E-string component localized over $z=d_1=0$ where the WSF sections vanish to orders $(4,6,12)$. To keep the discussion simple, we want to avoid those additional non-perturbative contributions in the following.
  \subsubsection{State counting in $\mathfrak{so}_8^{(3)}$ genus-one fibration}
Next we compute the number of BPS states that are non-trivially charged under $\fso_8^{(3)}$ from 
geometry. The discussion is analogous to that in the $\fe_6^{(2)}$ model. 

The starting point here are the 72 W-bosons of the adjoint representation of the  $\mathfrak{e}_6$ covering algebra as this is the geometric cover of $\fso_8^{(3)}$
. The M2 branes states wrap the curve 
  $\mathcal{C}$ that we are the linear combinations of the base curves $\mathcal{C}_i$ which intersecting as the affine $\fe_6$ Cartan matrix. We express those curves graphically as 
 \begin{align}
 \mathcal{C}= a_0 - a_1-  a_2  < \begin{array}{c} a_3 - a_4 \\ a_5 - a_6  \end{array} \,,
 \end{align} 
 with $a_i \in \mathbb{N}^0$ and start with all combinations that yield the $72$ W-roots of the $\fe_6$ adjoint.
 
 In the second step, we implement the $\mathbb{Z}_3$ monodromy action which is given as
 \begin{align}
     \mathbb{Z}_3: \{ \mathcal{C}_0 \rightarrow \mathcal{C}_{4} \rightarrow \mathcal{C}_6 \rightarrow \mathcal{C}_3 \}   \text{ and  } 
     \{ \mathcal{C}_1 \rightarrow \mathcal{C}_{3} \rightarrow \mathcal{C}_5 \rightarrow \mathcal{C}_1 \} \, .
 \end{align}  
 An  invariant basis of curves under the $\mathbb{Z}_3$ action is given by the combination\footnote{For more details, the very same fiber geometry has been discussed in \cite{Braun:2014oya} with a $\mathbb{P}^2$ base.}
 \begin{align}
   \{\mathcal{C}_2,  \mathcal{C}_1+ \mathcal{C}_3 + \mathcal{C}_5, \mathcal{C}_0+\mathcal{C}_4+\mathcal{C}_6     \} \, .
 \end{align}

We can now split up the $72$ base curves of $\fe_6$ into invariant curves under the $\mathbb{Z}_3$ action and compute their weights. The $\fg_2$ weights are then computed as
\begin{align}
    \mathcal{C} \cdot (\mathcal{C}_2,\mathcal{C}_1+ \mathcal{C}_3 + \mathcal{C}_5) \, . 
\end{align}
Furthermore it is important to identify shrinkable and non-shrinkable curves, given by a non-trivial combination of affine curves $a_0 + a_4+a_6 $ modulo 3. 
Those later three numbers also yield the shifted KK-momenta.  

Reducing the $72$ roots by the $\mathbb{Z}_3$ monodromy into invariant curves is depicted in Table~\ref{tab:G2tstates}.
  \begin{table}[t!] 
   \begin{align*}
 {\footnotesize
\begin{array}{|l|c|c|c|c|}\hline
\text{Mono} &   \# \text{Curves}& \text{Example Curve} &  \mathfrak{g}_2 \in \mathfrak{so}_8^{(3)} \text{weight} & \chi    \\ \hline 
Inv  &  6 &   {\footnotesize
   0 - 0 -1  <\begin{array}{c} 0-0 \\ 0-0 \end{array} } &  (-3,6)  & g    \\ \hline
 \mathbb{Z}_3 & 6 &  {\scriptsize   0 - 1 -0  <\begin{array}{c} 0-0 \\ 0-0 \end{array}+   0 - 0 -0  <\begin{array}{c} 1-0 \\ 0-0 \end{array}+   0 - 0 -0  <\begin{array}{c} 0-0 \\ 1-0 \end{array}} & (-2,-3) & g\prime\\ \hline 
   \mathbb{Z}_3  &6&   {\scriptsize   0 - 0 -1  <\begin{array}{c} 1-1 \\ 0-0 \end{array}+   1 - 1 -1  <\begin{array}{c} 0-0 \\ 0-0 \end{array}+   0 - 0 -1  <\begin{array}{c} 0-0 \\ 1-1 \end{array}}   & (-1,0) & g\prime  \\ \hline 
     \mathbb{Z}_3  & 6&   {\scriptsize   1 - 1 -1  <\begin{array}{c} 1-0 \\ 1-1 \end{array}+   0 - 1 -1  <\begin{array}{c} 1-1 \\ 1-1 \end{array}+   1 - 1 -1  <\begin{array}{c} 1-1 \\ 1-0 \end{array}}  & (-1,-3) & g\prime  \\ \hline 
 \mathbb{Z}_6  &1  &{\scriptsize \begin{array}{c}  0-0-1 <\begin{array}{c}  1-0\\ 1-1  \end{array} \,  + \,\,\,  1-1-1 <\begin{array}{c} 1-0\\ 0-0  \end{array} \,  + \, \,  \,  0-1-1 <\begin{array}{c} 1-1\\ 0-0  \end{array} \\
  0-0-1 <\begin{array}{c} 1-1\\ 1-0  \end{array}\, +\,\,\, \, 1-1-1 <\begin{array}{c} 0-0\\ 1-0  \end{array}\, + \, \,  \, 0-1-1 <\begin{array}{c} 0-0\\ 1-1  \end{array} 
 \end{array} }  & (0,0)   &\overline{g}\prime   \\ \hline  
  \mathbb{Z}_6  &1 & {\scriptsize
  \begin{array}{c}
  0-1-2 <\begin{array}{c} 2-1\\ 1-1  \end{array}\,   +\,\,\,   
  1-1-2 <\begin{array}{c} 2-1\\ 1-0  \end{array}   \,   +\,\,\,    0-1-2 <\begin{array}{c} 1-1\\ 2-1  \end{array}   \\
    1-2-2 <\begin{array}{c} 1-1\\ 1-0  \end{array}   \,   +\,\,\,    1-2-2 <\begin{array}{c} 1-0\\ 1-1  \end{array}   \,   +\,\,\,    1-1-2 <\begin{array}{c} 1-0\\ 2-1  \end{array}  \end{array} }
  & (0,0) &  \overline{g}\prime  \\ \hline  
   \end{array}
   }
\end{align*}
\caption{\label{tab:G2tstates}\textit{
Schematic decomposition of the 72 curves of $\mathfrak{e}_6$ roots into $\mathbb{Z}_3$ invariant curves,
their $\mathfrak{g}_2 \subset \mathfrak{so}_8^{(3)}$ charges and respective moduli space dimensions $\chi$ 
  computed in \eqref{eq:g2Z3branch} and \eqref{eq:g2Z6branch}. }}
\end{table}
In order to compute the moduli spaces, we first need to compute the intersection of the monodromy divisor with $\mathcal{Z}$. Before doing so, we need to restrict to the locus without E-strings by demanding  $[d_1]\cdot \mathcal{Z}=0$, which results in the condition
\begin{align}
\label{eq:E6U1nf}
(\mathcal{S}_7+\mathcal{S}_9) \cdot \mathcal{Z}=\mathcal{R} \cdot \mathcal{Z}= 3c_1 \cdot \mathcal{Z}-2 \mathcal{Z}^2 \, .
\end{align}
From this we can compute the number of ramification points  $R=2 \mathcal{R} \cdot \mathcal{Z}$. Using the 
RH theorem, we can compute the moduli spaces of each of such curve for $d=3$ for which we obtain
\begin{align}
\label{eq:g2Z3branch}
g\prime = 4(1-g)+ \mathcal{Z}^2 + g \, . 
\end{align}
Computing the moduli spaces $\overline{g}\prime$ on the other hand is a bit more involved. These elements have a trivial $S_3$ stabilizer and were for an $g=0, \mathcal{Z}^2=1$ case computed in \cite{Braun:2014oya}. In the following we propose to compute the general case as follows: We start with the order three ramified $\mathbb{P}^1$'s, given by $g\prime$ and take another order two cover, ramified at the same points. This then yields 
\begin{align}
\label{eq:g2Z6branch}
\overline{g}\prime=& (g\prime -1)+\frac12 R + g\prime \, ,\\
 =&13(1-g)+g+3 \mathcal{Z}^2 \, ,
\end{align}
which reproduces the result given in \cite{Braun:2014oya}. 
We are now in the position to identify the respective 5D states and their representations. First we have 
the states with trivial
weights under $\mathcal{C}_0 +\mathcal{C}_4 +\mathcal{C}_6$ which are hence shrinkable curves. Those $6+6$ states from the first two rows in Table~\ref{tab:G2tstates}, together with the two Cartan generators combine to the vector multiplets of $\fg_2$ as well as $g$ adjoint charged hypermultiplets. The moduli space of the shrinkable $\mathbb{Z}_3$ branched states however has dimension $g\prime$ and thus, additional hypermultiplets in the fundamental of $\fg_2$ are provided upon adding the neutral hyper. Subtracting the contribution to the $g$ adjoint hypers, we therefore get 
\begin{align}
n_{\mathbf{7}_0}=g\prime -g=4(1-g)+ \mathcal{Z}^2\, ,
\end{align}
fundamentals with trivial $U(1)_E$ charge.
In the last four rows of Table~\ref{tab:G2tstates} we find the non-shrinkable curves, i.e. states that have a non-trivial $U(1)_E$ charge which therefore stay massive at the origin of the $\fg_2$ Coulomb branch.

Curves in the third and fourth row of Table~\ref{tab:G2tstates} makeup  $\mathbf{7}_1$ and $\mathbf{7}_2$-plets respectively. Their multiplicities are given as 
\begin{align}
n_{\mathbf{7}_1}=n_{\mathbf{7}_2}=g\prime=4(1-g)+ \mathcal{Z}^2 + g \, .  
\end{align}
When turning to the singlets $\mathbf{1}_1$ and $\mathbf{1}_2$ we again need to take into account, that each of them is needed to complete the shifted $\mathbf{7}$-plets. Hence we find for both of them
\begin{align}
n_{\mathbf{1}_1}=n_{\mathbf{1}_2}= \overline{g}\prime -g\prime = 9(1-g)+2 \mathcal{Z}^2 \, .
\end{align}
We summarize the non-trivially charged spectrum under the $\fg_2$ subgroup as 
\begin{align}
\label{eq:so83Spectrum}
\begin{split}
    n_{\mathbf{14}_0}=& g \, , \, \quad n_{\mathbf{7}_1} =4(1-g)+ \mathcal{Z}^2\, \\
    n_{\mathbf{7}_1} =&n_{\mathbf{7}_2}=4(1-g)+ \mathcal{Z}^2+g\, .
\end{split}
\end{align} 
\subsubsection*{Comparison with $\fso_8 \rightarrow \fso_8^{(3)}$ reduction}
An $\fso_8^{(3)}$ theory can be obtained, by starting with an $\fso_8$ 6D theory and mod by the triality symmetry that interchanges vector, spinor and co-spinor representations upon the circle compactification. 
In geometric terms of an (unfrozen) 6D F-theory model, the mulitplicites of the respective states are given as 
\begin{align}
    n_{\mathbf{8}_V} =n_{\mathbf{8}_S} = n_{\mathbf{8}_C}  = 4(1-g)+\mathcal{Z}^2 \, , \quad   n_{\mathbf{28}}=g \, .
\end{align}
The twisted compactifiation breaks $\fso_8$ to its $\fg_2$ subgroup. The light states then branch to 
\begin{align}
     \mathbf{28} &\rightarrow \mathbf{14}_0 \oplus \mathbf{7}_{\frac13} \oplus \mathbf{7}_{\frac23} \, , \\
     (\mathbf{8}_V \oplus \mathbf{8}_S \oplus \mathbf{8}_V)&\rightarrow (\mathbf{7}_0 \oplus \mathbf{7}_{\frac13} \oplus \mathbf{7}_{\frac23} \oplus \mathbf{1}_0 \oplus \mathbf{1}_{\frac13} \oplus \mathbf{1}_{\frac23})\, ,
\end{align}
where the subscript denotes the usual shifted KK-charge.
Collecting   the multiplicities of $\fg_2$ charged states in terms of geometric data, we find 
\begin{align}
\label{eq:g2tspectrum}
    n_{\mathbf{7}_0}= 4(1-g)+\mathcal{Z}^2 \,, \qquad  
     n_{\mathbf{7}_\frac13}= 
      n_{\mathbf{7}_\frac23}= 4(1-g)+\mathcal{Z}^2 + g \, ,
\end{align}
which matches the geometric count upon rescaling the KK-charges to our geometric conventions.  
This perspective also gives a prediction for additional singlet states, counted as 
\begin{align}
    n_{\mathbf{1}_0} = 4(1-g)+\mathcal{Z}^2 \, .
\end{align}
In Section~\ref{sec:TVSUT} we will come back to those states and interpret them, as parts of the complex structure moduli space, not realized as polynomial deformations of the CY hypersurface equation.  
\subsubsection*{State counting in the $\ff_4^{(1)}$ Jacobian}
The Jacobian of the $\fso_8^{(3)}$ model lifts to the geometric cover, which is a generic Type $IV^{*,ns}$ singularity, as given in \eqref{eq:g2Jacfiber}.
This can be readily seen by noting, that the monodromy divisor of the genus-one model $\overline{D}_{3,u}$ appears at leading order at the discriminant as well as the $g$ coefficient of the discriminant. 
 Hence $g/z^4$ at $z=0$ is not a perfect square by construction and therefore the fiber is non-split. Note also once again, the appearance of the E-string points at   $z=d_1=0$ which we have switched off by fixing the base line bundle intersections accordingly.
 
We continue to compute the spectrum in a similar way as in the cases before. Here the appearing monodromy is only of degree $d=2$ and split the $\fe_6$ adjoint as 
\begin{align}
    \mathbf{78} \rightarrow \mathbf{52} \oplus \mathbf{26} \, .
\end{align}
The multiplicity of the adjoint is simply given by the genus $g$ of $\mathcal{Z}$ and the fundamentals originate from degree 2 ramified curves. Their moduli spaces are just computed as before, using the RH theorem
using $R=[\overline{D}_{3,u}] \cdot \mathcal{Z}$. Upon absorbing $g \times \mathbf{26}$-plets to complete the adjoint of $\ff_4$ we are left with the following spectrum  
\begin{align}
n_{\mathbf{52} }=g \, , \qquad n_{\mathbf{26}}=(g-1)+\frac12 R = 5(1-g) + \mathcal{Z}^2 \, ,
\end{align}
which is exactly what one expects from an anomaly free $\mathfrak{f}_4$ gauge group in 6D. Note also, that we can not infer the potential discrete charges for the $\ff_4$ fundamentals directly from the genus-one fibration. 

\subsubsection*{Toric examples}
We close by presenting two explicit toric examples. 
The first one is an $\mathfrak{so}_8^{(3)}$ on a $\mathcal{Z}^2=-1$ curve within a $dP_1$ base. The toric divisors are given in Appendix~\ref{app:Example3folds}.  
The Hodge numbers are given as 
\begin{align}
    (h^{1,1},h^{2,1}(h^{2,1}_{np}))= (5,71(6)) \, .
\end{align}
We find the $1+2+2$ expected K\"ahler parameters that originate from fiber, base and the $\fg_2$ Cartan generators as well as six non-toric polynomial deformations. The non-trivial $\fg_2$ charged spectrum is given as  
\begin{align}
n_{\mathbf{7}_0} = n_{\mathbf{7}_1} =  n_{\mathbf{7}_2} =3 \, .
\end{align}

Another interesting example, is to have a trivial $\fg_2$ hyerpmultiplet sector, which can be achieved by having $\mathcal{Z}$ being a curve of self-intersection
 $\mathcal{Z}^2=-4$ and genus $g=0$ (see \eqref{eq:g2tspectrum}).

Such geometry is nothing but the $\fso_8$ non-Higgsable cluster theory after a twisted reduction.  
The minimal compact base, that hosts such a curve is therefore $\mathbb{F}_4$. It is then straightforward to 
find a toric realization of $\fso_8^{(3)}$ over the $-4$ curve. 
\begin{table}[t!]
\begin{center}
\begin{tabular}{ccc}
$
\begin{array}{|c|c|}
\multicolumn{2}{c}{\text{ Generic Fiber }} \\ \hline 
u&(-1,-1,0,0)\\
v&(0,1,0,0)\\
w&(1,0,0,0)\\ \hline  
\end{array}$  & $ 
\begin{array}{|c|c|}
\multicolumn{2}{c}{\text{ $\mathbb{F}_4$ Base }} \\ \hline  
x_1 & (-1, -1,  0,  1) \\
y_0 & (8,  8, -1, -4) \\
y_1 & (-7, -8,  1,  0)\\ 
f_1&(1,0,0,-1) \\ \hline 
\end{array}$ &$
\begin{array}{|c|c|}
\multicolumn{2}{c}{\text{$\mathfrak{su}_3^{(2)}$ Fiber }} \\ \hline    
%f_3^*& (-2,-1,1,0)\\ 
g_1& (1, 0, 0, -2)\\
h_1& (1,  0,  0, -3) \\ \hline 
\hline
\end{array} $  
\end{tabular}
\caption{\label{tab:d43F4Rays}\textit{
Toric rays, of an $\mathfrak{so}_8^{(3)}$ genus-one fibration $\hat{X}^C$ over a $\mathbb{F}_4$ base with $f_1$ being the affine node.
}}
\end{center}
\end{table}
The toric rays that engineer the respective polytope are given in \eqref{tab:d43F4Rays}. The polytope data one can compute the Hodge numbers to
\begin{align}
    (h^{1,1},h^{2,1}(h^{2,1}_{np}))=(5,92(0))   \, .
\end{align} 
The five K\"ahler moduli can again be attributed to three classes coming from fiber and base, and two more from the $\fg_2$. Also note, that the geometry does not posses any non-polynomial complex structure deformations. We will come back to this point in Section~\ref{sec:TVSUT} and propose a physics explanation for that. 

\subsubsection{Transitions to $\fso_8^{(3)}$}
\label{sssec:f4z3}
In the following we want to use geometric transitions to connect other torus-fibered threefolds with enhanced gauge symmetries to the $\fso_8^{(3)}$ model.

In order to do so, we employ the toric tunings of the generalized Tate-vectors as summarized in  Table~\ref{tab:CubicTuningso83} and first start with an enhanced singularity which engineers an $\ff_4$ in the Jacobian and genus-one model with a three-section. It is straightforward to unhiggs the $\mathbb{Z}_3$ to an $\mathfrak{u}_{1,6d}$ by tuning the section $d_{10}$ to zero globally (see \cite{Klevers:2014bqa,Cvetic:2015moa} for more details). We will skip this part, for brevity and keep in mind that the $\ff_4$ singulariry is simply a spectator of this transition, similar to the unhiggsings of $\mathbb{Z}_2 \rightarrow \mathfrak{u}_{1,6D}$ we encountered in the examples before. Recall however, that the resulting 5D $\mathfrak{u}_{1,E}$ is a linear combination of the 6D $\mathbb{Z}_3$ with the KK $U(1)$. 

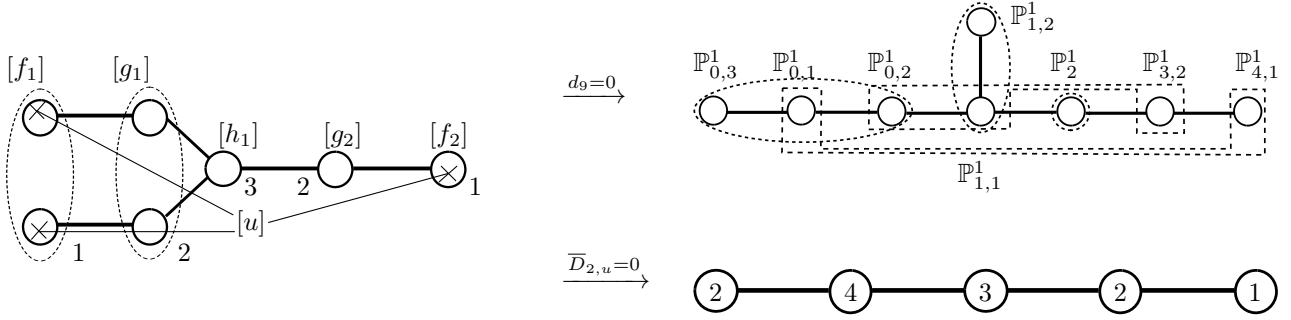
\begin{figure}[t!]
 \begin{center}
 {\footnotesize
 \begin{picture}(00, 140)
 \put(-240,40){\includegraphics[scale=0.5]{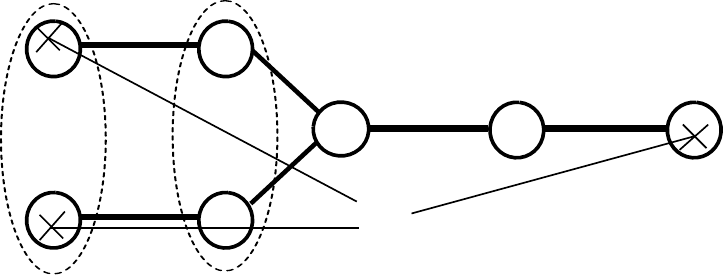} } 
    \put(-80,85){$[f_2]$} 
    \put(-120,85){$[g_2]$}  
       \put(-160,85){$[h_1]$} 
              \put(-200,110){$[g_1]$} 
              \put(-240,110){$[f_1]$} 
               \put(-153,52){$[u]$}
\put(-215,40){$1$}
\put(-175,40){$2$}

\put(-150,65){$3$}
\put(-130,65){$2$}
\put(-65,65){$1$}

                 \put(20,80){\includegraphics[scale=0.4]{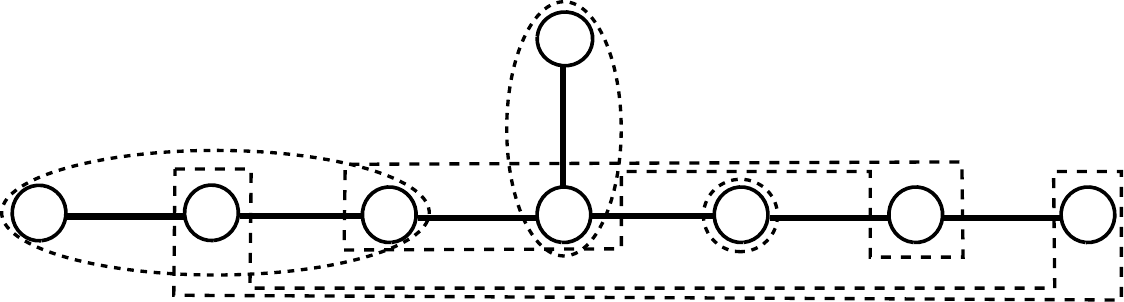} } 
    \put(20, 20){\includegraphics[scale=0.6]{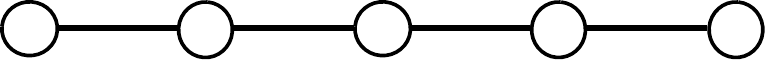} } 
                   \put(140,130){$\mathbb{P}^1_{1,2}$}
                    \put(120,70){$\mathbb{P}^1_{1,1}$}
                     \put(20,112){$\mathbb{P}^1_{0,3}$}
                    \put(50,112){$\mathbb{P}^1_{0,1}$}
                      \put(86,112){$\mathbb{P}^1_{0,2}$}
                      \put(155,112){$\mathbb{P}^1_{2}$}
                      \put(190,112){$\mathbb{P}^1_{3,2}$}
                      \put(225,112){$\mathbb{P}^1_{4,1}$}
           \put(-30,30){$\xrightarrow{\overline{D}_{2,u}=0}$}
           \put(-30,100){$\xrightarrow{d_9=0}$}
           \put(26,25){2}
           \put(77,25){4}
           \put(128,25){3}
           \put(179,25){2}
           \put(230,25){1}
                 
 \end{picture} }
 \caption{\label{fig:F4Z3Fiber}{\it Depiction of the $\mathfrak{f}_4^{(1)}$ fiber in the  cubic model and its intersections with the three-section $[u]$. There are two loci, where the fiber degenerates further, depicted on the right. Multiplicities of fibral curves are highlighted below.}}
 \end{center}
 \end{figure}

When compared to the $\fso_8^{(3)}$ geometry, the main  difference is the additional tuning 
\begin{align}
    d_{10} \rightarrow z d_{10} \,. 
\end{align}
Due to this, the type $IV^*$ singularity in the genus-one model is not intersected by the order three monodromy divisor $\overline{D}_{3,u}$ in \eqref{eq:Z3Monou} and hence, does not experience the full monodromy but only a subgroup thereof. This can be observed by noting, that the monodromy divisor $\overline{D}_{3,u}$ splits over $z=0$ as 
\begin{align}
    \overline{D}_{3,u} \rightarrow d_9^2  \,
\overline{D}_{2,u}  \, , \text{ with } \quad \overline{D}_{2,u} =d_7^2 - 4 d_4 d_9 \, .
\end{align} 
The appearance of the $\overline{D}_{2,u}$ divisor intersecting $z=0$, highlights the fact that the type $IV^*$ singularity is folded to an $\ff_4$. This can be explicitly seen in the resolved genus-one fibration which is given as 
  \begin{align}
p=& d_1 f_1^2 f_2 g_1 u^3 + d_2   f_1^2 f_2^2 g_1^2 g_2^2 h_1^2 u^2 v + 
 d_3  f_1 f_2^2 g_1 g_2^2 h_1 u v^2 + d_4   f_2^2 g_2^2 v^3 + 
 d_5  f_1^2 f_2 g_1^2 g_2 h_1^2 u^2 w \nonumber \\  &+ d_6   f_1 f_2 g_1 g_2 h_1 u v w + 
 d_7  f_2 g_2 v^2 w + d_8  f_1 g_1 h_1 u w^2 + d_9   v w^2 + 
 d_{10}  f_1 g_1^2 g_2 h_1^3 w^3 \, ,
\end{align}
where the additional exceptional coordinates $g_1, h_1, g_2, f_2$ are required. Computing their intersections can be done, when employing the Stanley-Reisner ideal  
 \begin{align}
 \mathcal{SRI}: \{  w f_2, w g_2, v f_1, v g_1, v h_1,  u g_1, u g_2, 
u h_1, f_1 g_2, f_1 h_1, f_2 h_1, w v u \} \, .
 \end{align}
 The fibral intersections are summarized in 
 Figure~\ref{fig:F4Z3Fiber}. We find that this figure corresponds to a folding along two legs of the $\mathfrak{e}_6$ cover. 
 The respective fibral curves are given as 
  \begin{align}
 \label{eq:f4Z3split}
 \begin{split}
 \mathbb{P}^1_{\alpha_0}&: f_2\cap  d_{10} f_1 g_1^2 g_2+d_8 f_1 g_1 u+d_9 v\, , \\ 
 \mathbb{P}^1_{\alpha_1}&: g_2\cap   d_1 f_2 g_1+d_8   g_1 h_1+d_9   v \, , \\
 \mathbb{P}^1_{\alpha_2}&: h_1\cap   d_1 g_1+d_4   g_2^2+d_7   g_2 w+d_9   w^2\, ,  \\
 \mathbb{P}^{1}_{\alpha_3,\pm}&: g_1\cap d_4 f_2^2 g_2^2+d_7 f_2 g_2 w+d_9 w^2 \, ,\\
 \mathbb{P}^{1}_{\alpha_4,\pm}&: f_1\cap d_4 f_2^2+d_7 f_2 w+d_9 w^2\, .
 \end{split}
 \end{align}
 Note that the fibral divisors $f_1=0$ and $g_1=0$ consist 
 both of two fibral curves, that are interchanged along $\overline{D}_{2,u}$. 
 Note again, that we took $f_1$ as the singular divisor upon which we tuned the model, although here $f_2$ yields the affine node. However, independent of which node we use to map into the singular Weierstrass model, 
  it leads in both cases to the following form
 \begin{align}
 \begin{split}
 f=&\frac12 z^3 d_1 (9 d_{10} d_4 d_6 - 6 d_{10} d_3 d_7 + 2 d_7^2 d_8 - d_6 d_7 d_9 - 
   6 d_4 d_8 d_9 + 2 d_3 d_9^2)+\mathcal{O}(z^4) \, ,\\
 g=&\frac14 z^4 d_1^2 d_9^2 \overline{D}_{2,u} + d_1 \mathcal{O}(z^5) \, , \\
\Delta=&\frac{27}{16} z^8 d_1^4 d_9^4 (\overline{D}_{2,u})^2  + d_1^3 \mathcal{O}(z^9) \, .
\end{split}
 \end{align}
It is straight forward to see, that also the Jacobian admits an  $\mathfrak{f}_4$ singularity
over $z=0$ as long as the polynomial $\overline{D}_{2,u}$ does not become a perfect square
at $z=0$. As the genus-one fibration does not posses multiple fibers, it admits the same type of fiber structure as its Jacobian. We can therefure use the Weierstrass form, to deduce codimension two singularities, which should lead to charged matter in both models.  
 \begin{table}[t!]
 \begin{align*}
\begin{array}{|c|ll|}\hline 
\text{node } & \multicolumn{2}{|c|}{$Irreducible components over $  d_9=0 }\\ \hline
 \mathbb{P}^1_{\alpha_0}: [f_2]& \mathbb{P}^1_{0,1}: f_2 \cap f_1\, , \quad \mathbb{P}^1_{0,2}: f_2 \cap g_1\, ,& \mathbb{P}^1_{0,3}: f_2\cap (d_{10} g_1 g_2 + d_8 u)  \\ \hline
 \mathbb{P}^1_{\alpha_1}:[g_2] &\mathbb{P}^1_{1,1}: g_2 \cap  g_1\, , & \mathbb{P}^1_{1,2}: g_2 \cap (d_1 f_2 + d_8 h_1) \\ \hline
 \mathbb{P}^1_{\alpha_2}:[h_1]& \mathbb{P}^1_2: h_1 \cap d_1 g_1 + d_4 g_2^2 + d_7 g_2 w &   \\ \hline
 \mathbb{P}^1_{\alpha_3}:[g_1] &  \mathbb{P}^1_{0,2}: g_1 \cap  f_2 \, , \quad \mathbb{P}^1_{1,2}: g_1 \cap g_2\, , \quad & \mathbb{P}^1_{3,2}:  g_1 \cap (d_4 f_2 g_2 + d_7 w)  \\ \hline
 \mathbb{P}^1_{\alpha_4}: [f_1] & \mathbb{P}^1_{0,1}: f_1 \cap f_2  \quad & \mathbb{P}^1_{4,1}: f_1 \cap (d_4 f_2 + d_7 w)  \\ \hline 
\end{array}
\end{align*}
\caption{\label{tab:f4gen1Fund}
{\it 
The codimension two irreducible components of the fibre at $d_9 = 0$, where $\ff_4$ funda-
mentals $\mathbf{26}_1$ are located. Their intersection picture is given on the right of Figure~\ref{fig:F4Z3Fiber}.
}}
\end{table}
First we find an E-string locus over $z=d_1=0$, which was also present in the $\fso_8^{(3)}$ model. The second component at the locus $z=d_9=0$ yields a vanishing order $(3,5,9)$ singularity, which is an $E_7$ point where we expect to find matter. Finally, we also have the monodromy divisor appearing. 
Indeed, apart from $d_1 = z=0$, we find the other two loci to yield reducible fibers whose intersections are depicted in Figure~\ref{fig:F4Z3Fiber}.
    
In order to compute the charges under $\mathfrak{u}_{1,E}$ in the genus-one geometry, 
we pick $u=0$ as the reference 3-section class and  orthogonalize the other 
generators to obtain the discrete Shioda map as
 \begin{align}
 \sigma (u)= [u]+2 [g_2]+4 [h_1]+ 3 [g_1]  +2[f_1] \, ,
 \end{align}
 which is constructed such, that all three intersection points of $\sigma(u)$ intersect only $f_2=0$.
 Over $z=d_9=0$ we expect to find charged fundamentals, which we will verify in the following. To do so, we impose $d_9=0$, which results in the fibral curves to split into various fibral $\mathbb{P}^1$s. The  irreducible curves are given in Table~\ref{tab:f4gen1Fund}.

The intersections of the curves are summarized in Figure~\ref{fig:F4Z3Fiber} and admits the shape of  $E_7^{(1)}$, just as suggested by the Jacobian. 
 These components are of importance as they get wrapped by M2 branes giving rise to multiplets that trigger the Higgsing in 5D to  the $\mathfrak{so}_8^{(3)}$ fiber geometry, by effectively folding the third leg of the $\mathfrak{e}_6^{(1)}$. Computing intersections of the reducible fibral curves and the divisors allows to deduce weights of the hypermultiplets at the Coulomb branch. We obtain the weights by intersection with the fibral divisors $(g_2, h_1, g_1, f_1)_{(\sigma)}$ .
  In the following we pick the curve $\mathbb{P}^1_{1,2}$ and compute the weights 
\begin{align}
\begin{array}{cl}
\mathbb{P}^1_{1,2} \cdot (g_2, h_1, g_1, f_1)_{(\sigma)} &:   (-1,0,1,0)_{(1)}\,, 
\end{array}
\end{align}
which indeed are weights of an $\mathbf{26}_1$-plet. The $U(1)_E$ charge is given by $\sigma(u)$ which lifts to the $\mathbb{Z}_3$ charge in the 6D F-theory lift. 

The second type of matter originates from the monodromy divisor $\overline{D}_{2,u}$, leading to non-localized $\ff_4$ fundamentals. Due to the geometric origin in the $\fe_6/\ff_4$ adjoint, they must have a trivial  $\mathfrak{u}_{1,E}$-charge.
 
We can then continue by counting the multiplicities of those states in terms of intersections of line bundles of the base. There are two line bundle classes $\mathcal{S}_7$ and $\mathcal{S}_9$ as before. 
Demanding absence of E-string points, fixes
the sum of  the intersections \eqref{eq:E6U1nf}, analogous to
the $\fso_8^{(3)}$ model.

To compute the multiplicity of $\mathbf{26}_1$-plets we simply take the intersection of $d_9=0$ with $\mathcal{Z}$ which hence results in  
\begin{align}
    n_{\mathbf{26}_1}= \mathcal{S}_9 \cdot \mathcal{Z} \, .
\end{align}
The second source of the $\mathbf{26}_0$-plets originates from the monodromy divisor that yields the ramification points  $\mathcal{R}=\overline{D}_{2,u} \cdot \mathcal{Z}$ which is used in the RH theorem, resulting in
\begin{align}
    n_{\mathbf{26}_0}=5(1-g) + \mathcal{Z}^2 - \mathcal{S}_9 \cdot \mathcal{Z}\, .
\end{align}
where we have  used \eqref{eq:E6U1nf}. 

The sum of the two contributions of fundamentals is given as 
\begin{align}n_{\mathbf{26}_0}+n_{\mathbf{26}_0}=5(1-g) + \mathcal{Z}^2 \, ,
\end{align}
consistent with the 6D gauge anomalies of $\ff_4$. This result is expected, as the theory admits no twisted algebras and hence a 6D SUGRA lift with the same gauge algebra structure. In that lift, only the $\mathfrak{u}_{1,E}$ charges turn into 
$\mathbb{Z}_3$ discrete charges. 

Similar to the $\fg_2$ model in the section before, the $\mathcal{S}_9 \cdot \mathcal{Z}$ intersection  
may result in different numbers of charged and uncharged fundamental multiplets depending on the chosen compactification.

Before discussing the Higgsing in the next section, we present a concrete toric example. The base is a 
$dP_1$, where the $\ff_4$ is placed over the $\mathcal{Z}^2=-1$ curve of genus $g=0$. The toric rays that of the 4D polytope that underlie the CY hypersurface
are summarized in Appendix~\ref{app:Example3folds}.
The Hodge numbers are given as 
\begin{align}
    (h^{1,1},h^{2,1}(h^{2,1}_{np}))=(7,67(4)) \, .
\end{align}
Indeed, we find three K\"ahler parameters of fiber and base, and four more that correspond to the $\ff_4$ Cartan generators. Analysis of the geometry reveals the intersection number $\mathcal{S}_9 \cdot \mathcal{Z}=2$ by finding the class of the section $d_9$ which yields the spectrum 
\begin{align}
n_{\mathbf{26}_0}=n_{\mathbf{26}_1}=2 \, .
\end{align} 
  \subsubsection*{Higgsing to $\fso_8^{(3)}$}
To Higgs to the twisted algebra $\fso_8^{(3)}$ and its $\fg_2$ finite sub-algebra, we can employ the group theory decomposition  
 \begin{align}
     \ff_4 \times \mathfrak{u}_{1,E} \rightarrow \fg_2 \times \fsu_2 \times \mathfrak{u}_{1,E} \, ,
 \end{align}
and then in the second step to break $\fsu_2 \times \mathfrak{u}_{1,E}$. A related type of breaking, that resembles the same type of breaking directly in the 5D Coulomb branch is given as follows: Starting off with the maximal enhanced gauge algebra $\ff_4 \times \mathfrak{u}_{1,E}$ that we constructed in the geometry before,
we move to a partial Coulomb branch by keeping the curves inside of the fibral divisor $f_1$ and $f_2$ finite. On this locus in the K\"ahler moduli space, there is a breaking  as
\begin{align}
    \begin{array}{|c|c|}\hline
\ff_4 \times \mathfrak{u}_{1,E} & \fso_7 \times  \mathfrak{u}_{1,A} \times  \mathfrak{u}_{1,E}   \\ \hline
\mathbf{52}_0 &   \mathbf{21}_{(0,0)} \oplus \mathbf{8}_{(1,0)} \oplus \mathbf{8}_{(-1,0)}  \oplus \mathbf{7}_{(2,0)} \oplus \mathbf{7}_{(-2,0)} \oplus   \mathbf{1}_{(0,0)} \\ \hline
\mathbf{26}_0 &   \mathbf{8}_{(1,0)} \oplus \mathbf{8}_{(-1,0)}  \oplus \mathbf{7}_{(0,0)} \oplus \mathbf{1}_{(-2,0)} \oplus   \mathbf{1}_{(2,0)}  \oplus   \mathbf{1}_{(0,0)} \\ \hline
\mathbf{26}_1 &   \mathbf{8}_{(1,1)} \oplus \mathbf{8}_{(-1,1)}  \oplus \mathbf{7}_{(0,1)} \oplus \mathbf{1}_{(-2,1)} \oplus   \mathbf{1}_{(2,1)}  \oplus   \mathbf{1}_{(0,1)} \\ \hline
    \end{array}
\end{align}
This group theory breaking is clear from geometry, simply
by observing that we shrunk an $\fso_7 \subset \ff_4^{(1)}$ sub diagram. 

On this more generic point in the CB moduli space, we may identify the $\mathbf{8}_{-1,1} \in \mathbf{26}_1$ as the Higgs candidate to break $\fso_7$ down to $\fg_2$. Note that this constrains the line bundle $\mathcal{S}_9$ and in particular its intersection with $\mathcal{Z}$ to be non-trivial for the respective Higgs to exist and the transition to be possible.

This path in the CB moduli space, allows to find the locus, where the $\mathbf{8}_{-1,1}$-plets become massless in order to allow for a non-trivial VEV that
triggers the respective Higgsing. 
At a generic point of the 2D Coulomb branch the 5D $\mathbf{8}$-plet  admits the mass 
\begin{align}
    m_{\mathbf{8}_{-1,1},n}=|- \xi_A + \tau(1 + 3n )  | \, ,
\end{align}
with $\xi_A$ and $\tau$ being the CB parameters of $\mathfrak{u}_{1,A} \times \mathfrak{u}_{1,E}$.
The respective state becomes massless, when choosing 
  $\xi_A = \tau$ and picking the $n=0$ KK-mode. Upon this choice of Coulomb branch parameters one obtains the new unbroken generator $\hat{\mathfrak{u}}_{1,E}=\mathfrak{u}_{1,A}+\mathfrak{u}_{1,E}$. Upon the Higgsing, we collect all states with non-trivial charges as 
\begin{align}
\begin{split}&n_{\mathbf{21}_0}=n_{\mathbf{52}_0}=g \, , \\ &n_{\mathbf{7}_0}= n_{\mathbf{26}_0}+n_{\mathbf{26}_1} -1 + n_{\mathbf{52}_0}= 4(1-g)+ \mathcal{Z}^2 \, ,\\
     &n_{\mathbf{7}_1}= n_{\mathbf{7}_2}= n_{\mathbf{26}_0}+n_{\mathbf{26}_1}+  2 n_{\mathbf{52}_0}= 5(1-g)+ \mathcal{Z}^2  + 2g \, ,
     \end{split}
\end{align}
which matches the geometric computation, but over counts the multiplicity of massive states by one. 

Note that the states recombine exactly in such a way, that
the dependence on the class $\mathcal{S}_9$ drops out.

Having identified the Higgs particles, we can also obtain predictions for the change in complex structure moduli upon the transition. Care needs to be taken
when doing so as the $\mathbf{26}_0$ contains already two uncharged states, just as $\mathbf{7}_0$ of $\fg_2$ that contribute to $h^{2,1}$. Taking this into account, we simply obtain
\begin{align}
    \Delta h^{2,1}= 3 n_{\mathbf{26}_1}-2 = 3 \mathcal{S}_9 \cdot \mathcal{Z} -2 \, .  
\end{align}
Another important detail, that already appeared in the examples before, is the number of neutral fields. I.e. 
we also find singlets states with $\mathfrak{u}_{1,E}$
$\mathbf{1}_3$ in the transition. However as there is full a $KK$-tower of such states present, we also should have states, such as $\mathbf{1}_{3+3n}$ with masses $m=|(3+3n)\tau|$. Therefore, for $n=-1$, there is a zero model that contributes a neutral massless hypermultiplet, which appears as a complex structure modulus in the geometry.  

Our general considerations are consistent with the toric examples. Consider for this, the change in Hodge numbers, given as  
\begin{align}
    (h^{1,1},h^{2,1})=(7,67) \rightarrow (5,71) \, .
\end{align}
First we find a rank reduction by $2$, consistent with the $\ff_4 \rightarrow \fg_2$ reduction in gauge symmetry. Second, there is a change of four complex structure moduli that originate from the $n_{\mathbf{26}_1}=2$ Higgs fields. 

\section{Decoupling Gravity and Non-Compact Limits}
\label{sec:TwistedDuality}
Up to this point, all the theories studied in this work were associated to compact geometries and hence gravitational theories. This proved useful in understanding symmetries and global constraints arising from the geometry. In this section however, we take a different approach and systematically decouple gravity by taking a decompactification limit of the threefold. The limits are chosen such that the non-compact threefold still admits compact divisors with twisted fibers.

The resulting theory is not a SUGRA, but merely a supersymmetric quantum field theory (SQFT) (which thanks to its origin, is guaranteed to have a consistent UV completion). In particular, we will chose the decompatification limit such that the SQFT flows to a little string theory (LST) in the UV. Analyzing this LST will the provide an additional cross check of our geometric construction.  

In order to do so, we exploit two of the striking features of little string theories. First, they admit a continuous global 2-group symmetry \cite{Cordova:2020tij} and second, they often exhibit T-duality. T-duality in this context refers to the feature that one or multiple theories can share a Coulomb branch upon circle compactification\footnote{See also \cite{Anderson:2016cdu} for related work.}. In  \cite{DelZotto:2020sop} it was proposed that the universal part of the 6D 2-group symmetries should match across T-duality\footnote{See also \cite{DelZotto:2022ohj, DelZotto:2022xrh,DelZotto:2023ahf} for a systematic exploration of heterotic LSTs and their T-duals exploiting the match of 2-group structures. }. This is very useful as the 
2-group symmetries depend on the
dual Coxeter numbers $h^v_\fg$ of gauge algebra factors that are coupled to the 6D tensors and therefore on the un/twisted affine extension. 

The match of the 2-group structure constants among two dual theories therefore provides another robust test of our geometric construction. 
 
\subsection{Supergravity phase and T-duality}
The starting point of our construction is a compact genus-one fibered threefold $X$ specified by a toric hypersurface in an ambient fourfold, obtained from a regular fine star triangulation of the toric rays given in Table~\ref{tab:example1}. 
The threefold $X$ admits the following Hodge numbers, that can be computed from the structure of the polytope as
\begin{align}
    (h^{1,1},h^{2,1}(h^{2,1}_{np}))(X)=(15,63 (3)) \, .
\end{align}
This geometry admits two toric fibrations that are inherited from the ambient space which can be identified by its 2D relexive sub-polytope structure\footnote{The 2D sub-polytope criterion \cite{Huang:2019pne} also allows to deduce a triangulation that respects the fibration of the ambient space, although it might lead to a non-reflexive but VEX polytope.} in $\Delta$. We are then guaranteed to have (at least) one regular fine star triangulation that respects that fibration structure. We will discuss the two fibrations in detail: one is a genus-one fibration and the other an elliptic fibration. The two corresponding 6D theories are therefore (twisted) T-duals. 

In the following we use the quiver notation of base curves and their fibers, which is commonly used in the literature (see e.g. \cite{Heckman:2018jxk,DelZotto:2023ahf}). A quiver is written as
  \begin{align}
 \ldots     \overset{\mathfrak{g}_i^{(s)}}{n_i} \ldots
  \end{align}
where $\mathfrak{g}^{(s)}_i$  an $(s)$ twisted affine algebra sitting over a curve $w_i$ of genus zero and self-intersection $w^2_i=-n_i$ and two neighbouring curves in the quiver intersect. For more details see \cite{Heckman:2018jxk,DelZotto:2023ahf} for recent reviews. The full structure can readily be obtained from the toric diagram of the base. 

\paragraph{The genus-one fibration:}The first fibration admits an $F_4$ ambient space in the $(a,b,0,0)$ plane in the enumeration used in \cite{Klevers:2014bqa}. The base is spanned by the primitive rays that have the $F_4$ ambient space over the generic point. This yields the base divisors $w_{0,0},w_{1,0},w_2, y_1, y_{1,0}, y_2, y_3 , y_{4,0}$ obtained from the primitive rays of $(-,-,c,d)$ vertices and then constructing a fan from those. The resulting base admits $h^{1,1}(B_2)=6$ or equivalently $T=5$ in six dimensions. In addition there are fibral divisors of $\mathfrak{su}_3^{(1)},\mathfrak{su}_2 ^{(1)},\mathfrak{su}_2^{(1)}$ and  $\mathfrak{e}_6^{(2)}$-type. This is consistent with the number of K\"ahler moduli
\begin{align}
    h^{1,1}(X)=2+T+\text{rk}(G)=2+5+8 = 15 \,. 
\end{align} 
 The full quiver of the model admits the following ring like structure
\begin{align}
\label{eq:quiver1}
    \begin{picture}(0,60)
    \put(-5,0){$\overset{\mathfrak{su}_2^{(1)} }{0}$}
    \put(20,0){\textcolor{red}{0}} 
    \put(35,0){$\overset{\mathfrak{su}_3^{(1)} }{2}$} 
    \put(35,25){$\overset{\mathfrak{su}_2^{(1)} }{2}$}
    \put(0,25){1} 
    \put(-5,45){$ \overset{\mathfrak{e}_6^{(2)}}{ 4}$}
    \put(40,45){2}
    \put(20,45){1}
    \end{picture}
\end{align}
Red marks the choosen curve, whose volume is send to infinity in the decompactification limit to a little string theory in the next section.  

As discussed in Section~\ref{sec:main_5D_example}, the $\fe_6^{(2)}$ twisted algebra admits a spectrum as follows
\begin{align}
\mathbf{26}_0 \oplus \mathbf{26}_1  \oplus \mathbf{1}_1 \oplus \mathbf{1}_0
\end{align}
This twisted reduction admits a Jacobian fibration with an $\mathfrak{e}_7$ gauge algebra and two full hypers in the $\mathbf{56}$ representation. 

\paragraph{The second elliptic fibration:}
The second torus fibration is given by the second 2D sub-polytope structure and found by projecting onto the
  $(a,0,0,d)$-slice in the ambient 4D polytope. The 2D sub-polytope that is given by $F_6$ yields a Morrison-Park model \cite{Klevers:2014bqa,Morrison:2012ei} which is elliptic and hence not twisted. As before, the base is found by inspection, resulting in the ring like-quiver
\begin{align}
\label{eq:quiver2}
    \begin{picture}(0,60)
    \put(-5,0){$\overset{\mathfrak{su}_3^{(1)} }{3}$}
    \put(20,0){\textcolor{red}{0}} 
    \put(35,0){$-1$} 
    \put(43,25){$2$}
    \put(0,25){1}  
    \put(-5,45){$ \overset{\mathfrak{f}_4^{(1)}}{ 5}$}
    \put(43,45){2}
    \put(27,45){2}
    \put(13,45){1}
    \end{picture}
\end{align}
In the LST decompactification limit, the volume of the curve in red is taken to infinite. This curve is dual to the one in the quiver \eqref{eq:quiver1}.
Note that there is also a curve with a geometric self-intersection $+1$, which we denote as $-1$ in the quiver convention. Due to the presence of a section, the resulting 5D compactification is untwisted and hence we only find untwisted algebras over each curve. The 6D theory admits $T=6$ and an $\mathfrak{su}_3\times \mathfrak{f}_4 \times \mathfrak{u}_1 $ gauge algebra, which matches the $15$ K\"ahler moduli of the geometry. 

\subsection{Little String limit and twisted T-duality}
In this subsection we will decouple gravity to obtain an LST.
At the level of the quiver, this can be achieved by removing the curve $w_2$ from the base, such that the resulting intersection form of the residual curves $w_I \cdot w_I = \Omega_{I,J}$ is positive semi-definite \cite{Bhardwaj:2015oru}. The respective base curve lifts to a toric divisor of the ambient space, given by the ray $(1,-1,-1,-1)$.

The residual toric rays lift to divisors on the full threefold and in fact also to the 4D ambient space \footnote{More details to the construction can be found in
\cite{DelZotto:2022xrh}.
}
which yields a non-compact ambient space.
This limit is exactly chosen such that it removes the red $0$ curves in the quiver diagrams \eqref{eq:quiver1} and \eqref{eq:quiver2}. While the red curve is completely removed, the neighbouring quiver nodes become non-compact and their fibers become flavor algebra factors that we denote by $[\mathfrak{g}^{(s)}_F]$.  The decompactification limit of the threefold is chosen such that the two torus-fibration structures are still part of the non-compact threefold which preserves T-duality for the LSTs.  

As there are two LSTs we can compute and match their 2-group structure constants. In order to do so, we first need to compute the little string charges $N_I$. These are given as the 
multiplicities of the unique null vector \cite{Bhardwaj:2015oru} of the intersection form $\Omega_{IJ}$ 
\begin{align}
    \Sigma_0=\sum_i N_I w^I \, ,
\end{align}
  We summarize the resulting non-compact quiver with its LS charges as the vectors 
 \begin{align}
 \label{eq:LSTQuiver1}
[\mathfrak{su}_2] \, \, 1\, \,  \overset{\mathfrak{e}_6^{(2)}}{4}   \, \,   1\, \, 2 \, \,  \overset{\mathfrak{su}_2^{(1)}}{2} \, \, [\mathfrak{su}_3]   \, , \qquad    N_I=(1,1,3,2,1) \, .
\end{align} 
The above LST admits four compact base curves and twisted fibers of total rank $5$ and hence a Coulomb branch of dimension  dim$(CB)=9$. We have excluded other flavor and matter factors from the quivers, as they are not important for our arguments. 
\begin{table} 
\begin{center}
\begin{tabular}{cc}
\begin{tabular}{|c|c|}\hline
\multicolumn{2}{|c|}{Genus 1 fiber} \\ \hline
X& (-1,1,0,0) \\  
Y& (-1,-1,0,0) \\
Z& (1,0,0,0) \\ \hline
$w_{0,0}$ &(-1,0,0,-1)  \\
$w_{0,1}$&(-1,-1,0,-1 ) \\
$w_{0,2}$ & (0,-1,0,-1)\\ \hline
$w_{1,0}$&(0,0,0,1) \\
$w_{1,1}$& (-1,0,0,1) \\ \hline
$w_2$ & (1, -1,-1,-1) \\ \hline
\end{tabular}
&
\begin{tabular}{|c|l|}  \hline
\multicolumn{2}{|c|}{Compact curves} \\ \hline
 $y_1$ & (-3,-1,1,1) \\ 
$y_{1,i} $&(-3,0,1,0),  (-5,-2,2,0) , (-7,-2,3,0),  (-9,-3,4,0) ,(-5,-1,2,0)  \\
$y_2 $ &(-7,-3,3,-1) \\
$y_3$ & (-5,-2,2,-1) \\
$y_{4,i}$& (-3,-2,1,-1) , ( -3,-1,1,-1) \\ \hline
\end{tabular}

\end{tabular}
\end{center}
\caption{\label{tab:example1}\textit{Toric rays of the compact threefold with inequivalent genus-one and elliptic fibration descending from the ambient space.}}
\end{table}
The second LST quiver is given as 
  \begin{align}
  \label{eq:QuiverDual2}
[\mathfrak{su}_3] \, \, 1\, \,  \overset{\mathfrak{f}_4^{(1)}}{5}   \, \,   1\, \, 2 \, \,  2 \, \,2 \, ,  \qquad    N_I=(1, 1, 4, 3, 2, 1) \, .
\end{align}
The theory admits five compact base curves and a rank four gauge group which also yields a Coulomb branch dimension dim$(CB)=9$ as expected.   

\subsubsection*{Consistency check from 2-groups}
As mentioned above, the fact that these two theories have the same compact geometric origin implies that the two LSTs are (twisted) T-duals. Beyond that however we can also match the 2-group structure constants characteristic to the LSTs as proposed in  \cite{DelZotto:2020sop}. 
 
In the following we focus on the universal 2-group structure constants obtained from the mixing of the $\mathfrak{u}_1$ LST 1-form symmetry and the Poincare and SU(2)-R-symmetry.  
Their structure constants are computed as \cite{Cordova:2020tij,DelZotto:2020sop}
\begin{align}
    \kappa_P =  \sum_I N_I (2-\Omega^{II}) \, , \qquad \kappa_R= \sum_I N_I h^v_{\mathfrak{g}^{(s)}_I} \, .
\end{align}
Here $\Omega^{II}=n_I$ and $h^v_{\mathfrak{g}^{(s)}_I}$ being the dual Coxeter number of the $s$ twisted gauge algebra $\fg^{(s)}$ coupled to the curve $w^I$ with string charge $N_I$. The dual Coxeter number can simply be computed from the sum of the Kac labels of the Dynkin diagram.
For $\mathfrak{su}_n^{(1)}$ algebras, the dual Coxeter is given as 
 \begin{align} h^{v}_{\mathfrak{su}_n^{(1)}}= n \, .
 \end{align}
 We formally assign the value $h^{v}=1$ for curves that are paired to the trivial gauge
 algebra.  Most importantly, the dual Coxeter number is sensitive to a twisted or untwisted affinization of a finite gauge algebra. For the case at hand, there is 
 \begin{align} 
     h^{v}_{\mathfrak{e}_6^{(2)}}=12 \, , \qquad h^{v}_{\mathfrak{f}_4^{(1)}}=9 \, .
 \end{align} 
 Taking the above difference into account, we obtain the matching 2-group structure coefficients in the two theories  
 \begin{align}
     (\kappa_P , \kappa_R)=(2,20) \, .
 \end{align}
 The above check is highly non-trivial and confirms that the two LSTs are indeed T-dual as implied from geometry.  

\subsubsection*{Untwisting the fiber}
At this point one might wonder how the threefold would differ if the $\fe_6^{(2)}$ fiber were to be replaced by a regular $\mathfrak{f}_4^{(1)}$ fiber over the same curve in a genus-one geometry. In such a case at least the number of K\"ahler moduli would agree, but what about the rest of the theory?

Fortunately, such a  geometry exists with polytope vertices given in Table~\ref{tab:LSTexample2} and is obtained simply by exchanging the vertex $ (-9,-3,4,0) $ for $(-2,-1,1,0)$. 
It should be noted that the resulting threefold has not only the same number of K\"ahler moduli, but also complex structure moduli
\begin{align}
    (h^{1,1},h^{2,1})(X_2)=(15,63 (2)) \, ,
\end{align}
as the one with $\fe_6^{(2)}$ fiber (however the number of complex structure moduli realized ``non-polynomially" has reduced in number).
\begin{table}[t!]
\begin{center}
\begin{tabular}{cc}
\begin{tabular}{|c|c|}\hline
\multicolumn{2}{|c|}{Genus 1 fiber} \\ \hline
X& (-1,1,0,0) \\  
Y& (-1,-1,0,0) \\
Z& (1,0,0,0) \\ \hline
$w_{0,0}$ &(-1,0,0,-1)  \\
$w_{0,1}$&(-1,-1,0,-1 ) \\
$w_{0,2}$ & (0,-1,0,-1)\\ \hline
$w_{1,0}$&(0,0,0,1) \\
$w_{1,1}$& (-1,0,0,1) \\ \hline
$w_2$ & (1, -1,-1,-1) \\ \hline
\end{tabular}
&
\begin{tabular}{|c|l|}  \hline
\multicolumn{2}{|c|}{Compact curves} \\ \hline
 $y_1$ & (-3,-1,1,1) \\ 
$y_{1,i} $&(-3,0,1,0),  (-5,-2,2,0) , (-7,-2,3,0),(-5,-1,2,0),(-2,-1,1,0)  \\
$y_2 $ &(-7,-3,3,-1) \\
$y_3$ & (-5,-2,2,-1) \\
$y_{4,i}$& (-3,-2,1,-1) , ( -3,-1,1,-1) \\ \hline
\end{tabular}
\end{tabular}
\end{center}
\caption{\label{tab:LSTexample2}\textit{Toric rays of the threefold where the $\fe_6^{(2)}$ fiber got exchanged for $\ff_4^{(1)}$. Both geometries share the same Hodge numbers but differ not only in curve structure but also the number of non-polynomial complex structures deformations.}}
\end{table}

In this new geometry, the $\mathfrak{e}_6^{(2)}$ fibration over the curve $\mathcal{Z}$ with self-intersection $\mathcal{Z}^2=-4$ has now changed to an untwisted $\mathfrak{f}_4^{(1)}$ fibration. Upon decoupling gravity the LST quiver for the first genus-one fibration is given as 
 \begin{align}
[\mathfrak{su}_2] \, \, 1\, \,  \overset{\mathfrak{f}_4^{(1)}}{4}   \, \,   1\, \, 2 \, \,  \overset{\mathfrak{su}_2^{(1)}}{2} \, \, [\mathfrak{su}_3]   \, , \qquad    N_I=(1,1,3,2,1) \, .
\end{align}
Since the Hodge numbers have not changed, the Coulomb branch dimension of the LST remains the same. As we will explore in more generality in Section~\ref{sec:TVSUT}, the $\mathfrak{f}_4^{(1)}$ fiber over the $\mathcal{Z}^2=-4$ curve 
admits a single $\mathbf{26}_0$-plet only. Although the change in fiber structure was very subtle, the 2-group structure constant $\kappa_R$ is able to detect it, as the dual Coxeter number of $\ff_4^{(1)}$ is lower than $\fe_6^{(2)}$.

Importantly, the change in the fiber structure has not eliminated the second elliptic fibration and as a result we can discuss the change in the T-dual geometry in this new context as well. 
 Here we find the LST quiver
 \begin{align}
 \begin{array}{ccc}
[\mathfrak{su}_3] \, \, 1 & \overset{\mathfrak{f}_4^{(1)}}{5}  &   1\, \, 2 \, \,  2 \, ,    \qquad    N_I=(1, 1, 3, 2, 1) \, \\
&1 & \qquad \qquad \quad \, \, \, \, \, 1,
\end{array} \,. 
\end{align}
The resulting quiver is similar to the original one given in \eqref{eq:QuiverDual2}. The only difference is that an E-string curve to the right of the $5$ curve has been moved down, making it a trivalent vertex.

The change in base curve structure has altered the LST charge vector $N_I$ exactly in such a way that the 2-group structure constants still match in both theories, given as
 \begin{align}
     (\kappa_P , \kappa_R)=(2,17) \, ,
 \end{align}

In the following we propose a physical interpretation of the two (dual) transitions as specific Higgs branch deformations. Indeed, in \cite{DelZotto:2022xrh}, the proposal was made  
that $\kappa_R$ should be thought of as a measure of the degrees of freedom of an LST which decrease monotonoically along Higgs branch deformations, analogously to the a-coefficient in an SCFT. This might at first seem puzzling as both 
geometries admit the very same number of complex structure coefficients. However we have observed that in the untwisted case, one of those deformations changed from non-polynomial to a polynomial deformation. This hints at the possibility that the twisted algebra lives on an enhanced point of symmetry in the the moduli space of the theory. In Section~\ref{sec:TVSUT} we will consider this type of transition in further generality. Similarly, for the T-dual theory we have an exotic type of E-string transition, which has changed the number of M-strings, i.e. 2 curves into an E-string.
\\\\ 
 
 We close this section by discussing how all of the above theories/geometries can be similarly used to construct twisted 5D SQFTs with an SCFT limit in the UV. The plan we employ is the same as for the LST phase. One simply needs to remove an arbitrary compact base curve from an LST quiver\footnote{As noted in \cite{DelZotto:2022xrh} any such removal destroys other inequivalent torus fibration in the geometry.} to obtain an SQFT that flows to an SCFT. 
 
The two simplest choices are to decompactify the left most $1$ curve or rightmost $2$ curve in the quiver \eqref{eq:LSTQuiver1} resulting in an SCFT quiver such as 
\begin{align}
    \overset{\mathfrak{e}_6^{(2)}}{4}\,\, 1 \, \, 2 \, \overset{\mathfrak{su}_2^{(1)}}{2} \,  .
\end{align} 
All of the constructions presented in this sections, straightforwardly generalize to the other twisted reductions that we discuss in Section~\ref{sec:MoreTwistedTheories}. In particular many of the explicit toric examples given in Appendix~\ref{app:Example3folds} have twisted algebras over $0$ curves or shrinkable curves that yield LSTs or SCFTs in their respective decompactification limits.   

\section{Twisted VS. Untwisted Genus-One Fibers}
\label{sec:TVSUT}

In the previous section we encountered the possibility of exchanging a twisted algebra with its untwisted counterpart 
\begin{align}
    \fg \in \hat{\fg}^{(n)}\, \,  \text{ and } \, \, \fg \in \fg^{(1)} \, .
\end{align}
In the $\overset{\fe_6^{(2)}}{4} \rightarrow \overset{\ff_4^{(1)}}{4}$ case, we noted that the two compact threefolds share some important features, in particular identical Hodge numbers. In this section we will compare twisted and untwisted fibers more systematically and observe that these relationships can be formulated in more generality. In fact in all the examples we have encountered, the geometries with twisted fibrations exhibit ``cousins" in the form of untwisted fibrations which share the very same Hodge numbers and can be connected through a geometric transition.  

These observations help to resolve a puzzle in the literature: In 
\cite{Braun:2014oya} the geometry of a genus-one fibration with a twisted  
$\fso_8^{(3)}$ algebra was considered and in that work it was expected to lift to a 6D F-theory with an $\fg_2$ gauge algebra. It was noted that all gauge and gravity anomalies could be cancelled. 
In view of the geometries we have seen in the present work, this conclusion may come as a surprise as the Jacobian admits a very different fiber structure and we therefore do not expect such a gauge algebra in the 6D F-theory lift (see Section~\ref{ssec:so83Geometry}).
The fact that the gauge algebra must change in the 6D lift is further signaled from the 5D BPS state count enumerated throughout this work, which deviates from  \cite{Braun:2014oya}. In the present work, we find that the twisted algebras do not solve the 6D anomalies if their finite gauge algebra is not altered. 
At first glance, this may appear as a very unlikely coincidence: How did the lifts of \cite{Braun:2014oya}, which assumed that the gauge algebra was unaltered between 5D/6D, lead to a seemingly consistent 6D SUGRA theory? 

This puzzle is resolved by the geometries we present in this section by noting that most of the twisted algebras we study are closely related to a similar geometry with an untwisted fiber. These untwisted algebras have the same fiber structure in the Jacobian and complex structure and BPS states consistent with the 6D anomalies as proposed in \cite{Braun:2014oya}. This section is therefore devoted to a comparison of these two type of geometries and the transitions among them.

Besides the subtle difference in the fiber structure of the two geometries, there is another similarly subtle difference in the complex structure moduli sector: We find that twisted fibrations generally admit complex structure deformations that are realized as non-polynomial deformations in the toric hypersurface description, as opposed to the untwisted one. Such terms can be conveniently computed via the Batyrev formula from the toric polytope. While this observation might first appear just as a technical detail, we propose a physical interpretation,  making those deformations relevant to understand the difference between twisted and untwisted fibers. In order to do so, we first discuss their physical significance in elliptic fibrations.

\subsection{Enhanced Symmetries and Non-Poly Defs}

The number of complex structure moduli and their non-polynomial contributions in toric hypersurfaces can be conventiently computed from
the pair of relexive polytopes $\Delta,\Delta^*$ in the Batryev construction \cite{Batyrev:1994pg}.
It is given via the combinatorial formula
\begin{align}
    h^{2,1}(X)=l(\Delta^*) - 4 -\sum_{\Gamma^*} l^\circ (\Gamma^*)  + \underbrace{\sum_\Theta l^*(\Theta) l^*(\Theta^*)}_{h^{2,1}_{np}} \, .
\end{align}
Here $(l^\circ)l$ counts (interior) points, with 
 $\Gamma^*$ being edges in $\Delta^*$ and $\Theta^*$ being codimension two faces of $\Delta^*$ with duals  $\Theta$. The part  $h^{2,1}_{np}$ is the main relevant contribution for us, that counts complex structures without a corresponding hypersurface monomial. 
 
 For elliptic fibration such deformations can be given a physical interpretation in the F/M-theory picture. For this we first introduce the neutral components of hypermultiplet fields
\begin{align}
    \delta h^{2,1}_{np}(\mathbf{R})=  \text{Dim}(\mathbf{R})-H_{ch}(\mathbf{R}) \, .
\end{align}
$H_{ch}(\mathbf{R})$ denotes the non-trivial weights of a hypermultiplet representation $\mathbf{R}$, called the charge dimension of a representation. In the case of the adjoint representation, $H_{ch}(\textbf{Adj})$ is simply the number of roots and $\delta  h^{2,1}_{np}(\textbf{Adj})$ the number of Cartan generators. As $\delta h^{2,1}_{np}(\mathbf{R})$ are themselves uncharged fields everywhere at the Coulomb branch, they are counted as complex structure moduli in the threefold. In fact there is a natural reason why those contributions should be non-polynomial when giving them a symmetry breaking vev. For adjoint hypermultiplets, such Higgs branches typically breaks to the maximal Cartan sub-algebra as in
\begin{align}
  \fsu_2 \rightarrow \mathfrak{u}_1\, ,
  \end{align}
  via $\mathbf{3} \rightarrow \mathbf{1}_1 \oplus\mathbf{1}_{-1} \oplus\mathbf{1}_0$. 
There are more options for non-simply laced algebras as those typically admit non-trivial (lower dimensional) representations with non-trivial  
$\delta h^{2,1}_{np}(\mathbf{R})$ besides the adjoint. When those representations acquire non-trivial vevs, a rank preserving Higgsing is possible such as in 
\begin{align}
   \fg_2 \rightarrow \fsu_3 \, , \quad  \ff_4 \rightarrow \fso_8\, ,
\end{align}
via the decompositions of the Hypermultiplet representations
\begin{align} 
\mathbf{7} \rightarrow \mathbf{3} \oplus \overline{\mathbf{3}} \oplus \mathbf{1}\,, \quad
\mathbf{26} \rightarrow \mathbf{8}_v \oplus \mathbf{8}_s  \oplus  \mathbf{8}_c \oplus  \mathbf{1} \oplus \mathbf{1} \, ,
\end{align}
that acquire the symmetry breaking vevs. Those  vevs sit precisely in the $\delta h^{2,1}_{np}(\mathbf{R})$ parts of $\mathbf{R}$. As the Higgsing is rank preserving, no Cartan generator becomes massive and the number of neutral fields is preserved, resulting in a conservation of Hodge numbers.

The above considerations are reflected in the geometry of the respective elliptic fibration: E.g. for each $\mathbf{26}$-plet representation of an $\ff_4$ fiber, one finds two $\delta h^{2,1}_{np}$, realized as non-polynomial complex structure deformations. When Higgsing to $\fso_8$, the threefolds admits the very same Hodge numbers. The difference however is, that the associated $\delta h ^{2,1}_{np}$ deformations became all realized as polynomial deformations. Similarly, when Higgsing $\fsu_2$ on $g\times \mathbf{3}$-plets, all $g$ non-polynomial deformations become polynomial and the geometry acquires a non-trivial Mordell-Weil rank \cite{Morrison:2014era}. 

Turning this logic upside down, the knowledge about non-polynomial deformations may allow us to track representation content\footnote{Care has to be taken from other sources of such deformations, such as adjoints or empty $-2$ curves. See for example \cite{Johnson:2016qar}.} of the gauge algebra, directly from geometry. Furthermore it suggests a direct physics interpretation. In the symmetry broken phase, the vevs are moduli that parameterize the (partial) Higgs branch of the two gauge algebras of same rank. As those vevs are moduli, they can take on any non-vanishing value matching the (real part of the) monomial coefficients in the hypersurface. At the locus of enhanced symmetry, the vevs must vanish, matching the absence of the respective polynomial coefficients, while the number of neutral fields is preserved.

We adopt this interpretation of the non-polynomial defs for the twisted algebras in what follows.
As already anticipated in Section~\ref{sec:TwistedDuality},
the presence of those non-polynomial deformation is a universal feature for (almost) all twisted fibers. Furthermore it suggests the possibility that they are part of a larger twisted affine type of representation, which can be Higgsed in a rank preserving way to a non-twisted one. Moreover we will use the field theory of twisted
compactifications, to match/predict non-poly defs in the compact threefolds. At the same time, their presence yields a criterion when geometric transitions to an untwisted algebra may be possible. 
These Higgsing transitions/singlets are generically present but with the exception of the twisted versions of 6D non-Higgsable cluster theories $\overset{\fe_6^{(2)}}{6},\overset{\fsu_3^{(2)}}{3}$ and $\overset{\fso_8^{(3)}}{4}$ as already alluded to in Section~\ref{sec:main_5D_example} and Section~\ref{sec:MoreTwistedTheories}.

\subsection{ $\fe_6^{(2)}$ vs $\ff_4^{(1)}$ genus-one fibers }
In the following we consider another singular genus-one model which is related to the $\fe_6^{(2)}$ twisted fibration of the quartic which we will call $X^D$. For this we consider the generalized Tate-vector  
\begin{align} 
\label{eq:TateF4G1}
    n_i=  \left\{3,2,2,1,0,2,1,0 \right\} \,,
\end{align}
which reduces the vanishing order of the $X^3 Y$ term in the quartic fiber model by one. From this perspective, the $\fe_6^{(2)}$ model appears as an enhanced singularity.

As we will discuss in the following, the resulting singularity will be of $IV^{*,ns}$ type both in the genus-one and Jacobian fibration.
Note  that the above tuning is such, that the monodromy divisor $\overline{D}_{2,X}=d_8^2 - 4 d_5 $ stays  un-affected, still intersecting the base divisor
$\mathcal{Z}$ non-trivially.
When mapping the singular genus-one model, specified by the Tate vector \eqref{eq:TateF4G1} to the Jacobian we obtain (at leading order in $z$) the expressions
\begin{align}
\begin{split}
\label{eq:WSFJacF4}
f= &z^3 \widehat{p} + \mathcal{O}(z^4) \, ,\\
g=&z^4 d_2^2  \overline{D}_{2,X} +  \mathcal{O}(z^5) \, ,\\
\Delta=&z^8  d_2^4  \overline{D}_{2,X}^2 +\mathcal{O}(z^9) \, . 
\end{split}
\end{align}
As long $\overline{D}_{2,X}|z=0$ does not become a perfect square, which is the case by construction\footnote{The polynomial in f is given as $\widehat{p}=d_2 d_7 d_8 - 2 d_1 d_8^2 - 2 d_2 d_4 + 8 d_1 d_5 $. }, we have a Type $IV^*$ non-split singularity i.e. an $\ff_4$ in the Jacobian as claimed. 
Thus we expect an $\ff_4 \times \mathbb{Z}_2$ gauge algebra in the 6D F-theory.
We begin by analyzing the spectrum in the Jacobian and then argue that this spectrum coincides with that in the genus-one fibration as well.  In order to do so, we first note that the discriminant admits two codimension two components over $z=0$: First there is the monodromy divisor $\overline{D}_{2,X}$ and the second component $d_2=0$. The later one yields a codimension two vanishing order $(3,5,9)$ which corresponds to an $E_7$ singularity. The hypermultiplets that are localized here yield  $\mathbf{26}_1$ matter multiplets which we will double check in the genus-one model momentarily.

When computing the multiplicity of states, we first start by being fully general.  
 For this we recall the line bundle classes of the various sections to be given as
\begin{align}
[d_2] \sim 2 c_1 - \mathcal{S}_9 - 2 \mathcal{Z} \, , \quad [d_8] \sim c_1 + \mathcal{S}_9 \, ,
\end{align} 
which are needed to compute the multiplicities of states. Those of the $\mathbf{26}_1$ states are simply evaluated via the intersections $\mathcal{Z} \cdot [d_2]$ whereas the neutral $\mathbf{26}_0$-plets 
are non-localized states over the loci where the fibral curves of $\mathcal{Z}$ are branched, which is fixed by the number of ramification points $\mathcal{R}=[\overline{D}_{2,X}] \cdot [\mathcal{Z}]$ in the RH theorem Eq.~\eqref{eq:RHT}. Taking the degree of the cover to be $d=2$, we find the following multiplicities
\begin{align}
\label{eq:F4realMultis}
n_{\mathbf{52}_0}= g \, , \quad n_{\mathbf{26}_1}=4(1-g)-\mathcal{S}_9 \cdot \mathcal{Z} \, , \quad n_{\mathbf{26}_0}=(1-g)+\mathcal{Z}^2+ \mathcal{S}_9 \cdot \mathcal{Z} \, ,
\end{align}
which admits the additional degree of freedom, to dial the relative multiplicities via the $\mathcal{S}_9 \cdot \mathcal{Z}$ intersections.
Notably, the combined multiplicity of the $\ff_4$ fundamentals yields
\begin{align}
\label{eq:f4Multi}
    n_{\mathbf{26}_1} + n_{\mathbf{26}_0}= 5(1-g)+ \mathcal{Z}^2 \, ,
\end{align}
which is consistent with 6D gauge algebra anomaly cancellation conditions (see e.g. \cite{Esole:2017rgz}). 

In order to compare the above model with an geometric transition to the $\fe_6^{(2)}$ geometry however, we also need to impose the extra condition that $[d_1]\cdot [\mathcal{Z}]=0 $ which resulted in $(4,6,12)$ points. These points are generally absent in the $\ff_4^{(1)}$ model, but would appear when performing the respective deformation to $\fe_6^{(2)}$. This extra condition removes the degree of freedom and  yields 
\begin{align}
    \mathcal{S}_9 \cdot \mathcal{Z}= 2(1-g)+ \frac12 \mathcal{Z}^2  \, .
\end{align}
Plugging  this restrictions into \eqref{eq:f4Multi} gives
\begin{align}
\label{eq:F4realMultis2}
n_{\mathbf{52}_0}= g \, , \quad n_{\mathbf{26}_1}=2(1-g) +  \frac12    \mathcal{Z}^2 \, , \quad n_{\mathbf{26}_0}= 3(1-g)+\frac12 \mathcal{Z}^2  \, .
\end{align}
When compared to the $\fe_6^{(2)}$ multiplicities that were computed in Section~\ref{ssec:E62}, we find the $\mathbf{26}_0$-plets to exactly coincide with the multiplicity there, whereas the number of $\mathbf{26}_1$ is reduced by one. 

This computation concludes the $\ff_4$ spectrum in the Jacobian. In the following however, we want to double check the claim, that the spectrum coincides with the genus-one model. 

\begin{figure}[t!]
 \begin{center}
 {\footnotesize
 \begin{picture}(170, 110)
 \put(-150,10){\includegraphics[scale=0.6]{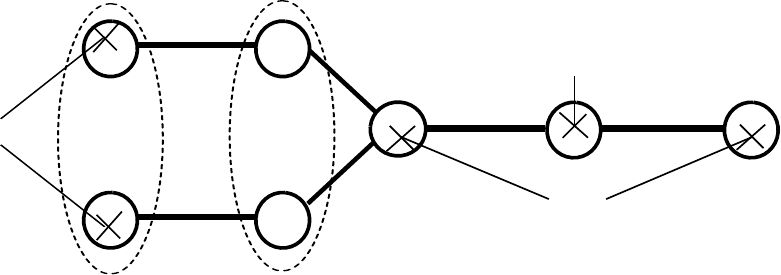} }
 
        \put(-160,47){$[X]$}
 \put(-125,95){$[f_2]$}
 \put(-78,95){$[g_2]$}
  \put(-40,65){$[h_1]$}

   \put(22,60){$[g_1]$} 
   \put(10,70){$[Y]$}  
    \put(10,26){$[Z]$}  
    %  \put(60,70){$[Y]$}  
    
       \put(25,40){$2$}

          %  \put(-110,30){$1$}
            \put(-108,80){$1$}
            
        %    \put(-60,30){$2$}
            \put(-55,80){$2$}

               \put(-35,35){$3$} 
      
                \put(100,90){$\xrightarrow{\overline{D}_{2,X}=0}$     }
      
      \put(150,83){\includegraphics[scale=0.5]{MultiF4.pdf} }
   
  \put(70,60){$[f_3]$}  
   \put(65,35){$1$}   
    \put(155,87){$2$}  
      \put(196,87){$4$}  
     \put(237,87){$3$}  
    \put(279,87){$2$}  
     \put(321,87){$1$}  
     
                 \put(100,40){$\xrightarrow{d_2=0}$     }
      
      \put(160,13){\includegraphics[scale=0.5]{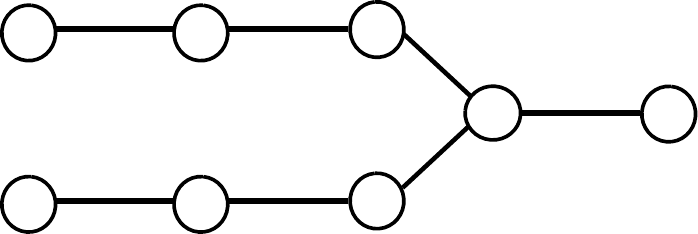} } 
      
    \put(165,17){$1$}  
      \put(165,58){$1$}

      \put(206,17){$2$} 
            \put(206,58){$2$} 
        
     \put(249,17){$3$}  
       \put(249,59){$3$}  
     
    \put(276,39){$2$}  
    
     \put(319,39){$1$}  
      
 \end{picture}
 }
 \caption{\label{fig:F4}{\it 
 Depiction of the $X^D$ genus-one fibration and its $\mathfrak{f}_4^{(1)}$ fibral structure. Choosing $f_2$ as the affine node results in an $\fso_9$ singularity while keeping $f_3$ finite results in an $\ff_4$ singularity. The two codimension two degeneration loci shown on the right.}} 
 \end{center}
 \end{figure}

For this we consider the resolved genus-one fibration $X^D$, given as 
 \begin{align}
 p=&d_1   f_2^3 g_2^2 h_1 X^4 + d_2   f_2^2 g_2 X^3 Y + 
 d_3   f_2^2 f_3 g_1^2 g_2^2 h_1^2 X^2 Y^2  + 
 d_4   f_2 f_3 g_1^2 g_2 h_1 X Y^3  \nonumber \\ & + d_5   f_3 g_1^2 Y^4 + 
 d_6  f_2^2 f_3 g_1 g_2^2 h_1^2 X^2 Z  + d_7   f_2 f_3 g_1 g_2 h_1 X Y Z +
  d_8 f_3 g_1 Y^2 Z +  f_3 Z^2 \, .
 \end{align}
 with fibral Stanley Reisner ideal 
 \begin{align}
 \mathcal{SRI}: \{ & Y X, Y f_2, Y g_2, Y h_1, e_1 Z, e_1 f_2, e_1 f_3, e_1 g_1, e_1 g_2, e_1 h_1, Z f_1, \nonumber \\ &
f_1 f_3, f_1 g_2, f_1 h_1, Z g_1, X f_3, f_2 f_3, f_3 g_2, X g_1, X g_2, X h_1, f_2 h_1 \} \, .
 \end{align}
The base divisor $\mathcal{Z}$ then pulls back to the reducible divisor  
\begin{align}
    \mathcal{Z} = f_2 g_2^2 h_1^3 g_1^2 f_3  \, ,
\end{align}
with fibral curves 
 \begin{align}
 \begin{array}{ll } 
%  \text{Fiberal curve}  & \text{vanishing ideal}   \\ \hline 
\mathbb{P}^1_{0,\pm}:&f_2 \cap d_5   g_1^2+d_8   g_1 Z+  Z^2  \, ,  \\ 
\mathbb{P}^1_{1,\pm}:& g_2 \cap d_5 g_1^2+d_8 g_1 Z+  Z^2  \, ,  \\  
\mathbb{P}^1_{2}:&h_1 \cap d_5 f_3 g_1^2+d_2 g_2+d_8 f_3 g_1 Z+  f_3 Z^2  \, ,  \\  
\mathbb{P}^1_{3}: & g_1 \cap   f_3+d_1   f_2^3 g_2^2 h_1+d_2  f_2^2 g_2 Y  \, ,   \\  
\mathbb{P}^1_{4}: & f_3 \cap d_1 h_1+d_2 Y \, .    \\ 
 \end{array} 
 \end{align}
In Figure~\ref{fig:F4} we have summarized the fibral intersections of the curves.  
The monodromy divisor $\overline{D}_{2,X}$ of the 2-section $X$ is again central as it is also the locus along which the two fibral curves in $f_2$ and $g_2$ get interchanged.  

 Note that the 2-section $X$ intersects the node $f_2$ which is the affine node we associate with the base coordinate $z$. This however might come as a surprise, as it is $f_3$ which lifts to an actual $F_4$ singularity in the genus-one geometry! In fact, when shrinking all but the $f_3$ node, we need to again orthogonalize the reference 2-section properly with respect to the unbroken gauge group.  

In order to account for both singular limits, we consider the two possible discrete Shioda map combinations
\begin{align}
\sigma_{2}(X) = [X] \, , \qquad \sigma^\prime_2(X) = [X] +2[f_2] +3 [g_2] +4 [h_1] +2 [g_1] \, ,
\end{align} 
for which we have the intersections
 \begin{align}
 \sigma_2(X) \cdot (f_2,g_2,h_1, g_1, f_3) = (2,0,0,0,0) \, , \, \text{ and } \,     \sigma^\prime_2(X) \cdot (f_2,g_2,h_1, g_1, f_3)=(0,0,0,0,2) \, .
 \end{align}
While the topology of the resolved fiber is that of $\mathfrak{f}_4^{(1)}$ 
the two singular limits, keeping either $f_2$ or $f_3$ finite, yield an $\fso_9$ or $\ff_4$ singularity respectively.\footnote{ 
 Note however, that both singular limits become identical when being mapped into the singular Weierstrass model.}

In order to cross check the computations made in the Jacobian, we investigate the matter loci in the resolved genus-one geometry next. 
First we recall the codimension two locus $d_2=0$. 
When investigating the fibral curves, we indeed find a split into an $E_7$ type of fiber, just as in the Jacobian. 
We can then compute the intersections with the other fibral divisors e.g. for the  $\mathbb{P}^1_{2,\pm}$ curve and obtain the intersections
\begin{align}
\mathbb{P}^1_{2,\pm} \cdot (f_2,g_2,h_1, g_1, f_3)_{(\sigma_2,\sigma_2^\prime)} = (0,1,-1,0,1)_{(0,-1)} \, .
\end{align}
These intersections numbers yield the charges of (massive) 5D particles obtained from M2 branes wrapping the respective curve. When taking $f_3$ as the affine node and hence finite, such that we obtain the $\ff_4$ limit we find weights of an $\mathbf{26}_1$ just as claimed in the Jacobian fibration. Similarly we find, over $\overline{D}_{2,X}=0$ to support the monodromy divisor that yields the $\mathbf{26}_0$-plets. As opposed to the $\fe_6^{(2)}$ model, the fiber does not become a degree two multiple fiber over such loci as the degree one curve  $\mathbb{P}^1_{3}$ does not split here.
Therefore we expect the Weil-Ch$\hat{\text{a}}$telet group to reduce to the Tate-Shafarevich group $\Sha(X^D)$, which we take as the main criterion, why Jacobian and genus-one fibration admit the very same $\ff_4^{(1)}$ fiber structure. From this perspective it is also clear, why the matter and its multiplicities in genus-one and Jacobian are the same, with the only difference being that the $\mathbf{26}_1$-states stay massive at the origin of the $\ff_4$ Coulomb branch. 

The spectrum of the $\ff_4^{(1)}$ genus-one geometry $X^D$ appears to be very similar to that of the $\ff_4 \in \fe_6^{(2)}$ geometry $X^A$
albeit the difference in the massive $\mathbf{26}_1$ multiplicities. One might then wonder, if and how the two fibrations might be related. This can be deduced by having a closer look at the precise geometry and recalling
that we kept different curves finite throughout the transition. In $X^A$, we kept curves in $f_2$ finte, which resulted in  an $\ff_4$ singularity, keeping $f_2$ finite in $X^D$ yields an  $\fso_9 \in \ff_4^{(1)}$ singularity. 
Using the methods from Section~\ref{sec:main_5D_example} and Section~\ref{sec:MoreTwistedTheories} and can indeed compute the shrinkable curves in $X^D$ while keeping $f_2$ finite resulting $32+4=\mathbf{36}$ vector multiplets, that make up the adjoint of $\fso_9$. Hence when performing the conifold transition from $\ff_4 \in \mathfrak{e}_6^{(2)}$ we rather obtain an $\mathfrak{so}_9$ via the breaking  
\begin{align*}
\mathbf{52} \rightarrow  \mathbf{36} \oplus \mathbf{16}\, , \qquad 
    \mathbf{26} \rightarrow  \mathbf{16} \oplus \mathbf{9} \oplus \mathbf{1} \, .
\end{align*}
The curious fact however is, that the resulting genus-one fibration admits another limit in the Coulomb branch moduli space of $\fso_9 \times \hat{\mathfrak{u}}_{1,E}$ such that an $\ff_4 \times \mathfrak{u}_{1,E}$ gauge algebra arises with a massless spectrum just as in the $\mathfrak{e}_6^{(2)}$ theory. This on the other hand suggests, that
not only Coulomb branch dimension but also the number of complex structure moduli of $X^D$ coincides with that of $X^A$. This is indeed the case, as we will demonstrate in an example in the following. 
\subsubsection*{Toric example}
Let us move from these general considerations to a concrete toric example. The toric rays defining the threefold are given in Appendix~\ref{app:Example3folds} and the Hodge numbers can be computed via the Batyrev construction.
In our example geometries, we have the same Hodge numbers 
\begin{align}
 (h^{1,1},h^{2,1})(X^A)=(h^{1,1},h^{2,1})(X^D)=(7,95) \, .
\end{align}
This seems to be at odds with the fact that we performed a tuning of the generalized Tate coefficients when relating the two theories. The resolution to this puzzle is the fact that not all complex structure moduli in the toric hypersurface equation of $X^A$ are realized as polynomial deformations.  
Computing the complex structure moduli, and the respective fractions of non-poly deformations of the two geometries, we obtain 
\begin{align}
    h^{2,1}(h^{2,1}_{np})(X^A)=95(9) \quad \text{ and } \quad h^{2,1}(h^{2,1}_{np})(X^D)=95(6)  \, .
\end{align}
Note that that there are several contributions of non-poly deformations in both geometries that need to be discussed.
The six deformations in the $\mathfrak{f}_4^{(1)}$ geometry $X^D$ are contributions from the 
$3 \times 2$ neutral components inside the three $\mathbf{26}_0$-plets. 

For the twisted fiber $\fe_6^{(2)}$ of geometry $X^A$ we have two contribtutions to the nine non-poly deformations: First there is the contribution of the $3 \times \mathbf{26}_0$-plets neutral fields, which is the same as in the $\mathfrak{f}_4^{(1)}$ model. We propose that the three missing neutral fields correspond to neutral localized singlets. These can be directly deduced from the twisted reduction of the $\fe_6$ theory, which we reviewed in Section~\ref{sec:main_5D_example}. From this 6D origin, we can obtain a prediction for the non-poly deformations as 
\begin{align}
  \delta h_{np}^{2,1}=n_{\mathbf{1}_0}+2 n_{\mathbf{26}_0}+4 n_{\mathbf{52}_0}=   9(1-g)+ \frac32 \mathcal{Z}^2+4g \, .
\end{align}
In the case at hand, the curve of consideration admits self-intersection $\mathcal{Z}^2=0$ and   $g=0$, such that we have a match of all non-poly deformations in $X^A$. Note that this also implies, that an $\fe_6^{(2)}$ on a $\mathcal{Z}^2=-6$ curve of genus zero does not posses any such singlets. Indeed, the respective genus one model, given in Section~\ref{ssec:E62} admits 151 complex structure moduli with $h^{2,1}_{np}=0$.

In analogy to the elliptic threefolds, discussed in the section before, it becomes apparent that those singlets are an integral part of a non-trivial twisted $\fe_6^{(2)}$ representation, and not only its $\ff_4$ subgroup. Hence, when breaking the $\fe_6^{(2)}$ symmetry via the $\ff_4 \rightarrow \fso_9$ chain, it is suggestive that those singlets are not fixed to a vanishing vev any longer which translates to polynomial deformations in the toric hypersurface. We wish to return to this specific point in future work. 

  \subsection{$\fso_8^{(3)}$ vs $\mathfrak{g}_2^{(1)}$ genus-one fibers}
  \label{sssec:g21Hodge}
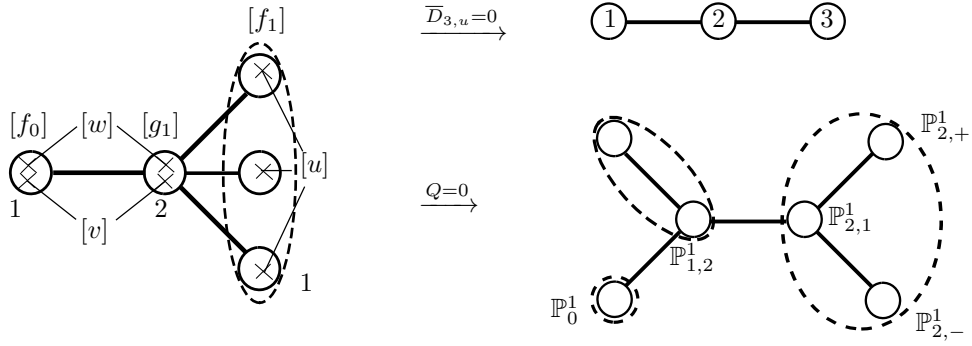
\begin{figure}[t!]
 \begin{center}
 {\footnotesize
 \begin{picture}(250, 150)
 \put(-50,30){\includegraphics[scale=0.6]{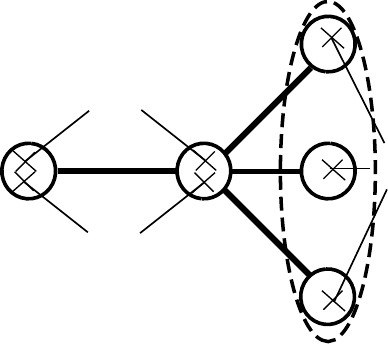} } 
        \put(60,80){$[u]$} 
  \put(-23,95){$[w]$}
    \put(-23,55){$[v]$}
   \put(-50,95){$[f_0]$}
   \put(-50,63){$1$}
   \put(5,63){$2$}
   \put(60,35){$1$}
      \put(0,95){$[g_1]$}
   \put(40,135){$[f_1]$}   
   
    \put(170,20){\includegraphics[scale=0.5]{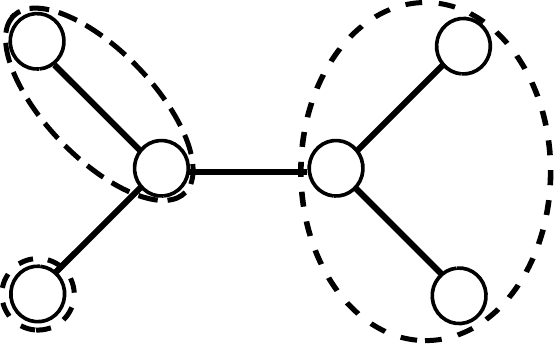} } 
      \put(155,25){$\mathbb{P}^1_0$} 
    \put(200,45){$\mathbb{P}^1_{1,2}$} 
        \put(260,60){$\mathbb{P}^1_{2,1}$} 
         \put(295,95){$\mathbb{P}^1_{2,+}$} 
               \put(293,20){$\mathbb{P}^1_{2,-}$} 
\put(105,65){$\xrightarrow{Q=0}$}
\put(170,130){\includegraphics[scale=0.5]{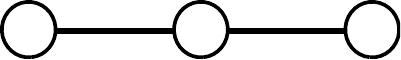} }
\put(105,130){$\xrightarrow{\overline{D}_{3,u}=0}$}
   \put(175,135){$1$}
  \put(216,135){$2$}
  \put(257,135){$3$}
 \end{picture}
 }
 \caption{\label{fig:G2hodge}{\it The fiber structure of the $\mathfrak{g}_2^{(1)}$ cubic genus-one fiber and its two degeneration loci.  }} 
 \end{center}
 \end{figure}

  We now turn to a genus-one fibration $X^D$ that admits a $\fg_2^{(1)}$ fiber which shares the same Hodge numbers as the genus-one fibration $X^A$ with twisted 
  $\mathfrak{so}_8^{(3)}$ fiber. The singular genus-one model with an $\fg_2^{(1)}$ fiber is obtained via the generalized Tate-vector 
 \begin{align} 
    n_i=  \left\{2,1,1,0,1,1,0,1,0,0\right\} \,.
\end{align} 
The $\fso_8^{(3)}$ model, is obtained upon a further tuning of the above vanishing orders to $d_2 \rightarrow z \hat{d}_2 \,, \, d_5 \rightarrow z \hat{d}_5 z\, ,\,  d_6 \rightarrow z d_6\, .$
When mapping the respective tuning into the singular Weierstrass model, we obtain at leading orders the Jacobian $J(X^D/B_2)$ 
\begin{align}
 \begin{split}
 f=&\frac13 z^2  (-9 d_{10} d_2 d_4 d_5 + 3 d_{10} d_2^2 d_7 - d_5^2 d_7^2 + 3 d_4 d_5^2 d_9 + 
   d_2 d_5 d_7 d_9 - d_2^2 d_9^2)+ \mathcal{O}(z^3) \, ,  \\ 
 g=&  z^3 P + \mathcal{O}(z^4) \, , \\
 \Delta =& z^6 Q^2 \overline{D}_{3,u} + \mathcal{O}(z^7) \, ,
 \end{split}
 \end{align}
 with $P$ some longer irrelevant polynomial and $Q$ given as
 \begin{align}
 Q=  d_{10} d_2^3 - d_4 d_5^3 + d_2 d_5^2 d_7 - d_2^2 d_5 d_9 \, .
 \end{align}
The resulting singularity is of $I_0^{*,ns}$ type i.e. an $\fg_2^{(1)}$ fiber upon resolution. Note that we also find here the order three monodromy divisor appearing at leading orders  of the discriminant. 
Lets compare this to the fiber structure in the genus-one model $X^D$ after sufficient resolution, which is given as 
 \begin{align}
p=& d_1 f_1^2 g_1 u^3 + d_2 f_1 u^2 v + d_3 f_0 f_1 g_1 u v^2 + d_4 f_0 v^3 + 
 d_5 f_1 u^2 w \nonumber \\ &+ d_6 f_0 f_1 g_1 u v w + d_7 f_0 v^2 w + d_8 f_0 f_1 g_1 u w^2 + 
 d_9 f_0 v w^2 + d_{10} f_0 w^3 \, ,
 \end{align}
   with Stanley-Reisner ideal
   \begin{align}
   \mathcal{SRI}: \{ u f_0, f_0 f_1, u g_1, w v u, w v f_1, w v g_1 \} \, .
   \end{align}
 The intersection picture of the fiber is given in Figure~\ref{fig:G2hodge}. 
 The fibral curves are given as
 \begin{align}
 \begin{split}
 \mathbb{P}^1_{\alpha_0}:& \quad f_0\cap d_1 g_1 + d_2 v + d_5 w \, , \\
 \mathbb{P}^1_{\alpha_1}:& \quad   g_1\cap d_2 f_1 v + d_4 f_0 v^3 + d_5 f_1 w + d_7 f_0 v^2 w +  d_9 f_0 v w^2 + d_{10} f_0 w^3 \, ,
  \\ 
 \mathbb{P}^1_{\alpha_2,i}:&\quad  f_1\cap d_4 v^3 + d_7 v^2 w + d_9 v w^2 + d_{10} w^3    \, .
 \end{split}
 \end{align}
 where $ \mathbb{P}^1_{\alpha_2,i}$ admits $3$ components that are interchanged along the $\overline{D}_{3,u}$ monodromy divisor. We conclude that we find a type $I_0^{*,ns}$ fiber in the genus-one model $X^D$ just as in its Jacobian.
  
 It is important to recall that the affine node $f_1$ in the $\fso_8^{(3)}$ model now hosts the three nodes of the $\fg_2^{(1)}$, while the new divisor $f_0$, yields the usual single affine node. Interestingly, we could have also blow-down $f_1,g_1$ and used $f_0$ to map the geometry into the Jacobian, which would result in the very same singularity type. As the fiber structure is the same in both models, we can use the Jacobian to find reducible fibers and similarly, we can use the genus-one model to compute the 6D discrete $\mathbb{Z}_3$ charges of the $\fg_2$ matter states. To do so, we first note that the discriminant over $z=0$ factorizes into two components in codimension two. The first component is given by  $z=Q=0$  where the fiber singularity enhances to vanishing orders  $(2,3,7)$, which is an $\fso_{10}$ singularity. The second component in the discriminant is the intersection locus with monodromy divisor $\overline{D}_{3,u}$. The later determines the ramification points that fold the covering $\fso_8$ to $\fg_2$ and provide non-localized mater. In this situation, the curves $\mathbb{P}^1_{1,i}$ degenerate to a single curve of degree three as depicted in Figure~\ref{fig:G2hodge}. Apparently, the associated genus-one fibration does not admit multiple fibers and hence we expect the WC$(X^D) $ group to reduce to the Tate-Shavarevich group $\Sha(X^D)$. We can now discuss these two loci in the genus-one model to deduce the respective matter charges.
 
 As we have selected $f_0$ as the affine component and $u=0$ as the reference 3-section, we need to orthogonalize the respective $\mathfrak{u}_1$ generators, which yields the discrete Shioda map
\begin{align}
\sigma(s_0^{(3)})= [u] + (2[f_1]+3 [g_1]) \, ,
\end{align}
which admits the intersections with the fibral divisors as 
\begin{align}
    \sigma(s_0^{(3)}) \cdot \{ f_0 , g_1 , f_1    \} = \{ 3,0,0 \} \, .
\end{align}
We can then impose $Q=0$ and deduce the reducible components when solving for $d_5$. The respective components are summarized in Table~\ref{tab:g2mattersplit}.
\begin{table}[t!]
   \begin{align*}
\begin{array}{|c|l |}\hline 
\text{node}  & \multicolumn{1}{c|}{$Irreducible components over $  Q=0}  \\ \hline
\mathbb{P}_{\alpha_0}: [f_0]&  \mathbb{P}^1_{0 }: f_0\cap (d_1 g_1+d_2 v+d_5 w),    \\ \hline
\mathbb{P}_{\alpha_1}:[g_1] &\mathbb{P}^1_{1,1}:  g_1\cap (d_2 v + d_5 w)\, , \quad \, \, \,   \mathbb{P}^1_{1,2}:   \begin{array}{l}g_1 \cap ( d_{10} d_2^2 f_0 v^2 + d_5^2 d_7 f_0 v^2 - d_2 d_5 d_9 f_0 v^2 \\ + d_5^3 f_1- 
  d_{10} d_2 d_5 f_0 v w + d_5^2 d_9 f_0 v w + d_{10} d_5^2 f_0 w^2)\end{array} \\ \hline
 \mathbb{P}_{\alpha_2,i}:[f_1]& \mathbb{P}^1_{2,1}: f_1\cap(d_2 v + d_5 w) \, \quad \, \, \, \, \, \,     \mathbb{P}^1_{2,\pm }: f_1\cap(v+ (a_1 \pm a_2) w)   \\ \hline  
\end{array}
\end{align*}
\caption{\label{tab:g2mattersplit}{\it 
The codimension two irreducible components of the fiber at $Q=0$ where $\fg_2$ charged matter resides. The intersections are depicted in Figure~\ref{fig:G2hodge}. 
}
}
\end{table}
A graphical representation of the intersections of these irreducible curves is given in Figure~\ref{fig:G2hodge}
and we them to have the shape of an affine $\fso_{10}^{(1)}$ Dynkin diagram, consistent with the Jacobian prediction. The weights of particles, obtained from M2 branes wrapping the curve $\mathbb{P}^1_{1,1}$ are computed as 
\begin{align}
\mathbb{P}^1_{1,1} \cdot (g_1, f_1)_{\sigma(s_0^{(3)}} =  (-1,1)_{-1} \ni \mathbf{7}_{-1} \, .
\end{align}
Hence upon the singular limit, those weights are completed to full $\mathbf{7}_{-1}$-plets with non-trival $\mathfrak{u}_{1,E}$ charge, which lifts to a $\mathbb{Z}_3$ charge in 6D. The non-localized states, obtained from the monodromy divisor on the other hand, are obtained from the $\fso_8$ adjoint and hence correspond to uncharged $\mathbf{7}$-plets. 

We turn now to the computation of the multiplicity of the states described above. To this end, we compute the intersections that lead to localized $\mathbf{7}_1$ plets and the $\mathbf{7}_0$-plets that originate from the monodromy divisor.  
Since we took $f_0$ as our new affine node, we need to change the factorization in the genus-one fibration above. For this we need the classes of the monodromy divisor in the new factorization\footnote{We read this off from $[d_{10}]\sim 2\mathcal{S}_9-\mathcal{S}_7-\mathcal{Z}$,  $[d_7]\sim \mathcal{S}_7-\mathcal{Z}$ and $[d_2]\sim 2 c_1- \mathcal{S}_9$.} geometry given as
\begin{align}
[\overline{D}_{3,u}]\sim 2 \mathcal{S}_9 + 2 \mathcal{S}_7 -4 \mathcal{Z} \, ,
\end{align}
as well as the class of
\begin{align}
[Q]\sim -\mathcal{S}_7 - \mathcal{S}_9 - \mathcal{Z} + 6 c_1 \, .
\end{align}
The number of charged $\mathbf{7}_1$ is given by
\begin{align}
n_{\mathbf{7}_1}=[Q]\cdot \mathcal{Z} =12(1-g)  -\mathcal{S}_7 - \mathcal{S}_9 +5 \mathcal{Z}^2
 \end{align} 
 The neutral $\mathbf{7}$-plets are computed from the monodromy divisor and the Riemann-Hurwitz theorem, which yields  
\begin{align}
n_{\mathbf{7}_0}= 2(g-1) + \frac12 R = 2(g-1) +\mathcal{S}_7 + \mathcal{S}_9 -2 \mathcal{Z}^2 \, .
\end{align}
Taking the sum of the two representations we obtain 
\begin{align}
n_{\mathbf{7}_1} + n_{\mathbf{7}_0} = 10(1-g)+3 \mathcal{Z}^2  \, , \quad n_{\mathbf{14}_0}=g \, ,
\end{align}
which satisfies the 6D anomalies as expected. Note again, that there is a degree of freedom to tune the relative number of charged and uncharged $\mathbf{7}$-plets. 

In the second step we want to relate the $\fso_8^{(3)}$ and $\fg_2^{(1)}$ models. 
In order to do so, we need to make sure that there are no $(4,6,12)$ points after tuning, which requires $d_1= z=0$ to be trivial (see  Section~\ref{ssec:so83Geometry}). 
This condition can be translates into the following intersection of the line bundle classes
\begin{align*}
(\mathcal{S}_7 + \mathcal{S}_9)\cdot \mathcal{Z} = 3 c_1 \cdot \mathcal{Z} \, .
\end{align*}
When using the above constraint and plugging them into the multiplicities of the states we obtain
\begin{align}
\label{eq:multig21}
n_{7_0}=4(1-g) + \mathcal{Z}^2 \, , \qquad n_{7_1}=6(1-g)+2 \mathcal{Z}^2 \, .
\end{align}
We can now compare the multiplicity of states
in the $\fg_2^{(1)}$ cubic genus-one fibration $X^D$ with that of the $\fso_8^{(3)}$ fibration $X^A$ (given in \eqref{eq:so83Spectrum}) we find perfect agreement for the massless $\mathbf{7}_0$ multiplicities.
However, just as in the $\fe_6^{(2)}$ case,
there is a difference in the massive $\mathbf{7}$-plet sector. We remark, that we do not find $\mathbf{7}_2$ states in the $\fg_2^{(1)}$ theory. This can be explained by noting, that the $\mathbf{7}$-plets are real representations. Thus, when conjugating a $\mathbf{7}_2$ state, it becomes a $\mathbf{7}_{-2}$. Recall that our KK tower is normalized with the three-section as $\mathbf{7}_{-2+3n}$. Hence, when shifting the KK-tower to $n\rightarrow n+1$, we find the $\mathbf{7}_2$ states to sit in the same KK-tower as the $\mathbf{7}_1$ states. Therefore the multiplicity given in \eqref{eq:multig21} yields the combined multiplicities of massive $\mathbf{7}_1$ and $\mathbf{7}_2$ states. 

Similar to the other twisted fibrations, we find the number of massive $\mathbf{7}$-plets to differ when engineered in their untwisted variants. In particular both the numbers of  $\mathbf{7}_1$ and $\mathbf{7}_2$ states are each enhanced in the $\fso_8^{(3)}$ model, when compared to those in $\fg_2^{(1)}$. 

Similar to the $\fe_6^{(2)}\rightarrow \fg_2^{(1)}$ case, there exists a chain of geometric transitions from $\fso_8^{(3)}$
to $\fg_2^{(1)}$. For this we first need to choose the 
the usual $\fg_2 \in \fso_8^{(3)}$ limit when shrinking the curves in $g_1,h_1$ that yields massless $\mathbf{7}_0$-plets which can be used as a Higgs resulting in the familiar breaking $\fg_2 \rightarrow \fsu_3$ with the branchings $\mathbf{14} \rightarrow \mathbf{8} \oplus \mathbf{3}\oplus \overline{\mathbf{3}}$ and $\mathbf{7} \rightarrow \mathbf{3} \oplus \overline{\mathbf{3}} \oplus \mathbf{1}$. After the deformation we replace the $\fg_2$ singularity by an $\fsu_3$, which is resolved by $f_0$ and $g_1$ which results in the $\fg_2^{(1)}$ diagram given in Figure~\ref{fig:G2hodge}. As $\fg_2 \rightarrow \fsu_3$ deformations preserve the Hodge numbers of the threefolds, we therefore claim that there exists an $\fg_2^{(1)}$
with exactly the same Hodge numbers. We give an example in the following.
 
\subsubsection*{Toric examples}
In the following we give toric example of a threefold geometry $X^D$ in which the gauge group is supported over a curve of self-intersection $\mathcal{Z}^2=-1$ and genus 0 within a $dP_1$ base. The toric
rays can be found in Table \ref{tab:appd43toric} of Appendix \ref{app:Example3folds}.
The $\fg_2$ spectrum is given as
\begin{align}
    n_{\mathbf{7}_0}=3, \qquad n_{\mathbf{7}_1}=2 \, , 
\end{align}
which indeed cancels the 6D $\fg_2$ anomalies. When 
compared to the 
$\fso_8^{(3)}$ model $X^A$, we find the very same Hodge numbers and a difference only in the non-polynomial complex structure deformations
\begin{align}
    (h^{1,1},h^{2,1}(h^{2,1}_{np}))(X^A)=(5,71(6)) \, , \text{ and } (h^{1,1},h^{2,1}(h^{2,1}_{np}))(X^D)=(5,71(3))  \, .
\end{align}
The three non-polynomial deformations in the $X^D$ geometry can be attributed to the singlet in the three $\mathbf{7}_0$-plets. The very same three $\mathbf{7}_0$-plets are also present in $X^A$ where three additional non-poly deformations are present. Analogously to the $\fe_6^{(2)}$ model, we propose those contributions to the additional singlets from the twisted reduction. In particular, we expect to find the total contribution of non-polynomial deformations 
\begin{align}
    \delta h^{2,1}_{np}= n_{\mathbf{1}_0}+n_{\mathbf{7}_0}+2n_{\mathbf{14}_0}=12(1-g)+ 3\mathcal{Z}^2 +2 g\, ,
\end{align}
 in the $\fso_8^{(3)}$ model $X^A$. Thus, we find three additional contributions, coming from $\mathbf{1}_0$ for the $\mathcal{Z}^2=-1\, , g=0$ case. 
  Those considerations also show, that those non-poly deformations should be absent for $\fso_8^{(2)}$ fibers over curves with $\mathcal{Z}^2=-4$  and
   $g=0$ as constructed in Section~\ref{sec:MoreTwistedTheories}.
   As expected from the 6D anomalies, this fiber can not be Higgsed to $\fg_2^{(1)}$.

\subsection{$\fsu_3^{(2)}$ vs $\mathfrak{su}_2^{(1)}$ genus-one fibers} 
Finally we discuss the relation between 
the twisted fibrations $X^A$ with fibers $\fsu_3^{(2)}$ 
and their untwisted cousins, $X^D$ with fibers 
$\fsu_2^{(1)}$. The later geometry is obtained  by the generalized Tate-vector
\begin{align}
    n_i = \left\{ 0,0,0,0,0,0,0,0,1\right\} \, .
\end{align} 
This theory admits an $\fsu_2^{(1)}$ fiber, both in the Jacobian and the genus-one fibration. The Jacobian $J(X^D/B_2)$ to leading orders in $z$ is given as 
\begin{align}
\label{eq:su2fiber}
\begin{split}
f=&Q^2 + \mathcal{O}(z)\,,  \\
g=&Q^3+Q \mathcal{O}(z)\, , \\
\Delta=&z^2  P  Q^2 +  Q\mathcal{O}(z^3)\, .
\end{split}
\end{align}
With $P$ and $Q$ polynomials in the $d_i$ 
\begin{align}
\label{eq:su2loci} 
Q=& -d_7^2 + 4 d_6 d_8 \, ,\\ 
P =&{\small d_5^2 d_6^4  - d_4 d_5 d_6^3 d_7 + d_3 d_5 d_6^2 d_7^2 - d_2 d_5 d_6 d_7^3 + 
 d_1 d_5 d_7^4 + d_4^2 d_6^3 d_8 - 2 d_3 d_5 d_6^3 d_8 - d_3 d_4 d_6^2 d_7 d_8 } \nonumber \\ & {\small +
 3 d_2 d_5 d_6^2 d_7 d_8 + d_2 d_4 d_6 d_7^2 d_8 - 4 d_1 d_5 d_6 d_7^2 d_8  - 
 d_1 d_4 d_7^3 d_8 + d_3^2 d_6^2 d_8^2 - 2 d_2 d_4 d_6^2 d_8^2 + 
 2 d_1 d_5 d_6^2 d_8^2 } \nonumber  \\& {\small - d_2 d_3 d_6 d_7 d_8^2 + 3 d_1 d_4 d_6 d_7 d_8^2  + 
 d_1 d_3 d_7^2 d_8^2 + d_2^2 d_6 d_8^3 - 2 d_1 d_3 d_6 d_8^3 - d_1 d_2 d_7 d_8^3 + 
 d_1^2 d_8^4} \, . 
\end{align}

The fully resolved genus-one fibration $X^D$ on the other hand is given as 
\begin{align}
p=&d_1 f_0 X^4 + d_2 f_0 X^3 Y + d_3 f_0 X^2 Y^2 + d_4 f_0 X Y^3 + d_5 f_0 Y^4 \nonumber \\ & + 
 d_6 X^2 Z + d_7 X Y Z + d_8 Y^2 Z + d_9 f_1 Z^2 \, ,
\end{align}
with the fibral Stanley-Reisner ideal
\begin{align}
\mathcal{SRI}: \, \{ Y X,  Z f_0 \} \, .
\end{align}
The two fibral curves are given as 
\begin{align}
\begin{split}
\mathbb{P}^1_{\alpha_0}&:  f_1\cap    d_1 f_0 X^4 + d_2 f_0 X^3 Y + d_3 f_0 X^2 Y^2 + d_4 f_0 X Y^3 \\ & \qquad \qquad \qquad  \, + d_5 f_0 Y^4 + 
 d_6 X^2 Z + d_7 X Y Z + d_8 Y^2 Z \, , \\ 
    \mathbb{P}^1_{\alpha_1}&: f_0\cap  d_6 X^2 + d_7 X Y + d_8 Y^2 + d_9 f_1 \, .  
    \end{split}
\end{align}
This results in an intersection picture as given in Figure~\ref{fig:SU2Fiber}:
\begin{figure}[t!]
 \begin{center} 
 \begin{picture}(00, 150)
 \put(-160,40){\includegraphics[scale=1]{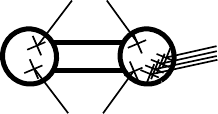} } 
 
  \put(-124,95){$[X]$} 
    \put(-124,35){$[Y]$}
         \put(-55,65){$[Z]$}
               \put(-80,85){$[f_1]$}
               \put(-80,45){$1$}
               \put(-160,45){$1$}
               
                \put(-170,85){$[f_0]$}
                \put(80,0){\includegraphics[scale=1]{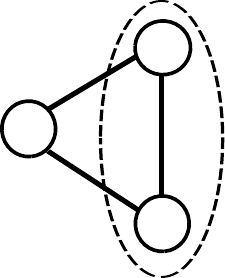} } 
              \put(10,60){  $\xrightarrow{P=0}$}
\put(85,68){$\mathbb{P}^1_{\alpha_1}$}
\put(150,110){$\mathbb{P}^1_{0,1}$}
\put(150,23){$\mathbb{P}^1_{0,2}$}
\put(150,140){$[f_1]$}
 \end{picture}  
 \caption{\label{fig:SU2Fiber}Depiction of the $\mathfrak{su}_2$ fiber in the quartic. Intersections with the fibral curves in the divisors $f_0$ and $f_1$ with respect to the two 2-section classes $X$ and $Y$ are depicted, as well as with the 4-section $Z$.}
 \end{center}
 \end{figure}
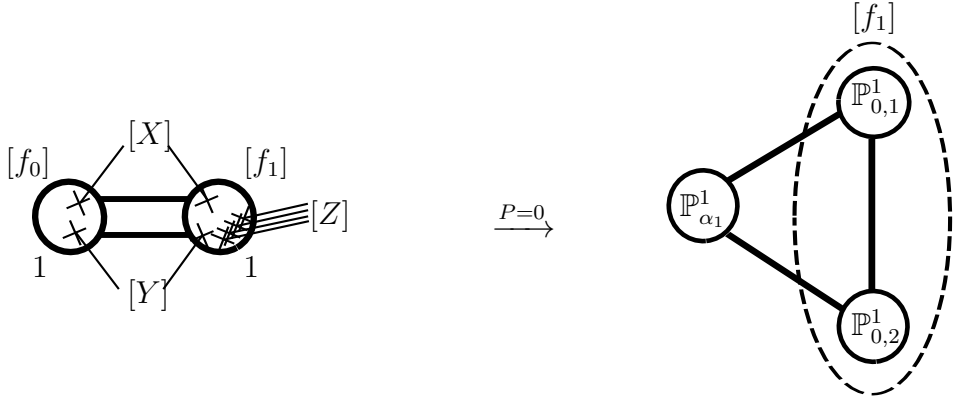 
Hence again, we have to orthogonalize our reference two-section $X$ with respect to the $\fsu_2$ divisor that we want to shrink, that is $f_0$ which yields the discrete Shioda map
\begin{align}
\sigma(X) = [X] + \frac12 [f_0] \, .
\end{align}
Note that the $\fsu_2$ model is precisely the same as the most general quartic fibration which makes our discussion fully analogous to the one in \cite{Klevers:2014bqa} (see page 32).

To discuss the matter loci, it is again sufficient to simply deduce them from the Jacobian $J(X^D/B_2)$ and then investigate them further in the respective genus-one geometry. First there is the discriminant locus $Q=0$ for which we find the vanishing orders $(1,2,3)$ in the Weierstrass model and hence a
type $III$ enhancement where no additional matter resides. Over the codimension two locus $P=0$ instead, we find a type $I_3$ fiber and hence fundamental matter. Over this locus, the fibral curve $\mathbb{P}^1_{\alpha_0}$ splits into $\mathbb{P}^1_{0,1}+\mathbb{P}^1_{0,2}$ which gives rise to  $\mathbf{2}_{\frac12}$-plets. Note that the $\frac12$ charge highlights a non-trivial mixing of $\mathfrak{u}_{1,E}$ with the
$\mathbb{Z}_2$ center of the $\fsu_2$, similar to the $\fe_7$ case in Section~\ref{sec:main_5D_example}, resulting in a $G=\frac{SU(2) \times \mathbb{Z}_4}{\mathbb{Z}_2}$ gauge group in the F-theory lift. 

In order to compute the multiplicities, we use Appendix~\ref{app:LBData} to deduce the classes of the polynomial $P$ which is 
\begin{align}
    [P] \sim [2 d_1 + 4 d_8]\quad  \text{ with } \quad [d_8] \sim c_1 + \mathcal{S}_9 - \mathcal{Z} \, .
\end{align} This results in the multiplicity
\begin{align}
n_{\mathbf{2}_{\frac12}}= \mathcal{Z} \cdot (8 c_1 -2 \mathcal{Z})= 16 (1-g) +6 \mathcal{Z}^2 \, ,
\end{align}
consistent with the 6D gauge anomalies. 

Finally we are in the position to comment on the relation of the $\mathfrak{su}_2^{(1)}$ theory to that of $\fsu_3^{(2)}$. As noted above, the $\fsu_2^{(1)}$ geometry $X^D$ does not posses multiple fibers and admits a consistent 6D F-theory uplift, highlighted by a matter spectrum that solves the 6D anomalies.
The $\fsu_3^{(2)}$ geometry $X^A$ on the other hand admits two more massive $\mathbf{2}_\frac12$-plets. We expect again, that there exists a geometric transition between the two threefolds, that preserves gauge algebra rank and number of flat direction, i.e. the exact Hodge numbers. The exact type of conifold transition between the twisted and the untwisted theory is more obscure than in the other cases as there is not much room for a rank preserving Higgsing in the $\fsu_2 \times \mathfrak{u}_{1,E}$ theory, apart from using adjoints. Similarly to the other examples discussed in this section, we expect such fields to exist and return to this question in future works. 

\subsubsection*{Toric example}
We exemplify the general considerations with a toric example. The details of the toric rays, from which the threefolds $X^A$ and $X^D$ are constructed are summarized  Appendix~\ref{app:Example3folds}. The threefold $X^D$ with $\fsu_2^{(1)}$ fiber is engineered over a $\mathcal{Z}^2=+1$ curve with $g=0$ in a $\mathbb{P}^2$ base. 
The massless $\fsu_2\in \fsu_2^{(1)}$ charged spectrum is then given by the multiplicities
\begin{align}
    n_{\mathbf{3}_0} = 0 \, , \qquad n_{\mathbf{2}_\frac12 }= 22 \, ,
\end{align}
consistent with the 6D anomalies. 
The corresponding $\fsu_3^{(2)}$ model $X^A$ on the other hand, admits 24 massive doublets states but leads to a threefold with exactly the same Hodge numbers, given as  
\begin{align}
    (h^{1,1},h^{2,1}))=(3,107) \, .
\end{align} 
Besides the fiber structure, another difference shows up in the number of non-polynomial complex structure deformations:
\begin{align}
    h^{2,1}_{np}(X^C)=12\,, \quad  h^{2,1}_{np}(X^D)=0  \, .
\end{align} 
As before, we use the twisted reduction of the 6D $\fsu_3$ theory, to give   a prediction of such non-poly deformations.
Those contributions come from singlets, and adjoint hypermultiplet representations as discussed in Section~\ref{ssec:su32Geometry} 
that sum up to 
\begin{align}
    \delta h^{2,1}_{np}= n_{\mathbf{1}_0}+n_{\mathbf{3}_0}= 9(1-g)+3\mathcal{Z}^2 + g \, .
\end{align}
Thus, for our toric example we expect to have 12 singlet state contributions in $X^A$. Similarly we find that the $\fsu_3^{(2)}$ fiber over a $-3$ curve has a trivial spectrum. Therefore, there is no transition to an $\fsu_2^{(1)}$ consistent with the fact that there are no non-polynomial complex structure deformations.

\section{Summary and Outlook}\label{sec:summary}
In this work we have initiated a detailed
exploration of the geometry of genus-one fibrations with twisted algebras and their F/M-theory physics. We have focused explicitly on the cases of  $\fe_6^{(2)},\fso_8^{(3)}$ and $\fsu_3^{(2)}$ realized over an arbitrary 2D base. For these cases the geometry exhibits multiple fibers over smooth points in the base and hence a non-trivial Weil-Ch$\hat{\text{a}}$talet group is a prerequisite for these twisted algebras to exist. 
We compute the multiplicities of the lightest  BPS states charged under the $\ff_4, \fg_2$ and $\fsu_2$ finite sub-algebras  purely from geometry and find agreement with prior results on twisted field theory reductions of 6D gauge algebras $\fe_6, \fso_8$ and $\fsu_3$ respectively. To gain additional insight into these M-theory backgrounds, we consider 5D geometric transitions from $\fe_7^{(1)}, \ff_4^{(1)}$ and $\fg_2^{(1)}$ respectively, to the twisted algebras.

While we do not fully determine the 6D background that leads to these twisted M-theory compactifications upon circle reduction, we do identify a range of features and symmetries that it must possess. In particular we give the explicit Jacobian fibration associated to each genus-one fibration. In each case, the fiber structure we find in the Jacobian differs from that found in the genus-one geometries and from what would be expected from a typical, field-theoretic twisted reduction. Instead of 
$\fe_6, \fso_8$ and $\fsu_3$ we find the generic geometric cover of the twisted algebras, that is $\fe_7, \ff_4$ and $\fg_2$ respectively. As a result we find different dimensionality in the K\"ahler moduli spaces associated to the genus-one fibrations and their Jacobians. We further discuss several applications of the geometry of twisted algebras in SUGRAs, LSTs and SCFTs. In particular we discuss twisted T-dualities and  match the generalized symmetries, following the proposal in \cite{DelZotto:2020sop}, to provide another cross-check of our construction.

The connection between untwisted and twisted algebras with the same finite sub-algebra are also investigated. We find evidence of (chains of) geometric transitions among them. Surprisingly, the connected geometries admit the very same Hodge numbers, but differ not only in the fibral curve structure but also in the number of non-polynomial complex structure deformations of the CY hypersurface equation. We propose a physics interpretation for those non-polynomial deformations and further use the twisted field theory reduction as a way to understand them.  

The results outlined here open up several exciting directions that deserve further exploration in the future. In particular, limits of the geometries in consideration here connect with several topics within the Swampland program \cite{Ooguri:2006in} (see \cite{Palti:2019pca,vanBeest:2021lhn,Agmon:2022thq} for recent reviews) such as the weak gravity \cite{Arkani-Hamed:2006emk} and
emerging string conjecture \cite{Lee:2019wij}.
First, we find that a twisted compactification in a SUGRA theory does not only require a compatible gauge algebra but also a discrete zero-form gauge symmetry to embed the twist into. Secondly, these twisted dimensional reductions make 6D discretely charged states generally massive, and gives them a fractional 5D $U(1)$KK charge, which hence makes them relevant in the context of the (sub-lattice) weak gravity conjecture (see e.g. \cite{Harlow:2022gzl} and reference therein). Finally, twisted compactifications display an interesting feature in that the decompactification limit not only results in an extra dimension but also leads to a gauge algebra enhancement. 

This work has also raised a number of interesting open questions.
First there is the gauge symmetry of the Jacobian fibration, which is associated with the untwisted circle reduced theory. There the gauge algebra seems to always be enhanced compared to the purely field theoretic expectation (e.g. an $\mathfrak{e_7}$ symmetry instead of $\mathfrak{e}_6$ for the geometries of Section \ref{sec:main_5D_example}, for example). This enhancement is also directly related to puzzles regarding the twisted reduction from 6D to 5D as outlined in Section \ref{sec:Circle_reduc_spec}. One potential explanation might involve the non-polynomial complex structure deformations present in the twisted genus one fibrations studied in this work. It is possible that in the presence of such degrees of freedom the mapping to the Jacobian geometry might be modified in a way that fixes some complex structure moduli, potentially breaking the naive symmetry group to that expected from ordinary gauge theory reductions. 

As a further intriguing observation, in our geometric analysis the multi-section monodromy divisors played a key role in describing the twisting of the fibers and also appears to be linked to co-dimension 2 structure in the Jacobian fibration. This co-dimension 2 structure is crucial in the cancellation of the 6D discrete gauge anomalies \cite{Dierigl:2022zll} (which were beyond the scope of the present work to explore). This observations hints at a non-trivial interplay between the 6D discrete gauge anomalies and the existence of twisted dimensional reductions. Moreover, the explicit form of the twisted fibrations and their Jacobians here would also in principle allow for the computation of twisted/twined elliptic genera. It would be interesting to do this and compare with the results of \cite{Lee:2022uiq} in non-compact scenarios as well as  for those for twisted T-dualities \cite{DelZotto:2020sop, Bhardwaj:2022ekc,Braun:2021lzt} and 5D SCFTs \cite{Bhardwaj:2019fzv}.
 
Finally it should be noted that fibrations exhibiting twisted algebras are not yet fully mathematically explored. 
While we have discussed only three different types of twisted fibers, a full Kodaira/Tate type of classification of singular genus-one fiberd threefolds would be desirable. This could in particular include series such as 
 $\fsu_{n}^{(2)}$ or $\fso^{(2)}_{k}$. 
As those are only degree two twisted algebras, one might succeed in constructing these in relatively simple 2-section models such as the quartic. This is linked of course, to the question of whether all possible n-section geometries can be classified. Thus far, some models for $n<6$ \cite{Schimannek:2019ijf} have been constructed and analyzed explicitly, with an expectation that $n$ should not be much larger than 6 \cite{Lee:2022uiq}. While such high multi-section degrees might not be directly required to explore the unexplored twisted algebra series', they are still relevant 
in cases where $\fsu_n$ gauge algebras are twisted to \textit{nothing} as e.g. explored in \cite{Oehlmann:2016wsb,Baume:2017hxm,Oehlmann:2019ohh,Anderson:2019kmx}. We hope to return to some of these questions in future work.

\section*{Acknowledgements} 
We thank Markus Dierigl, Zhihao Duan, Jonathan Mboyo Esole, Antonella Grassi,  David Morrison, Kimyeong Lee, Nikhil Raghuram, Xin Wang and in particular Thorsten Schimannek for illuminating discussions. Furthermore we thank Jie Gu, Ling Lin, Guglielmo Lockhart, Marcus Sperling and Michele Del Zotto for related discussions.
  L.A. and J.G. are supported by NSF grant PHY-2014086.  P.O. has received funding from  the NSF CAREER grant PHY-1848089 and startup funding from Northeastern University by Fabian Ruehle.

 \appendix
 
 \newpage
\section{Toric Data for Threefold geometries}
 \label{app:Example3folds}
 In this appendix we present concrete threefold examples and give their concrete spectra together with their Hodge numbers. The associated vertices of the toric variety are listed below.  
 \subsection{$\fsu_3^{(2)}$ example geometries}
 \begin{table}[h!]
    \begin{center}
   \begin{tabular}{|c|c|c|c|c|} \hline
    \begin{tabular}{c}Base with \\ $(g,\mathcal{Z}^2, \mathcal{S}_9 \cdot \mathcal{Z})$ \end{tabular} & 5D  Gauge Algebra &$ h^{1,1} $& $h^{2,1}$ & Reps    \\ \hline  
  \multirow{5}{*}{$\begin{array}{c}\mathbb{P}^2 \text{ with}\\ (0,1,1) \end{array}$}& $\fg_2 \times \mathfrak{u}_{1,F} \times \mathfrak{u}_{1,kk}$   &5 & 89 & $ \begin{array}{c} 5\times \mathbf{7}_{1}    \\ 8 \times \mathbf{7}_{0}  \end{array}$ %&  $10 \times 1_2$ 
  \\ \cline{2-5} 
  & $\fg_2 \times \mathfrak{u}_{1,E} $&  4 & 98 & $ \begin{array}{c} 5\times \mathbf{7}_{1}    \\ 8 \times \mathbf{7}_{0}  \end{array}$   \\ \cline{2-5} 
    & $ \fsu_2 \times \mathfrak{u}_{1,E} \in  \fsu_3^{(2) }$&  3 & 107 & 24 $\times \mathbf{2}_{\frac12}$    \\ \cline{2-5} 
   & $\fsu_2 \times \mathfrak{u}_{1,E} $  & 3 & 107 &  22 $\times \mathbf{2}_{\frac12}$  \\ \cline{2-5} 
     & $\fsu_2 \times \mathfrak{u}_{1,F} \times \mathfrak{u}_{1,kk}$     &  4 &93 & 22 $\times \mathbf{2}_{\frac12}$   \\ \hline  \hline 
 \multirow{5}{*}{$\begin{array}{c}\mathbb{P}^2_{1,1,2} \text{ with}\\ (0,-2,-2) \end{array}$
  }& $\fg_2 \times \mathfrak{u}_{1,F} \times \mathfrak{u}_{1,kk}$ &  6 & 96 & $\begin{array}{c} 2\times \mathbf{7}_{1}    \\ 2 \times \mathbf{7}_{0}  \end{array}$   \\ \cline{2-5}
  
 & $\fg_2\times \mathfrak{u}_{1,E} $&  5 &109 & $\begin{array}{c} 2 \times \mathbf{7}_{1}   \\ 2 \times \mathbf{7}_{0}  \end{array}$  \\ \cline{2-5} 
   & $ \fsu_2 \times \mathfrak{u}_{1,E} \in \fsu_3^{(2)} $&   4 & 112 &  6 $\times \mathbf{2}_{\frac12}$    \\ \cline{2-5} 
  & $\fsu_2 \times \mathfrak{u}_{1,E} $     &  4 & 112 & 4 $\times \mathbf{2}_{\frac12}$  \\  \cline{2-5} 
    & $\fsu_2 \times \mathfrak{u}_{1,F} \times \mathfrak{u}_{1,E} $     $ $&  5 & 97 & 10 $\times \mathbf{2}_{\frac12}$  \\ \hline 
 \end{tabular}
     \caption{\label{tab:su32Summary}\textit{A chain of 5D theories, obtained from genus-one fibrations over a $\mathbb{P}^2$ and a an $\mathbb{F}_2$ base with gauge algebras over a $\mathcal{Z}^2=0$ curve of genus $g=0$ and their light BPS states.}}
   \end{center} 
    \end{table}
   \begin{table}[h!]
 
   \begin{center}
   \begin{tabular}{ccccc}
  \begin{tabular}{|c|c|}
   \multicolumn{2}{c}{Generic Fibers}\\ \hline
  X & (-1,1,0,0)\\
  Y & (-1,-1,0,0) \\
  Z & (1,0,0,0) \\ \hline
  U & (0,1,0,0) \\ \hline 
  \end{tabular}
  &
    \begin{tabular}{|c|c|} 
    \multicolumn{2}{c}{$\fg_2^{(1)}$ Fiber}\\ \hline
 $ f_1$ & (1,0,1,0)\\
 $ f_0 $& (0,0,1,0) \\
$  g_1$ & (1,0,2,0) \\ 
 $ f_2 $ & (0,1,1,0) \\ \hline 
  \end{tabular}
  
    &
    \begin{tabular}{|c|c|} 
    \multicolumn{2}{c}{$\fsu_3^{(2)}$ Fiber}\\ \hline
 $ f_1$ & (1,0,1,0)\\
 $ f_0$ & (0,0,1,0) \\
  $g_1$ & (1,0,2,0) \\ \hline 
  \end{tabular}
  & 
    \begin{tabular}{|c|c|} 
    \multicolumn{2}{c}{$\fsu_2^{(1)}$ Fiber}\\ \hline
 $ f_1$ & (1,0,1,0)\\
 $ f_0$ & (0,0,1,0) \\ \hline
  \end{tabular}
  
    & 
    \begin{tabular}{|c|c|} 
        \multicolumn{2}{c}{$\mathbb{P}^2$  Base 1}\\ \hline
     $x_0 $ & (-3,0, -1,-1)\\
 $x_1$ & (0, 0, 0, 1) \\   \hline 
    \multicolumn{2}{c}{$\mathbb{P}^2_{112}$ Base 2}\\ \hline
 $x $ & (0,0, 1,1)\\
 $y$ & (0, 0, 1, -1) \\  
  $z$ & (-2,0,-1, 0) \\ \hline 
\multicolumn{2}{c}{ } \\  
  \end{tabular}
  
  \end{tabular}
  \caption{\label{tab:appa22toric}\textit{The collection of toric rays that can be combined to the five threefolds summarized in Table~\ref{tab:su32Summary}: First 
  pick the base and elliptic fibers is picked from the leftmost column. Second a non-trivial fiber top is added from either of the other columns. 
  }}
   \end{center}
  \end{table}

 \subsection{$\fe_6^{(2)}$ example geometries}
 \label{app:E62Example}
 \begin{table}[h!]
    \begin{center}
   \begin{tabular}{ |c |c|c|c|} \hline 5D  Gauge Algebra & $h^{1,1}$ & $h^{2,1}$ & Reps %& Singlets
    \\ \hline  
    $\fe_7 \times \mathfrak{u}_{1,F} \times \mathfrak{u}_{1,KK}$ &11 & 59 & 4  $ \times \mathbf{56}_{1/2}$ %& 24 $\times \mathbf{1}_2 + 70 \times \mathbf{1}_1$    
    \\ \hline
 $\fe_7 \times \mathfrak{u}_{1,E}$&  10 & 82 & 4  $ \times \mathbf{56}_{1/2}$ %&  70$ \times \mathbf{1}_1$      
 \\ \hline
% $E_6^{(1)}$ & 9 & 85 & $\begin{array}{c} 2\times \mathbf{27}_{1/3}?   \\ 4 \times \mathbf{27}_{-2/3}? \end{array} $ & $74 \times \mathbf{1}_1$   \\ \hline
    $\ff_4 \times \mathfrak{u}_{1,E} $&  7 & 95 & $\begin{array}{c} 2\times\mathbf{ 26}_{1}   \\ 3  \times \mathbf{26}_{0} \end{array} $ %& 76$\times \mathbf{1}_1$ 
    \\ \hline
  $\ff_4 \times \mathfrak{u}_{1,E} \in \fe_6^{(2)}$  & 7 & 95 & $\begin{array}{c} 2\times\mathbf{26}_{1}    \\ 3  \times \mathbf{26}_{0} \\ 2\times \mathbf{1}\times 3 \end{array} $ %& -
   \\ \hline  
 
 \end{tabular}
   \caption{\label{tab:e62Summary}\textit{A chain of 5D theories, obtained from genus-one fibrations over a $\mathbb{F}_0$ base with gauge algebras over a $\mathcal{Z}^2=0$ curve of genus $g=0$ and their light BPS states.}}
   \end{center} 
    \end{table}
    
    \begin{table}[h!]
\begin{center}
\begin{tabular}{cccc}
$
\begin{array}{|c|c|}
\multicolumn{2}{c}{\text{ Generic Fiber }} \\ \hline 
X&(-1,1,0,0)\\
Y&(-1,-1,0,0)\\
Z&(1,0,0,0)\\ \hline
e_1&(0,1,0,0) \\ \hline
\multicolumn{2}{c}{ } \\  
\multicolumn{2}{c}{\text{ $\mathbb{F}_0$ Base }} \\ \hline 

x_1 & (1,0,-1,0) \\
y_0 & (0, 0, 0, 1) \\
y_1 & (-2,-2,0,-1)\\ \hline
\end{array}$ &$
\begin{array}{|c|c|}
\multicolumn{2}{c}{\text{$\fe_7$ Fiber }} \\ \hline  
f_2&(-3,0,1,0) \\
  f_4& (-2,0,1,0)\\ 
g_1& (-5,-2,2,0)\\
g_2  & (-5,-1,2,0)\\
 g_3&(-4,-1,2,0) \\
h_1 &(-7,-2,3,0)\\
 h_2 & (-6,-2,3,0)\\
 k_1 & (-9,-3,4,0) \\ \hline
\end{array} $&$
\begin{array}{|c|c|}
\multicolumn{2}{c}{\text{$\fe_6^{(2)}$ Fiber }} \\ \hline    
f_2&(-3,0,1,0) \\
g_1& (-5,-2,2,0)\\
g_2  & (-5,-1,2,0)\\
h_1 &(-7,-2,3,0)\\
 k_1 & (-9,-3,4,0) \\ \hline
\end{array}$
&$
\begin{array}{|c|c|}
\multicolumn{2}{c}{\text{$\ff_4 $ Fiber }} \\ \hline   
f_2&(-3,0,1,0) \\
f_3 & (-2,-1,1,0)\\ 
g_1& (-5,-2,2,0)\\
g_2  & (-5,-1,2,0)\\ 
h_1 &(-7,-2,3,0)\\ \hline
\end{array}$
\end{tabular}
\caption{\label{tab:appe62toric}\textit{The collection of toric rays that can be combined to the five threefolds summarized in Table~\ref{tab:e62Summary}: First 
  pick the base and elliptic fibers is picked from the leftmost column. Second a non-trivial fiber top is added from either of the other columns. 
  }}
   \end{center}
  \end{table}
 \newpage
  \subsection{ $\fso_8^{(3)}$ example geometries}
   \begin{table}[h!]
   \begin{center}
   \begin{tabular}{|c|c|c|c|} \hline
   5D  Gauge Algebra& $ h^{1,1} $& $h^{2,1}$ & Reps %& charged singlets
      \\ \hline   $\fe_6 \times  \mathfrak{u}_{1,F} \times \mathfrak{u}_{1,kk}  $& 9 & 60 & $ \begin{array}{c} 2\times \mathbf{27}_{1}    \\ 3 \times \mathbf{27}_{0}   \end{array}$ %&
     \\ \hline
           $\ff_4 \times  \mathfrak{u}_{1,F} \times \mathfrak{u}_{1,F}  $&  8 & 62 & $ \begin{array}{c} 2\times \mathbf{26}_{1}    \\ 2 \times \mathbf{27}_{0}   \end{array}$ %&
           \\ \hline
                    $\ff_4   \times \mathfrak{u}_{1,E}  $&  7 & 67 & $ \begin{array}{c} 2\times \mathbf{26}_{1}    \\ 2 \times \mathbf{27}_{0}   \end{array}$ %&
                    \\ \hline
                                     $\fg_2 \times \mathfrak{u}_{1,E} \in \fso_8^{(3)}  $&  5 & 71 & $ \begin{array}{c} 3\times \mathbf{7}_{1} \\ 3 \times \mathbf{7}_{2}   \\ 3 \times \mathbf{7}_{0}    \end{array}$ %&\
                                    \\ \hline 
  $\fg_2 \times \mathfrak{u}_{1,E}$     &5 & 71 & $\begin{array}{c} 2\times \mathbf{7}_{1} \\ 2 \times \mathbf{7}_{2}   \\ 3 \times \mathbf{7}_{0}    \end{array}$ %&
      \\ \hline   
 \end{tabular}
    
   \end{center}
   \caption{\label{tab:s083Summary}\textit{A chain of 5D theories, obtained from genus-one fibrations over a $dP_1$ base with gauge algebras over a $\mathcal{Z}^2=-1$ curve of genus $g=0$ and their light BPS states.
   }}
   \end{table}
   
   \begin{table}[h!]
    \begin{center}
   \begin{tabular}{cc cc}
  \begin{tabular}{|c|c|}
   \multicolumn{2}{c}{Generic Fibers}\\ \hline
  w & (-1,0,0,0)\\
  v & (0, 1,0,0) \\
  u & (1,-1,0,0) \\ \hline  
 $ e_1$ &(1,0,0,0)\\ \hline
  \multicolumn{2}{c}{}\\
  \multicolumn{2}{c}{dP$_1$ Base}\\ \hline
     $x_0 $ & (0,0, -1,0)\\
 $x_1$ & (-1,  2, 1, -1) \\  
  $x_2$ & (0,0,0, 1) \\ \hline  
  \end{tabular}
  &
    \begin{tabular}{|c|c|} 
    \multicolumn{2}{c}{$\fe_6^{(1)}$ Fiber}\\ \hline
 $ f_0$ & (-1,1,1,0)\\
 $ f_1 $& (0,0,1,0) \\
$  f_2$ & (-1,0,1,0) \\ 
 $ f_3 $ & (0,1,1,0) \\ 
  $ g_1 $ & (-1,1,2,0) \\ 
   $ g_2  $ & (-2,1,2,0) \\ 
    $ g_3 $ & (-1,2,2,0) \\ 
   $ h_1 $ & (-2,2,3,0) \\ \hline
  \end{tabular}
  
  &
      \begin{tabular}{|c|c|} 
    \multicolumn{2}{c}{$\ff_4^{(1)}$ Fiber}\\ \hline
 $ f_0$ & (-1,1,1,0)\\
 $ f_1 $& (0,0,1,0) \\ 
 $ f_2 $ & (0,1,1,0) \\ 
  $ g_1 $ & (-1,1,2,0) \\  
    $ g_2 $ & (-1,2,2,0) \\ 
   $ h_1 $ & (-2,2,3,0) \\ \hline
  \end{tabular}
    &
    \begin{tabular}{|c|c|} 
    \multicolumn{2}{c}{$\fso_8^{(3)}$ Fiber}\\ \hline
 $ f_1$ & (-1,1,1,0)\\
 $ f_0$ & (0,0,1,0) \\
  $g_1$ & (-1,1,2,0) \\  
    $h_1$ & (-2,2,3,0) \\ \hline 
\multicolumn{2}{c}{ }\\    
       \multicolumn{2}{c}{$\fg_2^{(1)}$ Fiber}\\ \hline
 $ f_1$ & (-1,1,1,0)\\
 $ f_0$ & (0,0,1,0) \\
  $g_1$ & (-1,1,2,0) \\  \hline   
  \end{tabular}  
   % & 
  \end{tabular}
  \caption{\label{tab:appd43toric}
\textit{The collection of toric rays that can be combined to the five threefolds summarized in Table~\ref{tab:s083Summary}: First 
  pick the base and elliptic fibers is picked from the leftmost column. Second a non-trivial fiber top is added from either of the other columns. 
  }}
   \end{center}
  \end{table}
  
%%%%%%%%%%%%%%%%%%%%%%%%%%%%%%%%%%%%%%%

\newpage
\section{Line Bundle Data for Genus-One fibrations}
\label{app:LBData}
The various $\mathbb{F}_2$ and $\mathbb{P}_2$ ambient fiber types imply that the coordinates are are certain projective bundles whose data we will list here. This data is important to fix the base divisor classes of the generalized Tate-coefficients that is used throughout this work.
First the Quartic $\mathbb{P}^2_{1,1,2}$ coordinates transform as given in Table~\ref{tab:quarticSections}.
\begin{table}[h!]
\begin{center}
\begin{tabular}{cc}
\begin{tabular}{c|c}
Section & Bundle\\ \hline
X& $H-E_1 +\mathcal{S}_9 - c_1$\\
Y& $H-E_1 - \mathcal{S}_7+ \mathcal{S}_9  $ \\
Z&  $2H-E_1 + \mathcal{S}_9 - c_1 $\\
$e_1$&  $E_1$ 
\end{tabular}
&
\begin{tabular}{c|c}
Section & Divisor Class  \\ \hline
$s_1$    &$3 c_1 - \mathcal{S}_7 - \mathcal{S}_9 $ \\
$s_2$ & $2c_1 -\mathcal{S}_9$\\
$s_3$ & $c_1 + \mathcal{S}_7 - \mathcal{S}_9$ \\ 
$s_4$ & $2 \mathcal{S}_7 - \mathcal{S}_9$ \\ 
$s_5$ & $-c_1 + 3 \mathcal{S}_7 - \mathcal{S}_9 $\\
$s_6 $& $2c_1 - \mathcal{S}_7$ \\ 
$s_7$ & $c_1 $ \\
$s_8$ & $\mathcal{S}_7$\\
$s_9$ & $c_1 - \mathcal{S}_7 + \mathcal{S}_9$
\end{tabular}
\end{tabular}
\caption{\textit{\label{tab:quarticSections} Summary of the line bundle classes of a generic quartic fibration given in \eqref{eq:GenericQuartic}. Throughout most of this work, we fix $\mathcal{S}_7 \sim c_1+\mathcal{S}_9$ such that $s_9$ is just a constant.
}}
\end{center}
\end{table} 
The second model relevant for this work is the cubic fibration with a three-section, which admits the bundle data as summarized in Table~\ref{tab:cubicBundles}.
\begin{table}[h!]
\begin{center}
\begin{tabular}{cc}
\begin{tabular}{c|c}
Section & Bundle\\ \hline
$u$& $H +\mathcal{S}_9 - c_1$ \\
$v$& $H  - \mathcal{S}_7+ \mathcal{S}_9  $  \\
$w$& $H  $ 
\end{tabular}
&
\begin{tabular}{c|c}
Section & Divisor Class  \\ \hline
$s_1$    &$3 c_1 - \mathcal{S}_7 - \mathcal{S}_9 $ \\
$s_2$ & $2c_1 -\mathcal{S}_9$\\
$s_3$ & $c_1 + \mathcal{S}_7 - \mathcal{S}_9$ \\ 
$s_4$ & $2 \mathcal{S}_7 - \mathcal{S}_9$ \\ 
$s_5$ & $2 c_1 - \mathcal{S}_7  $\\
$s_6 $& $c_1  $ \\ 
$s_7$ & $\mathcal{S}_7$ \\
$s_8$ & $c_1 + \mathcal{S}_9+ \mathcal{S}_7$\\
$s_9$ & $  \mathcal{S}_9$\\
$s_{10}$ & $  2\mathcal{S}_9-\mathcal{S}_7$
\end{tabular}
\end{tabular}
\end{center}
\caption{\textit{\label{tab:cubicBundles} Summary of the line bundle classes of a generic cubic fibration given in \eqref{eq:singcubic}. 
}}
\end{table}

\bibliographystyle{ytphys}
\bibliography{refs.bib}

\end{document}